\documentclass[compsoc,conference,a4paper,10pt,times]{IEEEtran}
\IEEEoverridecommandlockouts
\usepackage{cite}
\usepackage[ruled,vlined]{algorithm2e}

\usepackage{amsmath,amssymb,amsfonts}
\usepackage{graphicx}
\usepackage{textcomp}
\usepackage{xcolor}
\usepackage{url}   
\usepackage{booktabs}
\usepackage{mdframed}
\usepackage{tabularx}
\usepackage{multirow} 
\usepackage{subcaption}
\usepackage{ulem}
\usepackage{pifont}
\usepackage{amsthm}

\newtheorem{definition}{Definition}

\newcommand{\JIE}[1]{{{\textcolor{black}{\textbf{JIE:}}}{\textcolor{red}{\textbf{#1}}}}}
\newcommand{\WW}[1]{{{\textcolor{black}{\textbf{WW:}}}{\textcolor{purple}{\textbf{#1}}}}}

\newcommand{\targetdata}{\mathcal{D}}
\newcommand{\shadowdata}{\mathcal{D}_s}

\newcommand{\targetmodeloriginal}{f_{\theta}}
\newcommand{\targetmodelunlearned}{f_{\theta^-}}
\newcommand{\simtargetmodeloriginal}{f}
\newcommand{\simtargetmodelunlearned}{f^-}
\newcommand{\shadowmodeloriginal}{f_s}
\newcommand{\shadowmodelunlearned}{f^-_s}

\newcommand{\targetdataTrain}{\targetdata^{train}}
\newcommand{\learningalgo}{A}
\newcommand{\unlearnalgo}{U}
\newcommand{\para}{\theta}
\newcommand{\unlearnpara}{\theta^-}
\newcommand{\attack}{\mathcal{C}}

\newcommand{\nop}[1]{}

\newcommand{\targetdataforget}{\targetdata^F}
\newcommand{\targetdataretain}{\targetdata^R}
\newcommand{\targetdataunseen}{\targetdata^U}
\newcommand{\shadowdataForget}{\shadowdata^F}
\newcommand{\shadowdataRetain}{\shadowdata^R}
\newcommand{\shadowdataUnseen}{\shadowdata^U}

\newcommand{\system}{\textsf{TC-UMIA}}

\newcommand{\reducedstrut}{\vrule width 0pt height .9\ht\strutbox depth .9\dp\strutbox\relax}
\newcommand{\gray}[1]{%
  \begingroup
  \setlength{\fboxsep}{0pt}%
  \colorbox{lightgray}{\reducedstrut#1\/}%
  \endgroup
}

\newcommand{\white}[1]{%
  \begingroup
  \setlength{\fboxsep}{0pt}%
  \colorbox{white}%
  \endgroup
}

\newenvironment{ditemize}{%
\begin{list}{{\bf $\bullet$}}{%
\setlength{\itemsep}{0pt}\setlength{\rightmargin}{0pt}%
\setlength{\leftmargin}{1.2em}\setlength{\parsep}{0pt}}}{
\end{list}}

\makeatletter
\newenvironment{game}[1][htb]{%
  \renewcommand{\ALG@name}{Game}
  \begin{algorithm}[#1]%
  }{\end{algorithm}}
\makeatother

\usepackage[colorlinks=true,urlcolor=black]{hyperref}
\def\BibTeX{{\rm B\kern-.05em{\sc i\kern-.025em b}\kern-.08em
    T\kern-.1667em\lower.7ex\hbox{E}\kern-.125emX}}
\begin{document}

\title{Revisiting Privacy Leakage in Machine Unlearning: Membership Inference Beyond the Forgotten Set}

\author{
Jie Fu$^{\dagger}$, Nima Naderloui$^{\ddagger}$, Da Zhong$^{\S}$, Yuan Hong$^{\ddagger}$, Wendy Hui Wang$^{\dagger}$\\
$^{\dagger}$Stevens Institute of Technology, Hoboken, NJ, USA\\
$^{\ddagger}$University of Connecticut, Storrs, CT, USA\\
$^{\S}$Meta Inc., CA, USA\\
$^{\dagger}$\{jfu13, hwang4\}@stevens.edu, $^{\ddagger}$\{nima.naderloui, yuan.hong\}@uconn.edu, $^{\S}$dazhong0516@gmail.com
}

\maketitle

\begin{abstract} 
Machine unlearning (MU) has emerged as a key mechanism for ensuring data privacy and regulatory compliance by enabling models to forget specific training samples. However, recent studies have shown that the removal of data can inadvertently introduce privacy leakages to the \textit{retain set}, i.e., data that remain in the model after unlearning. 
In this paper, we extend the scope of privacy analysis in unlearning to the often-overlooked retained data. 
We introduce \system, the first tri-class {\itshape unlearning membership inference attack}. \system\ is a population-level inference framework that leverages model predictions before and after unlearning to distinguish among the \textit{forget}, \textit{retain}, and \textit{unseen} set. Extensive experiments on five state-of-the-art unlearning algorithms and six real-world datasets demonstrate that:  (i) unlearning can introduce additional privacy risks to the retain set, making it more susceptible to membership inference attacks; (ii) \system\ is effective across a wide range of model architectures, datasets,  and MU  approaches.  
Beyond launching the attack, we rigorously evaluate three defense mechanisms, namely \textit{label-only outputs}, \textit{dropout}, and \textit{differential privacy}, to mitigate the privacy risks posed by \system. Our results reveal a fundamental trade-off between privacy protection and model accuracy, with the dropout approach offering the most favorable balance.\footnote{
Our code is available at: \url{https://github.com/JeffffffFu/TC-UMIA}.} 

\end{abstract}



\section{Introduction}\label{sec-Introduction}


As machine learning (ML) models become increasingly integrated into real-world applications ranging from recommendation systems to personalized healthcare, they often rely on large-scale user data. However, with the growing emphasis on data privacy and user control, there is a rising demand for allowing individuals to retract their data after it has been used to train a model. 
To address this need, data regulations such as the General Data Protection Regulation (GDPR) \cite{gdpr} and the California Consumer Privacy Act (CCPA) \cite{CCPA}  have
been established in various countries and regions. 

In response to these regulations, \textit{machine unlearning} (MU) has emerged as a technical paradigm to ensure the ``right to be forgotten'' (RTBF) to users. Efficient MU methods have been developed for various types of machine learning models, including Deep Neural Networks \cite{sekhari2021remember,kurmanji2024towards,thudi2022unrolling}, Graph Neural Networks \cite{wu2023certified,wu2023gif}, recommend systems \cite{chen2022recommendation,li2024making}, and large language models \cite{liu2025rethinking,yao2024large}. 

Although MU focuses on removing the influence of designated data points, its privacy implications extend beyond the deleted samples. Unlearning may  inadvertently alter the representations and decision boundaries of the model, introducing  collateral privacy risks of the retained data. For instance, prior work \cite{onion} has identified the \textit{``privacy onion''} effect, where removing the ``layer'' of outlier points that are most vulnerable to a privacy attack make a new layer of previously-safe points susceptible to the same attack. 
Our study further reveals that such privacy leakage in the retained data can also manifest at the population level. As shown in Table~\ref{tab:attack_results}, \textit{ML-leaks} \cite{salem2018ml}, a state-of-the-art membership inference attack, achieves higher accuracy in identifying the retained samples from the unlearned model than from the original model. The attack performance, measured as the accuracy of predicting retained samples as members, increases by up to 11.13\%. 
These findings highlight the amplified privacy risks of retained data after unlearning, thus motivating the need for a comprehensive privacy framework for MU that addresses not only the removed samples but also the retained data.

While the ``privacy onion'' effect \cite{onion} primarily examines vulnerabilities of retained samples at the individual-example level\footnote{The evaluation of the ``privacy onion'' effect first computes a \textit{privacy score} for each training example, measured as the \textit{attack success rate}  under the Likelihood Ratio Attack (LiRA) \cite{carlini2022membership}. It then removes the samples with the highest privacy scores and reevaluates the privacy scores of the remaining samples that were previously associated with the lowest scores.}, the broader privacy risks affecting the full retained set remain largely unexplored. Despite recent advances in studying the privacy of unlearning models \cite{chen2021machine,kurmanji2024towards,golatkar2020forgetting,hayes2024inexact,naderloui2025rectifying}, existing work has mainly examined the risks associated with the removed samples. To date, no study has systematically analyzed the privacy risks of retained samples at the population level, leaving a critical gap in understanding the privacy vulnerabilities of MU models. 


\begin{table}[!t]
\centering
\caption{Percentage (\%) of samples in the retain set correctly predicted as members by \textit{ML-leaks} \cite{salem2018ml} before and after unlearning (ResNet-18 model, with retraining as the unlearning method). The results demonstrate that the samples in the retain set are more susceptible to ML-leaks after unlearning. }
\label{tab:attack_results}
\begin{tabular}{l|c|c}
\toprule
\textbf{Dataset} & \textbf{Pre-unlearning} & \textbf{Post-unlearning} \\
\hline
CIFAR-10      & 80.27 & 90.87 \\\hline
CIFAR-100      & 82.00 & 90.65 \\\hline
CINIC-10       & 76.04 & 82.59 \\\hline
TinyImageNet   & 72.30 & 83.43 \\
\bottomrule
\end{tabular}
\end{table}

{\bf Challenges.} Intuitively, extending the privacy analysis of MU models to the retain set can be viewed as a problem of distinguishing among three groups: the \textit{forget}, \textit{retain}, and \textit{unseen} sets. A straightforward method is to run a \textit{membership inference attack} (MIA) \cite{shokri2017membership,hu2022membership,carlini2022membership} in two rounds: one against the original model and another against the unlearned model. Samples whose predicted status changes from members to non-members are classified into the forget set; those that remain members belong to the retain set; and those that remain non-members fall into the unseen set. However, as our empirical study shows, this two-round attack performs poorly because it neglects the relationship between the original and unlearned models. \footnote{More details on the performance of the two-round attack can be found in Section \ref{sec-Evaluation}.} 

An alternative strategy is to adapt existing {\itshape unlearning membership inference attacks}  (\textit{U-MIA})
\cite{chen2021machine,hayes2024inexact,naderloui2025rectifying}, which leverage the outputs of both the original and unlearned models when making predictions, to our setting.    
However, as these attacks mainly focus on the forget set, they can only handle the two-class inference (\textit{forget vs. unseen}). Extending them to our  three-class setting (\textit{forget vs. unseen vs. retain}) is far from trivial. The main challenge lies in constructing an attack feature space that can reliably separate all three classes. As we will show, naive adaptations of the features of these existing U-MIAs fail to distinguish the retain set from the forget set effectively. 

{\bf Our Contributions.} 
We propose \system, the first U-MIA that evaluates the privacy risks of the MU models beyond the forget set.  
\system\ is a black-box attack whose features are  extracted from the differences in model predictions before and after unlearning. We propose two compact feature sets, derived respectively from the difference and the sum of posterior probabilities for a given class, and show that both are effective for separating the three sets.  Using these features, a three-class classifier is trained to distinguish among the forget, retain, and unseen sets. 
The key differences between \system\ and existing U-MIAs are summarized in Table \ref{table:comparison}. 

We conduct extensive experiments across three widely-used DNNs, one language model, six real-world datasets, and five state-of-the-art unlearning algorithms (covering both exact and approximate methods). Our results show that \system\ is consistently effective under all settings. For instance, when attacking a ResNet-18 model trained on the TinyImageNet dataset with SISA \cite{bourtoule2021machine} as the unlearning method, \system\ achieves an overall attack accuracy (micro F1-score) of 95.6\%, with per-class F1-scores of 95.99\% (unseen set), 96.53\% (forget set), and 94.32\% (retain set), respectively. Moreover, \system\ consistently outperforms two baselines, including the state-of-the-art U-MIA \cite{chen2021machine}, in both overall and per-class accuracy. Furthermore, we observe that across all five MU approaches, the retain set becomes more susceptible to privacy leakage in the post-unlearning model. Notably, \system\  achieves higher attack accuracy on the retain set than attacks applied to either the pre-unlearning or post-unlearning model alone, with up to 22.7\%  improvement over the best single-model attack. Finally, \system\ remains effective even when its shadow model architecture, training data, or the MU algorithm differ from those of the target model. This demonstrates \system's strong transferability across diverse settings. 

Beyond attacks, we evaluate three defense mechanisms, namely {\itshape label-only output}, {\itshape dropout}, and {\itshape differential privacy}, to mitigate the privacy risks introduced by \system. Our empirical analysis reveals that, while all the three methods can reduce the effectiveness of the attack, the dropout strategy offers the best trade-off between model utility and privacy protection.


\newcolumntype{C}[1]{>{\centering\arraybackslash}p{#1}}
\begin{table}[!t]
\centering
\caption{Comparison between \system\ and existing U-MIAs ($\mathbf{f_\theta}$/$\mathbf{f_{\theta^-}}$: Output of original/unlearning model; \textit{Pop.}: Population-level; \textit{Ex.}: Example-level).}
\label{table:comparison}
\scalebox{0.9}{
\begin{tabular}{C{23mm}||C{8mm}|C{8mm}||C{5mm}|C{5mm}||C{5mm}|C{5mm}}
\hline
\multirow{2}{*}{\textbf{Work}}                                       & \multicolumn{2}{c||}{\textbf{Privacy Analysis}}             & \multicolumn{2}{|c||}{\textbf{Adv. access}}                                        & \multicolumn{2}{c}{\textbf{Granularity}}                    \\ \cline{2-7} 
&\textbf{Forget set} & \textbf{Retain set} & $\mathbf{f_\theta}$& $\mathbf{f_{\theta^-}}$& \textbf{Pop.} & \textbf{Ex.}  \\\cline{2-7}
\hline
~\cite{chen2021machine}&\ding{51}&&\ding{51}&\ding{51}&\ding{51}&\\ \hline
~\cite{hayes2024inexact,naderloui2025rectifying}&\ding{51}&&\ding{51}&\ding{51}&&\ding{51}\\ \hline
\cite{ma2022learn,kurmanji2024towards,golatkar2020forgetting,graves2021amnesiac}&\ding{51}&&&\ding{51}&\ding{51}&\\ \hline
\system\ &\ding{51}&\ding{51}&\ding{51}&\ding{51}&\ding{51}&\\ \hline
\end{tabular}
}
\end{table}

In summary, our contributions are as follows:
\begin{ditemize}

\item We extend privacy analysis of machine unlearning to the retained data, and formulate a tri-class membership inference game for the analysis. 

\item We propose \system, the first black-box tri-class U-MIA, and extensively
evaluate its effectiveness on both deep neural networks (DNNs) and language models.  


\item  We evaluate the performance of three defense mechanisms against \system\ and demonstrate their trade-offs between model accuracy and defense power. 
\end{ditemize}

\section{Background and Related Work}
\label{sec-Preliminary}



\subsection{Machine Unlearning} \label{subsec-unlearing}

Machine unlearning refers to the process of removing the influence of specific training examples on an already trained machine learning model \cite{bourtoule2021machine,chen2021machine}. Formally, given a model trained on a dataset $\targetdata$ using a learning algorithm 
$\learningalgo$, let $\theta$ be its parameters, and $\targetdataforget \subset \targetdata$ be a set of examples to be removed from the model, the machine unlearning algorithm $\unlearnalgo(\para,\targetdata,\targetdataforget)$ aims to obtain a new model with  parameters $\unlearnpara$ by removing the influence of $\targetdataforget$ while preserving model performance on $\targetdata \setminus \targetdataforget$~\cite{nguyen2022survey,shaik2024exploring}. 
Essentially, existing machine unlearning solutions can be categorized into two types: (1) {\itshape Exact unlearning} ensures that the requested instances are completely removed from the model. A straightforward solution is to retrain the model  from scratch, which can be computationally expensive \cite{thudi2022unrolling}. An alternative solution is {\itshape Sharded, Isolated, Sliced, and Aggregated training} ({\itshape  SISA})~\cite{bourtoule2021machine}, which 
 partitions the data into shards and slices. For each shard, it is used to train a model. The models trained over all the shards are aggregated. During unlearning, only the models whose shards contains removed instances are retrained;
 (2) {\itshape Inexact unlearning} aims to obtain a model whose output is approximately the same as that of the exact unlearning model but with much cheaper computational overhead. The existing inexact unlearning approaches include adjusting the model parameters \cite{graves2021amnesiac,guo2020certified,sekhari2021remember,kurmanji2024towards,thudi2022unrolling}, modifying the model architecture \cite{liu2024model}, and filtering the outputs \cite{baumhauer2022machine}. 

\subsection{Membership Inference Attacks}  \label{subsec-mia}

The \textit{Membership inference attack} (MIA) is one of the most common attacks against machine learning model. It aims to infer whether a specific data instance was part of the training set of a target   model~\cite{shokri2017membership,salem2018ml}. Based on the granularity of privacy, the existing MIAs can be categorized to two types \cite{hayes2024inexact}: {\itshape population-level MIAs}  and {\itshape example-level MIAs}.

{\bf Population-level MIA.} This type of MIA instantiates  the same attack for all instances. A typical attack is to train an attack model as a binary classifier from the output of a set of samples from the data distribution~\cite{salem2018ml,liu2022membership,long2018understanding,truex2019demystifying,shokri2017membership}. 
It employs a shadow dataset to train a set of  shadow models, and utilizes the outputs of the member and non-member data of the shadow models as inputs to train the attack classifier.

{\bf Example-level MIA.} 
This type of MIA instantiates a dedicated attack for each example. \textit{LiRA}~\cite{carlini2022membership}, one of the state-of-the-art example-level MIA, is a likelihood ratio attack which compares the model’s behavior across multiple retrained models. It trains multiple shadow models on datasets with and without the target point and observes how much the prediction confidence varies across those versions. Then it computes the likelihood of the target point's predictions assuming it was a member vs. a non-member, using the ratio to decide which is more likely. 
\subsection{Unlearning Membership Inference Attacks}

 Adapting MIAs to unlearning models introduces  a new type of MIAs named {\itshape unlearning membership inference attack} ({\bf U-MIA}) \cite{hayes2024inexact}.  
Similar to MIAs, U-MIAs can be categorized into two types: {\itshape population U-MIA} and {\itshape example-level U-MIA} \cite{hayes2024inexact}.

{\bf Population-level U-MIA.} 
Several studies have applied population-level MIAs to unlearning models in order to evaluate unlearning effectiveness at the population level~\cite{goel2022towards,liu2024model,ma2022learn,graves2021amnesiac}. These attacks typically rely solely on the outputs of the unlearning model to distinguish samples in the unseen set from those in the forget set. Chen et al.~\cite{chen2021machine} proposed {\itshape U-Leak}, a black-box MIA that leverages posterior probabilities from both the original and unlearned models. U-Leak trains a binary attack classifier to infer whether specific samples have been removed from the model.

{\bf Example-level U-MIA.} The LiRA attack~\cite{carlini2022membership} has been adapted to unlearning models to evaluate unlearning efficacy at the example level. Kurmanji et al.~\cite{kurmanji2024towards} applied LiRA using only the unlearning model, while Hayes et al.~\cite{hayes2024inexact} extended it with access to both the original and unlearned models. In addition to the standard two-way hypothesis (\textit{forget vs. unseen}), they also briefly explored a three-way hypothesis (\textit{retain vs. forget vs. unseen}), similar to our setting. Their results showed that while retain-set instances are easily identifiable, forget-set and unseen-set instances cannot be distinguished based on the logits of the unlearning model alone. By contrast, we demonstrate that these instances can be separated by \system\ through a set of carefully designed features.
More recently, Naderloui et al.~\cite{naderloui2025rectifying} introduced {\itshape RULI}, a dual-objective attack that jointly measures unlearning efficacy and privacy risk at the per-sample level. However, none of these approaches readily extend to privacy evaluation at the population level.

The existing U-MIAs also can be classified into the categories of \textit{attack-driven} and \textit{evaluation-driven} ones. The distinction between these two categories, while subtle, is crucial. First, because they serve fundamentally different purposes, their expectations of the attack performance are opposite: the evaluation-driven U-MIAs prefer lower attack success on removed samples, whereas in the attack scenarios, higher attack success exposes the privacy risk. Second, they differ in their assumptions about the adversary’s capabilities. Evaluation-driven U-MIAs typically assume a powerful attacker with extensive knowledge, which often includes white-box access to the model. By contrast, attack-driven U-MIAs often operate under more realistic constraints, such as black-box access. In this paper, we focus on attack-driven U-MIAs under the black-box setting.

\nop{
\subsection{Membership Inference Attacks based on Model Updates}

Few works have been spent on investigating the unintended privacy leakage caused by model updates, in particular, whether the different 
outputs of an ML model’s two versions queried with the same set of data samples leak information of the corresponding 
updating set. Salem et al. \cite{salem2020updates} designed {\itshape Update-Leak} which enables the adversary to reconstruct the updated samples (either insertion or deletion) by exploiting information from two versions of the target ML model (before and after the updating). 
Chen et al.~\cite{chen2021machine} proposed a new MIA named {\itshape U-Leak} to expose the unintended privacy risks of the deleted samples. The key idea of U-Leak is to train a binary attack classifier whose attack features are derived from the posterior probabilities output by both original model and unlearned model. While these works focus on the privacy leakage of the updating set only (i.e., the forget set in the unlearning setting), we target at the privacy leakage of all the three sets (i.e., forget set, retain set, and unseen set). 
}

\section{Problem Setup}\label{sec-Problem}

\nop{
\begin{table}[t!]
\small
\centering
\caption{Notations used in the paper.}
\label{tab:notations}
\begin{tabular}{r|l} 
\toprule
\textbf{Notation} & \textbf{Description} \\ \midrule
$\targetdata$/$\shadowdata$& Original/shadow dataset\\\hline
$\targetdataforget$/$\targetdataretain$/$\targetdataunseen$ & Forget/retain/unseen set \\\hline
$\shadowdataForget$/$\shadowdataRetain$/$\shadowdataUnseen$ & Shadow forget/retain/unseen set \\\hline
$\simtargetmodeloriginal$/$\simtargetmodelunlearned$& Original model/unlearning model \\\hline
$\shadowmodeloriginal$/$\shadowmodelunlearned$ & Original/unlearning shadow model \\\hline
$\mathbb{P}$/$\mathbb{P}^-$ & Posterior probabilities output by $\simtargetmodeloriginal$/$\simtargetmodelunlearned$\\
\bottomrule
\end{tabular}
\end{table}
}

Consider a dataset $\targetdata$ that contains a set of samples, where each sample is represented by input features $x$  and a class label $y$. We consider a Deep Neural Network (DNN) with trainable parameters $\theta$ (denoted as $\targetmodeloriginal$) which is trained in a supervised manner via empirical risk minimization over $\targetdata$.
\nop{
\begin{align} \label{equ:empirical-risk}
       \begin{split}
	\begin{aligned}
\mathcal{L}(w)=\frac{1}{N} \sum_{i=1}^{N} L\left(f\left(x_{i} ; \theta\right), y_{i}\right),
		\end{aligned}
	\end{split}
\end{align}
where $L$ denotes the cross-entropy loss between the predicted probability distribution $f(x_i;\theta)$ and the true class label $y_i$. 
}We consider classification as the downstream task of learning. 
Thus for any given sample $x$, the model outputs a \textit{posterior probability vector}, in which the $i$-th entry indicates the probability that $x$ is associated with the $i$-th label.

In this paper, we consider a Machine-Learning-as-a-Service (MLaaS) setting, where a service provider (server) hosts a machine learning model and offers black-box access to users (clients). Once the model is trained and deployed on the MLaaS platform, it may be subject to various unlearning requests for reasons such as privacy preservation or security enhancement \cite{xu2024machine}. 
To accommodate these requests, the server applies a machine unlearning mechanism, resulting in an updated version of the model - referred to as the unlearning model. After unlearning, the server continues to offer users black-box access of the unlearning model.  
This paradigm has been supported by real-world AI platforms. For instance, both DataRobot \cite{datarobot} and H2O.ai \cite{h2oai} offer end-to-end AI platforms that support model versioning, a foundational capability for enabling machine unlearning in practice. 

\subsection{Threat Model}

In the unlearning setting, the server maintains two models: (1) the original model $\targetmodeloriginal$, with parameters $\theta$, trained on the full dataset $\targetdata$; and (2) the unlearning model $\targetmodelunlearned$, obtained by removing a subset of data $\targetdataforget$ from $\targetmodeloriginal$, resulting in updated parameters $\theta^-$.
For simplicity, we use the notations $\targetmodeloriginal$ and $\simtargetmodeloriginal$ interchangeably to refer to the original model, and likewise, $\targetmodelunlearned$ and $\simtargetmodelunlearned$ to refer to the unlearning model throughout the remainder of this paper. 

{\bf Attack Goal.}  The adversary attempts to determine the membership of a specific data instance $(x, y)$ in the training dataset of $\simtargetmodeloriginal$ and $\simtargetmodelunlearned$. 
Specifically, there are three types of membership sets: 
\begin{ditemize}
    \item {\bf Forget set} $\targetdataforget$, i.e., the set of instances to be removed from the trained model.  
    \item {\bf Retain set} $\targetdataretain=\targetdata \setminus \targetdataforget$, i.e., the set of instances remained in the trained model. 
    \item {\bf Unseen set} $\targetdataunseen$, which is the set of instances that were not included in the training dataset of  $\simtargetmodeloriginal$ (and thus $\simtargetmodelunlearned$). 
\end{ditemize}

\begin{figure}[t!]
	\begin{center}
  \includegraphics[width=0.98\linewidth]{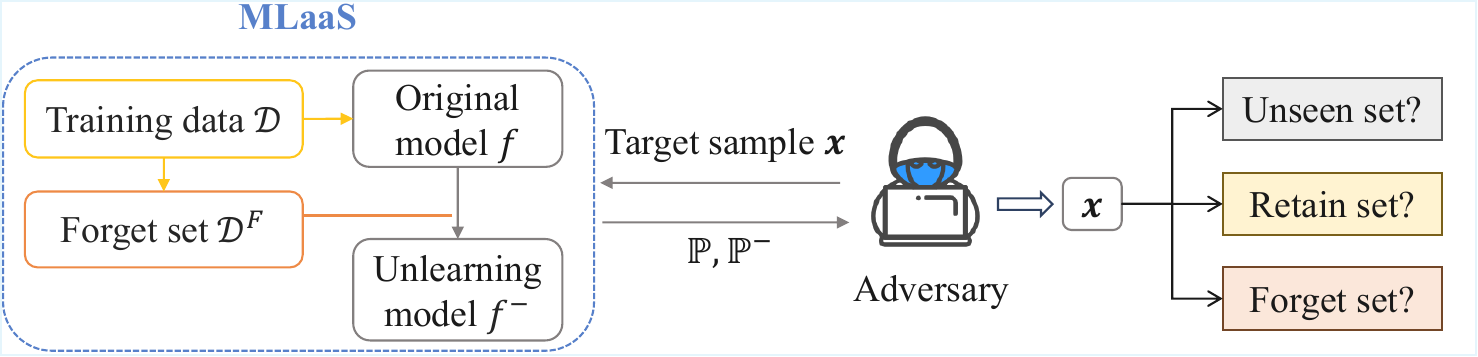}
		\caption{Illustration of threat model.}
        \label{figure:problem}
	\end{center}
\end{figure}

\textbf{Adversary Knowledge and Capabilities.} 
The adversary is granted black-box access to both the original and unlearning models, i.e., they can query each model with specific inputs and receive the corresponding output in the form of posterior probability vectors. However, the adversary has no knowledge of either model’s internal architecture, hyperparameters, or the unlearning algorithms. 
Additionally, we assume the adversary possesses a shadow dataset, whose data distribution may differ from that of the target model’s training dataset. Importantly, the adversary cannot interfere with the unlearning process; for example, they are unable to poison the training data and then submit a removal request for the poisoned samples. Figure \ref{figure:problem} illustrates the threat model. 

\subsection{Problem Formulation}

We formalize the problem as an indistinguishability game between a challenger and adversary, where the challenger is responsible for model training and unlearning, and the adversary aims to determine the membership of a particular sample.  We adapt the membership inference game on machine unlearning  \cite{hayes2024inexact}, which is a two-way hypothesis test, to our 3-class setting. 

\begin{definition}[{\bf Tri-class Unlearning Membership Inference Game}]
Let $\learningalgo$ and $\unlearnalgo$ be a learning algorithm and an unlearning algorithm, respectively, and $\pi$ be the underlying data distribution. The game between a challenger and an adversary proceeds as follows.

\begin{enumerate}
\item[{\bf i.}] The challenger samples a training dataset $\targetdata \sim \pi$, and trains a model $\targetmodeloriginal\sim \learningalgo(\targetdata)$. 

\item[{\bf ii.}] The challenger chooses a subset of samples $\targetdataforget \subseteq \targetdata$, and obtains the unlearning model using $\unlearnalgo$: $\targetmodelunlearned\sim\unlearnalgo(\theta, \targetdata, \targetdataforget)$. 

\item[{\bf iii.}] The challenger randomly selects a number $b \in \{0,1,2\}$, with equal probability. If $b=0$, the challenger samples a data point $z \notin \targetdata$ from $\pi$. If $b=1$, the challenger samples a data point $z$ from $\targetdataforget \subseteq \targetdata$. If $b=2$, the challenger samples a data point $z$ from $\targetdataretain \subseteq \targetdata$. Then, the challenger sends $z$ to the adversary.

\item[{\bf iv.}] The adversary infers a rule $\attack$: $\big(x, \targetmodeloriginal(x), \targetmodelunlearned(x)\big) \to \{0,1,2\}$, and outputs $ \attack(z) \to \hat{b} $.

\item[{\bf v.}] If $\hat{b}=b$, output 1 (success). Otherwise, output 0. 

\end{enumerate}

\end{definition}

The inference game models a scenario in which the adversary can only issue queries on randomly selected samples. While an adversary could, in theory, construct a forget set composed of the most vulnerable instances to evaluate worst-case privacy leakage, our goal is not to assess such a scenario. Instead, we focus on evaluating the privacy leakage of a given unlearning model under more realistic conditions. In this setting, the adversary has no prior knowledge of, nor control over, the composition of the forget set.

\nop{
We formalize the attack as the problem of finding a mapping between the data instance $x$ and one of the three labels (0, 1, 2): 
\begin{align} \label{equ:attack goal}
       \begin{split}
	\begin{aligned}
\attack:(x, \targetmodeloriginal(x), \targetmodelunlearned(x)) \to \{0,1,2\}, 
		\end{aligned}
	\end{split}
\end{align}
where $0, 1, 2$ denote the $x$'s is a non-member, a member in the forget set, and a member in the retain set, respectively. 
}

\subsection{A Straw-man Approach: Two-round Attack}
\label{sc:double}

\begin{table}[t!]
\caption{Decision rules of the Two-round Attack.}
\centering
\label{tab:2roundattack}
\begin{tabular}{c|c|c}
\hline
{\bf Membership in } & {\bf Membership in } & \multirow{2}{*}{\bf Inference} \\ 
{\bf original model} & {\bf unlearning model} & \\
\hline
Non-member            & Non-member              & Unseen set     \\ \hline
Member             & Member              & Retain set     \\ \hline
Member             & Non-member               & Forget set     \\ \hline
Non-member             & Member               & N/A        \\ \hline
\end{tabular}
\end{table}

A straightforward approach to implementing the three-class attack is to perform a standard binary MIA in two rounds. In each round, the goal is to infer whether a given instance is a member (i.e., part of the model’s training dataset) or a non-member. 
In the first round, the attack is conducted on the original model $\simtargetmodeloriginal$, and in the second round, it is performed on the unlearned model $\simtargetmodelunlearned$. The adversary then combines the outcomes from both rounds to determine the final membership class of the target instance.
Table \ref{tab:2roundattack} summarizes the decision rules used by the two-round attack. 

The two-round attack approach has two key limitations: 
(1) It requires executing the attack twice, once on each model, which can be resource-intensive;
(2) It treats the two models independently, ignoring their inherent relationship. This oversight can lead to inconsistent classifications, such as labeling an instance as a non-member in the original model but as a member in the unlearned model, which is invalid in the unlearning context. Such inconsistencies degrade the overall accuracy of the attack, as we will demonstrate in our empirical evaluation (Section \ref{sec-Evaluation}).

\section{Pre-attack Analysis}
\label{sec-pre-attack}

\begin{figure*}[t!]
    \centering
      \begin{subfigure}[b]{0.19\textwidth}
        \includegraphics[width=\textwidth]{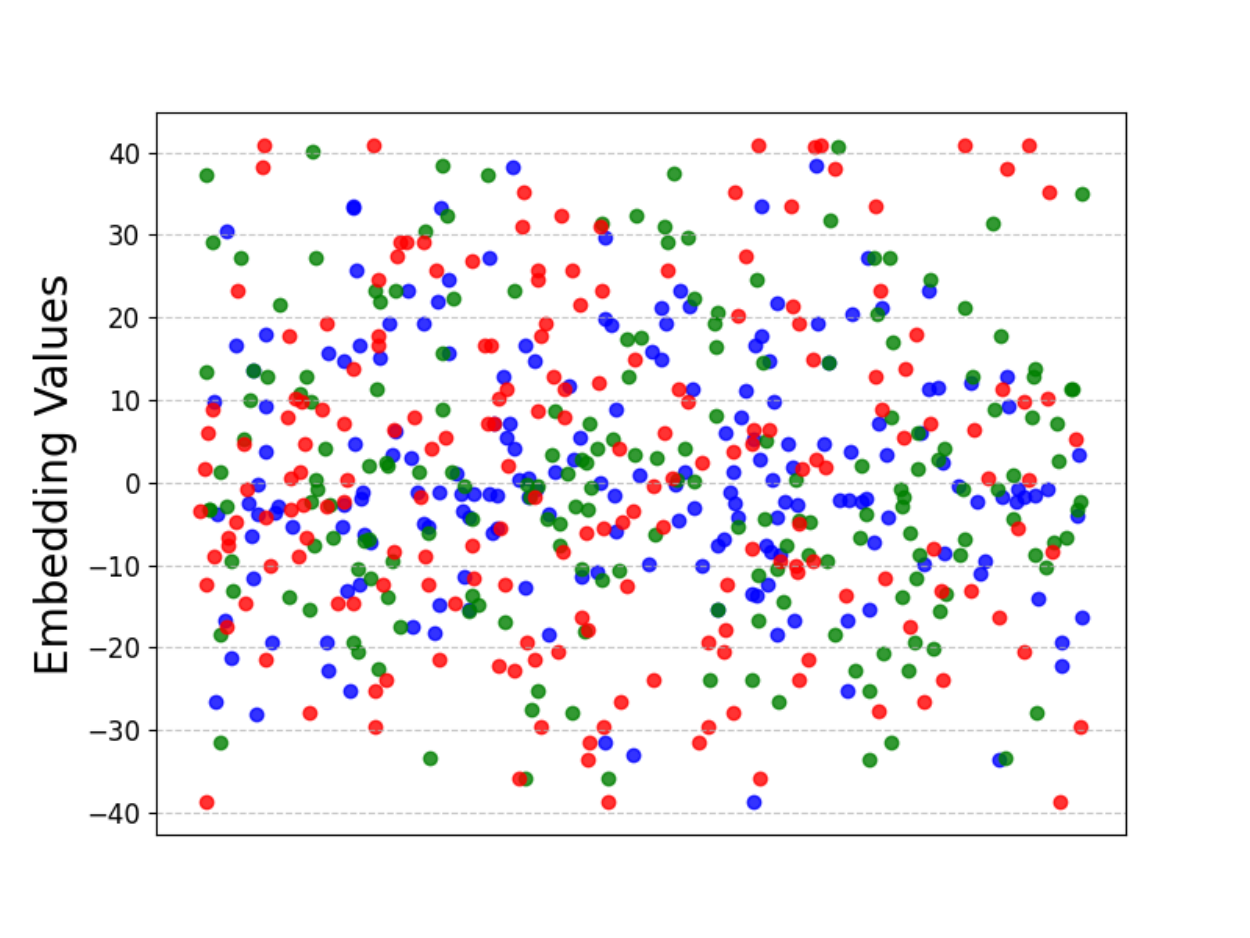} 
        \caption{$\mathbb{P} || \mathbb{P}^-$}
        \label{fig:concate}
    \end{subfigure}
        \begin{subfigure}[b]{0.19\textwidth}
       \includegraphics[width=\textwidth]{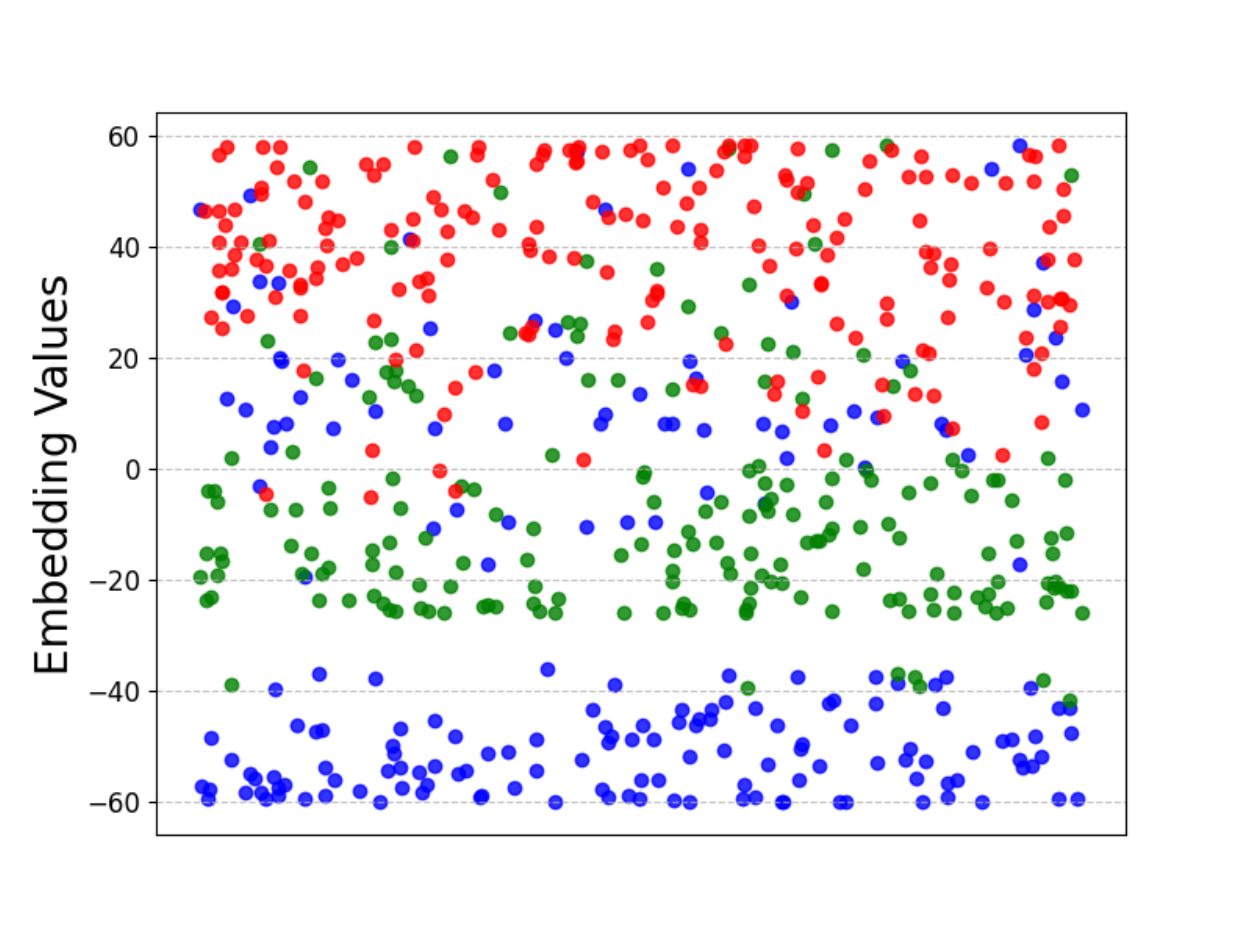} 
        \caption{$\mathbb{P}_{y} || \mathbb{P}^-_{y}$} 
        \label{fig:conprob}        
    \end{subfigure}
        \begin{subfigure}[b]{0.19\textwidth}
        \includegraphics[width=\textwidth]{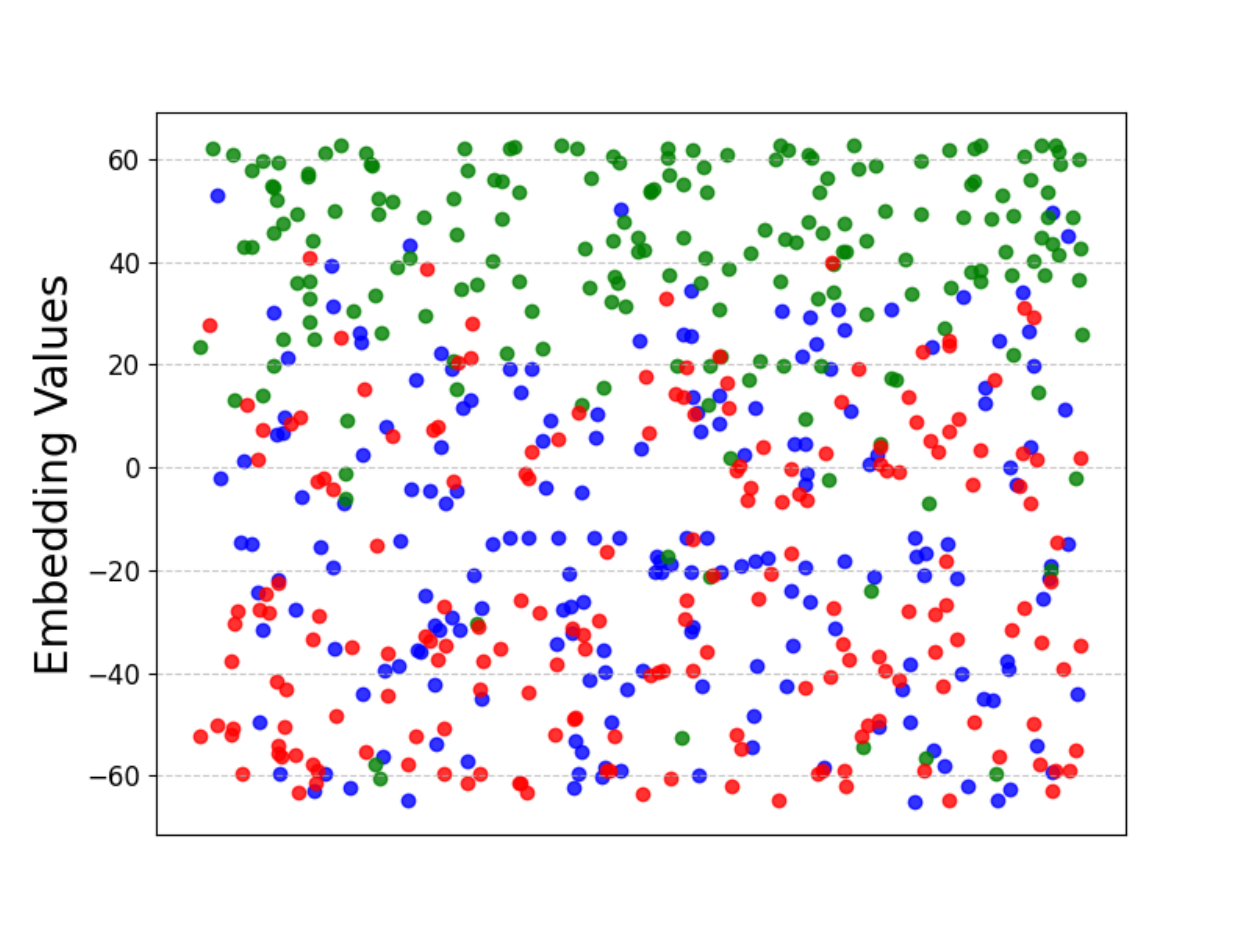} 
        \caption{$\mathbb{P}_{y} -  \mathbb{P}^-_{y}$}        
        \label{fig:diftruprob}        
    \end{subfigure}
           \begin{subfigure}[b]{0.19\textwidth}
       \includegraphics[width=\textwidth]{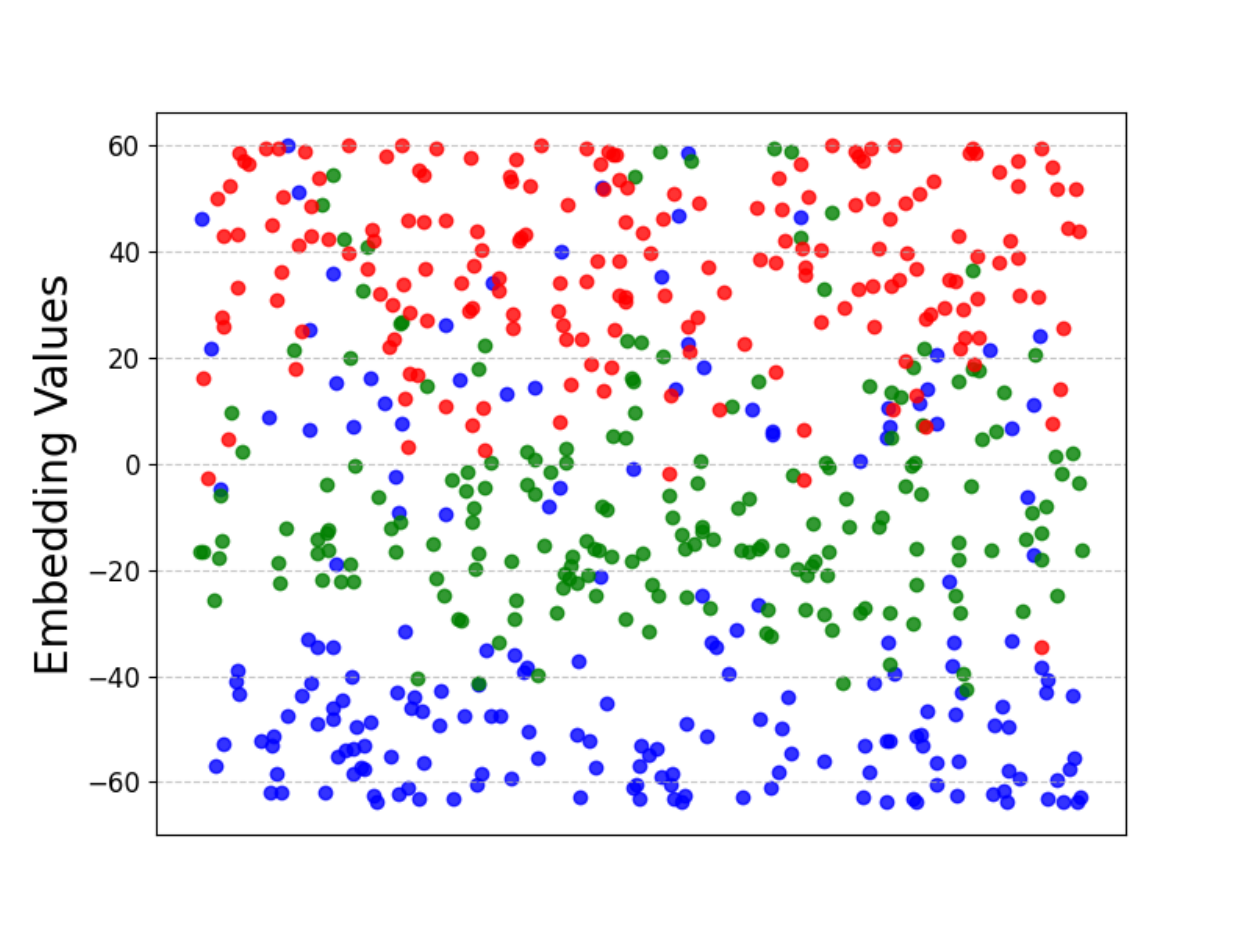} 
        \caption{$\mathbb{P}_{y} + \mathbb{P}^-_{y}$}        
        \label{fig:sumtruprob}        
    \end{subfigure}
            \begin{subfigure}[b]{0.19\textwidth}
       \includegraphics[width=\textwidth]{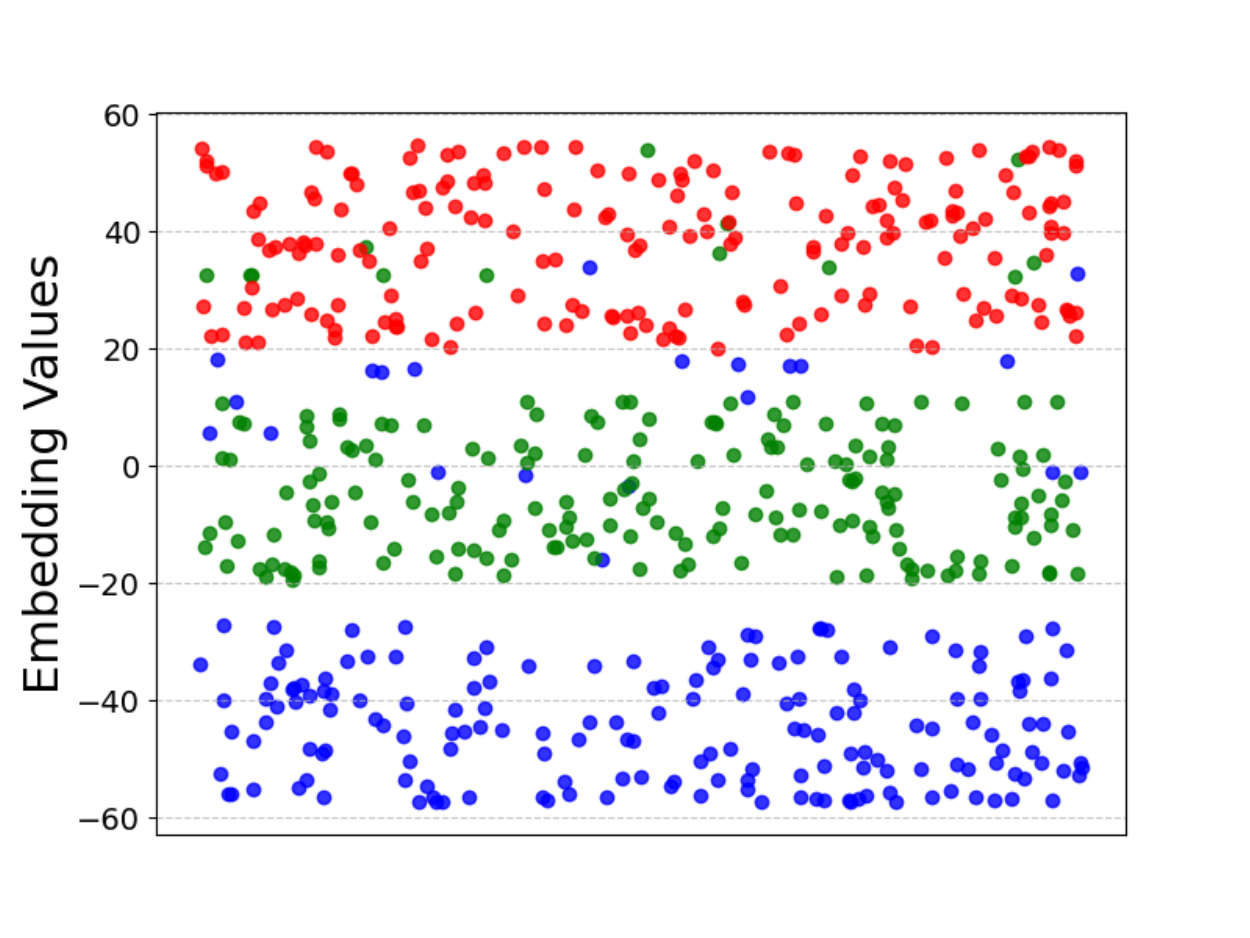} 
        \caption{$\big(\mathbb{P}_{y} -  \mathbb{P}^-_{y}\big) ||\big(\mathbb{P}_{y} +  \mathbb{P}^-_{y}\big)$}       
        \label{fig:difsum}        
    \end{subfigure}
    \caption{\small Visualization of the instances across three membership classes by various types of features (CIFAR-100 dataset). The \textcolor{blue}{unseen set}, \textcolor{green}{forget set}, and \textcolor{red}{retain set} are colored in blue, green, and red, respectively. $\mathbb{P}$ and $\mathbb{P}^-$ are the posterior probability vector output by the models before and after unlearning, respectively. $\mathbb{P}_{y}$ and $\mathbb{P}^{-}_{y}$ are the posterior probability of the ground-truth label $y$ predicted by the models before and after unlearning, respectively.}
    \label{figure:vis-feature}
\end{figure*}

Before designing the attack, we conduct an empirical analysis to characterize the three membership classes. The insights gained from this analysis serve as the foundation for the design of \system. Prior work~\cite{chen2021machine} has shown that the membership status of any instance in the original training set can be leaked through its predictions by both models (i.e., the models before and after unlearning). 
We will extend this reasoning and investigate how to leverage the predictions by both models to further distinguish between retain set and the forget set, besides the unseen set, through the empirical evaluation.

We train a 18-layer ResNet (ResNet-18) model on the CIFAR-100 dataset to obtain the original model $\simtargetmodeloriginal$.\footnote{The details of the model and dataset can be found in Appendix \ref{appendix:expsetup}. } We randomly select 1\% of samples from the training set as the forget set $\targetdataforget$.
The unlearned model $\simtargetmodelunlearned$ is obtained by retraining $\simtargetmodeloriginal$ on the dataset that excludes $\targetdataforget$. 
Next, we randomly select 192 data instances from both retain and unseen sets, respectively.
For each data instance in the three sets, we compute its posterior probability vectors output by both $\simtargetmodeloriginal$ and $\simtargetmodelunlearned$, denoted as $\mathbb{P}$ and $\mathbb{P}^-$. 

Figure \ref{figure:vis-feature} illustrates the distribution of the three membership classes using different types of features derived from the posterior probability vectors produced by both $\simtargetmodeloriginal$ and $\simtargetmodelunlearned$. For visualization purposes, we apply t-SNE to project the feature representations into a one-dimensional embedding. 

We start with a straightforward approach to constructing features: concatenating the posterior probability vectors produced by the original and unlearned models, i.e., $\mathbb{P} || \mathbb{P}^-$, which is a  technique previously employed in two-class attack-driven U-MIAs \cite{chen2021machine}. As shown in Figure \ref{fig:concate}, $\mathbb{P} || \mathbb{P}^-$ fails to clearly separate the three membership classes, indicating that they are heavily intertwined. This leads to the following key observation: 
 \vspace{0.1in}
 \begin{mdframed}[align=left]
\noindent {\bf Finding 1}. $\mathbb{P}||\mathbb{P}^-$ cannot distinguish the three classes effectively.  
\end{mdframed} 


Since using the full posterior probability vectors fails to effectively distinguish the three membership classes, we instead consider leveraging the posterior probability corresponding to the ground-truth label - a technique commonly adopted in prior MIAs \cite{shokri2017membership,ye2022enhanced,carlini2022membership,zarifzadeh2024low}. In the following discussion, we refer to this value as the {\itshape true-label probability}.

Intuitively, instances in the forget set are expected to exhibit a more significant change in prediction after unlearning compared to those in the retain set. Furthermore, since both forget and retain sets were seen by the model during training, they should generally have higher prediction confidence than instances in the unseen set. 
Motivated by these observations, we explore features based on the difference and sum of the true-label probabilities before and after unlearning. Specifically, let $\mathbb{P}_{y}$ and $\mathbb{P}^{-}_{y}$ denote the posterior probabilities of the ground-truth label $y$ predicted by $\simtargetmodeloriginal$ and $\simtargetmodelunlearned$, respectively. Based on these, we consider the following four alternative features:

\begin{ditemize}
    \item {\itshape Concatenation of true-label probabilities}, i.e., $\mathbb{P}_{y} || \mathbb{P}^-_{y}$.
    \item {\itshape Difference between true-label probabilities}, i.e., $\mathbb{P}_{y} -  \mathbb{P}^-_{y}$.
    \item {\itshape Sum of true-label probabilities}, i.e., $\mathbb{P}_{y} +  \mathbb{P}^-_{y}$. 
    \item {\itshape Concatenation of both difference and sum of true-label probabilities}, i.e., $\big(\mathbb{P}_{y} -  \mathbb{P}^-_{y}\big) ||\big(\mathbb{P}_{y} +  \mathbb{P}^-_{y}\big)$. 
\end{ditemize}

We illustrate the distribution of the three classes by four alternative methods in Figure \ref{fig:conprob} - \ref{fig:difsum}, respectively.
We have the following observation: 
\vspace{0.1in}
\begin{mdframed}[align=left]
\noindent {\bf Finding 2}: Either $\mathbb{P}_{y} || \mathbb{P}^-_{y}$ or $\big(\mathbb{P}_{y} -  \mathbb{P}^-_{y}\big) ||\big(\mathbb{P}_{y} +  \mathbb{P}^-_{y}\big)$ can effectively distinguish among the three classes.
\end{mdframed} 

The detailed analysis of why $\mathbb{P}_{y} -  \mathbb{P}^-_{y}$ or $\mathbb{P}_{y} +   \mathbb{P}^-_{y}$ alone fails to distinguish the three sets can be found in Appendix \ref{appendix:disprob}.  
We will follow Finding 2 to design the attack features of \system. Our empirical evaluation (Section \ref{sec-Evaluation}) will reveal that the two alternative features exhibit disparate effectiveness across different models.

\section{Methodology}
\label{sec-Method}

\begin{figure*}[t!]
	\begin{center}
  \includegraphics[width=0.9\linewidth]{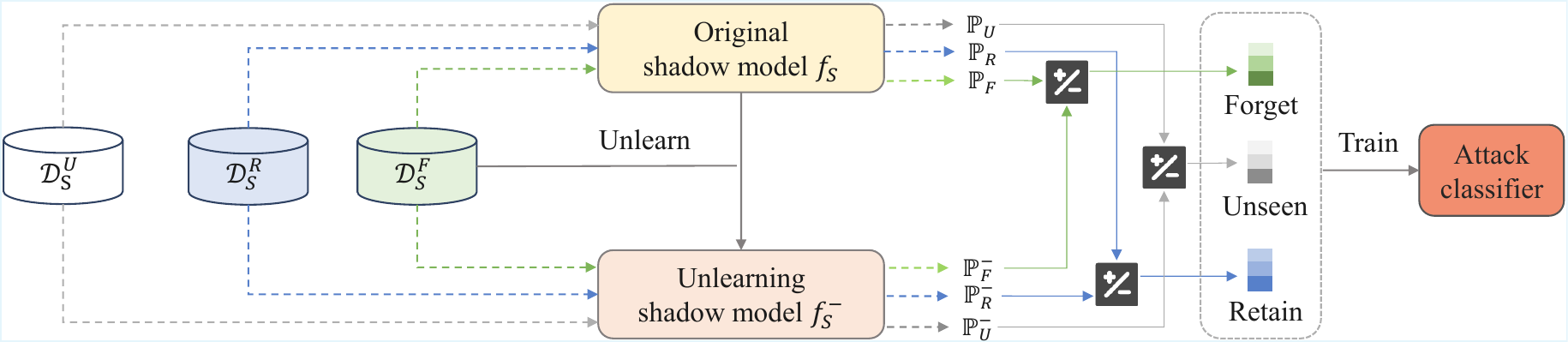}
		\caption{\small The training process of \system. $\mathbb{P}_U$ ($\mathbb{P}_U^-$), $\mathbb{P}_F$ ($\mathbb{P}_F^-$), and $\mathbb{P}_R$ ($\mathbb{P}_R^-$) denote the posterior probabilities of the ground-truth labels produced by the original (unlearned) model for samples in the unseen, forget, and retain sets, respectively.} 
        \label{figure:attack model}
	\end{center}
\end{figure*}

In this section, we present the details of our attack model, 
\system. \system\ consists of three components: 
(1) \textit{Shadow model training}: We first train an original shadow model to mimic the behavior of the target model. Then we train a unlearning shadow model from the original shadow model. 
(2) \textit{Attack classifier training}: We derive the attack features from the output of both shadow models and train a three-class classifier; (3) \textit{Inference}: We utilize the trained classifier to infer the membership of given data samples. 
Below, we explain the details of each component. The pseudo code of \system\ can be found in Appendix \ref{appendix:pseudo}.

\subsection{Shadow Model Training}
Recall that the attacker does not have access to the target model. To approximate the behavior of the original target model $\simtargetmodeloriginal$, the attacker trains an original shadow model $\shadowmodeloriginal$. Notably, the architecture and hyperparameters of $\shadowmodeloriginal$ may differ from those of the target model. To construct $\shadowmodeloriginal$, the attacker samples a dataset $\shadowdata$ from the shadow dataset available in their adversarial knowledge. This dataset is then used to train $\shadowmodeloriginal$.
Once the original shadow model is trained, the attacker randomly selects a subset of $\shadowdata$ to form the shadow forget set $\shadowdataForget$. A corresponding shadow unlearned model $\shadowmodelunlearned$ is then obtained by removing $\shadowdataForget$ from $\shadowmodeloriginal$. Since the adversary does not know which unlearning algorithm is used by the server, they may apply a different unlearning mechanism. This mismatch can introduce behavioral discrepancies between the shadow and target unlearning models, potentially affecting the performance of \system. We will empirically assess the transferability of \system\ across different unlearning algorithms in Section~\ref{sec-Evaluation}.

\nop{
\begin{color}{blue}
The computational effort of shadow model training depends on several factors, including the complexity of the shadow model and the size of the shadow dataset. As we assume the architecture of the shadow model can be different from that of the target model, the shadow model can be significantly smaller than the target model in terms of both the number of neurons and layers. Similarly, the shadow dataset can be much smaller than that of the target dataset in terms of both sample size and feature dimensions. Training on such a smaller shadow model and dataset can result in substantially lower training costs. As demonstrated in our transferability analysis (Figure~\ref{figure:factor-analysis}), \system\ remains effective under these settings. 
\end{color}
}

\subsection{Training Attack Classifier} \label{subsec:Training Attack Classifier}

The training process of the attack classifier is illustrated in Figure~\ref{figure:attack model}. It consists of two major steps, namely \textit{deriving attack features} and \textit{classifier training}. Below, we will discuss the details of each step.

{\bf Deriving attack features.} 
Following Finding 2 (Section \ref{sec-pre-attack}), the attack features  of \system\ are derived from the posteriors output by both the original shadow model and the unlearning shadow model. 
Specifically, for a given instance $(x, y)$, 
let $\mathbb{P}_{y}$ and $\mathbb{P}^-_{y}$ be the posterior probability of the ground-truth label $y$ output by the original shadow model $f_s$ and the unlearning shadow model $f_s^-$, respectively. We have two strategies to derive the attack features.

\begin{ditemize}
\item  {\bf ConProb}: The attack feature of the instance $(x, y)$ is derived as the concatenation of $y$'s posterior probability output by $f_s$ and $f_s^-$, respectively: 
\begin{equation}
\label{eqn:feature1}
   \mathbb{X} = \mathbb{P}_{y} || \mathbb{P}^-_{y}.   
\end{equation}
\item  {\bf ConSumDif}: The attack feature of the instance $(x, y)$ is derived as the concatenation of both difference and sum of $y$'s posterior probability output by $f_s$ and $f_s^-$, respectively: 
\begin{equation}
\label{eqn:feature2}
    \mathbb{X} = (\mathbb{P}_{y} - \mathbb{P}^-_{y})|| (\mathbb{P}_{y} + \mathbb{P}^-_{y})
\end{equation}
\end{ditemize}

With the feature derivation strategies in place, the attacker proceeds to construct the training dataset $\attack_{Train}$ for training the attack classifier. The process involves the following steps:
\underline{First}, the attacker randomly selects a subset of instances, denoted as $\shadowdataForget$, from $\shadowdata$, and removes them from the training data of the original shadow model $\shadowmodeloriginal$ to obtain the corresponding unlearned shadow model $\shadowmodelunlearned$.
\underline{Second}, the attacker samples three disjoint sets for labeling: (1) The shadow forget set, drawn from $\shadowdataForget$;
(2) The shadow retain set, sampled from $\shadowdata \setminus \shadowdataForget$; and (3) The shadow unseen set, composed of instances not included in $\shadowdata$. 
An equal number of instances is selected for each set to ensure class balance in the attack dataset.
\underline{Third}, the attacker queries both $\shadowmodeloriginal$ and $\shadowmodelunlearned$ to obtain the posterior probability vectors for all instances in the three sets. Based on these outputs, features are derived using either Equation~\eqref{eqn:feature1} or Equation~\eqref{eqn:feature2}, forming the input features for $\attack_{Train}$. Each feature is labeled as 0, 1, or 2, corresponding to whether the associated sample belongs to the shadow unseen, retain, or forget set, respectively. Target data is inferred as a member of the unseen set, retain set, or forget set when the attack model outputs 0, 1, or 2, respectively.

{\bf Training Attack Classifier.} The adversary trains a 3-class classifier on the constructed dataset. In this paper, we construct a Multilayer Perceptron (MLP) classifier. More details of the model setup can be found in the empirical evaluation (Section \ref{sec-Evaluation}).

\subsection{Inference}
Given a data sample, the adversary derives its attack features (Eqn~\eqref{eqn:feature1} -  \eqref{eqn:feature2}) from its posterior probabilities output by both $\simtargetmodeloriginal$ and $\simtargetmodelunlearned$, and feeds them to the trained attack classifier $\attack$ for membership inference. 

\section{Evaluation}
\label{sec-Evaluation}
This section presents the results of our empirical evaluation, aiming to address seven research questions: 
\begin{ditemize}
\item $\mathbf{RQ1}$ - How does $\system$ perform against the state-of-the-art machine unlearning models?

\item $\mathbf{RQ2}$ - Does the “privacy onion" effect \cite{onion} extend to the population level, and can \system\ more effectively identify privacy risks in the retained data compared to existing MIAs?

\item $\mathbf{RQ3}$ - How does the unlearning-induced change in model overfitting impact \system's performance?

\item $\mathbf{RQ4}$ - Can $\system$ be transferred across the settings where the adversary knowledge of the shadow model, shadow dataset, and unlearning algorithms, are different from those used by target model? 

\item $\mathbf{RQ5}$ - How does $\system$ perform under various parameter configurations? 

\item $\mathbf{RQ6}$ - Which attack features of \system\ yield the best attack performance?

\end{ditemize}

\subsection{Setup} \label{subsec-setup}

All the experiments are executed on a server with 4 NVIDIA A100
GPUs, each with 40GB of memory. All the algorithms are implemented in Python along with PyTorch. All the experimental results are obtained as the average of three trials. 

\begin{table}[t!]
\centering
\footnotesize
\caption{Statistics of datasets}
\label{tab:dataset_stats}
\scalebox{0.9}{
\begin{tabular}{cccccc}
\toprule
\textbf{Type} & \textbf{Dataset}     & \textbf{{\begin{tabular}[c]{@{}c@{}}Feature\\  dim.\end{tabular}
} } & \textbf{\# labels} & \textbf{{\begin{tabular}[c]{@{}c@{}}\# train\\ samples \end{tabular}
} } & \textbf{{\begin{tabular}[c]{@{}c@{}}\# test \\samples\end{tabular}
} } \\ \hline
\multirow{4}{*}{Image}&CIFAR10                   & 32*32*3                   & 10                  & 50,000 & 10,000             \\ 
& CIFAR100                   & 32*32*3                   & 100                 & 50,000 & 10,000              \\ 
& CINIC-10                      & 32*32*3                   & 10                  & 90,000 & 90,000             \\ 
& TinyImageNet                      & 224*224*3                   & 200                  & 100,000   & 10,000           \\ \hline
\multirow{2}{*}{Text}&SST5            & 512     & 5                & 8,544              & 2,210              \\
&News20   & 512     & 20               & 11,314             & 7,532    \\
\bottomrule
\end{tabular}
}
\end{table}

\textbf{Datasets and Models.} 
In our experiments, we set up three widely-used Convolutional Neural Network (CNN) models: \textit{SimpleCNN}~\cite{chen2021machine}, \textit{DenseNet}~\cite{huang2017densely}, and \textit{ResNet-18}~\cite{he2016deep}, with image classification as the downstream task. We use four image datasets ({\itshape CIFAR-10} ~\cite{krizhevsky2009learning}, {\itshape  CIFAR-100}~\cite{krizhevsky2009learning}, {\itshape CINIC-10}~\cite{darlow2018cinic}, and {\itshape TinyImageNet}~\cite{deng2009imagenet}) to train these models. 
Table \ref{tab:dataset_stats} reports the statistics of the datasets used in the experiments. More details of models are provided in Appendix~\ref{appendix:expsetup}. 

Besides DNNs trained over image data, we evaluate \system\ over language models trained over text data, and include the results in Section \ref{exp:lm}. 


\textbf{Unlearning Algorithms.} We consider two types of unlearning approaches: 
\begin{ditemize}
    \item {\itshape Exact unlearning algorithms}: We consider both \textit{retraining} and  \textit{SISA}~\cite{bourtoule2021machine}); 
\item {\itshape Approximate unlearning algorithms}: We employ three state-of-the-art inexact unlearning algorithms (\textit{Sparsity}~\cite{liu2024model}, \textit{SCRUB}~\cite{kurmanji2024towards},  and \textit{Gradient Ascent (GA)}~\cite{graves2021amnesiac,thudi2022unrolling,golatkar2020eternal}. Sparsity utilizes model sparsification via weight pruning to achieve the unlearning effect. SCRUB employs a teacher-student formulation by which the original model is treated as the teacher model, aiming to train a student model that inherits knowledge from the teacher model about $\targetdataretain$ while forgetting $\targetdataforget$. 
GA reverses the
model training on $\targetdataforget$ by adding the correspond gradients back to the weights of the original model weights.  
\end{ditemize}
The unlearning performance of these approaches can be found in Appendix~\ref{appendix:expsetup}.


\textbf{Attack Setup.} We employ a three-layer fully connected Multilayer Perceptron (MLP) as the attack classifier. The first hidden layer contains 32 neurons, followed by a second hidden layer with 16 neurons; both layers use the ReLU activation function. The final output layer consists of three neurons with a softmax activation function. Due to limited space, we include the details of the attack training and testing datasets in Appendix \ref{appendix:expsetup}.



\textbf{Evaluation Metrics.}  
We consider three metrics to evaluate the attack performance: 
\begin{ditemize}
\item {\bf Micro F1-score (F1)}~\cite{grandini2020metrics}: It is a metric that has been widely used to evaluate  the performance of  multi-class classification. It is calculated as follows:
\begin{equation}
\label{eqn:F1}
     F1_{\text{Micro}}=2 \times\left(\frac{\text{Precision}_{\text{Micro}} * \text{Recall}_{\text{Micro}} }{\text{Precision}_{\text{Micro}} + \text{Recall}_{\text{Micro}} }\right),
\end{equation}
\nop{Where 
\begin{align}
    \text{Precision}_{\text{Micro}} &= \frac{\sum TP}{\sum TP + \sum FP} \\
    \text{Recall}_{\text{Micro}} &= \frac{\sum TP}{\sum TP + \sum FN}
\end{align}
}
where $\text{Precision}_{\text{Micro}}$ is computed as the total number of true positives across all classes divided by the total number of predicted positives across all classes, and $\text{Recall}_{\text{Micro}}$ is computed as the total number of true positives across all classes divided by the total number of actual positives across all classes. 
We use the micro F1-score to measure the overall performance of \system\ across all the samples in the attack testing data. 
A high micro F1-score value indicates that the attack is more effective. 
We do not consider the macro F1-score (i.e., the average of F1-scores of all classes) as it can be easily computed from the per-class F1-scores.

\item {\bf Per-class F1-score}: It measures the accuracy of \system\ for each class. Specifically, the F1-score of   class $k$, denoted as $F1_k$, is computed as follows:
\begin{equation}
\label{eqn:F1k}
    F1_k=2 \times \frac{\text{Precision}_k\times \text{Recall}_k}{\text{Precision}_k+\text{Recall}_k},
\end{equation}
where $\text{Precision}_k$ measures the proportion of true positives - i.e., samples correctly predicted as class $k$ - among all samples predicted as class $k$, and $\text{Recall}_k$ measures the proportion of samples in class $k$ that are correctly predicted by the classifier. 

\item{\bf TPR@5\%FPR}: We measure the true positive rate at the false positive ratio of 5\% for each membership class. 
\end{ditemize}


\textbf{Baseline Approaches.} 
Since we are the first to explore three-class U-MIAs, there is no existing method that can be directly compared with. Therefore, we consider the following two approaches as the baselines:

\begin{ditemize}
    
    \item \textbf{Two-round Attack}: We implemented the two-round attack (Section \ref{sc:double}) by launching {\itshape ML-leaks}~\cite{salem2018ml}\footnote{\url{https://github.com/AhmedSalem2/ML-leakss}}, a state-of-the-art population-level MIA, to attack the original and unlearning models independently. 

    \item \textbf{U-Leak}: We adapt  U-Leak~\cite{chen2021machine}\footnote{\url{https://github.com/MinChen00/UnlearningLeaks}}, the only attack-driven U-MIA in the literature, to our setting. \footnote{Evaluation-driven per-example U-MIAs \cite{kurmanji2024towards,hayes2024inexact,naderloui2025rectifying} are fundamentally different from the population-level U-MIAs in their scope and evaluation methodology and thus cannot be readily adapted to our setting.} Since U-Leak was originally designed to distinguish between the forget set and unseen set, we adapt it to our setting by relabeling the samples in the attack training set with three membership classes, while keeping its original features unchanged.\footnote{U-Leak has five variants, each leveraging a distinct set of attack features. We report the best performance among them.}  
\end{ditemize}


\definecolor{light-gray}{gray}{0.8}

\begin{table*}[!t]
\centering
\caption{Overall and per-class F1-score (\%) of the attacks (ResNet-18) over CIFAR-100, CINIC-10, and TinyImageNet datasets. The best results of the three attacks per evaluation metric are highlighted in \gray{gray}. }
\label{tab:Performance}
\scalebox{0.93}{
\begin{tabular}{c|c|cccc|cccc|cccc}
\toprule
\multirow{2}{*}{\textbf{\begin{tabular}[c]{@{}c@{}}Unlearning\\ method\end{tabular}}} & \multirow{2}{*}{\textbf{Dataset}} & \multicolumn{4}{c|}{\textbf{\system\ (Ours)}}                                                                                           & \multicolumn{4}{c|}{\textbf{U-Leak (Best performance)}}                                                                               & \multicolumn{4}{c}{\textbf{Two-round Attack}}                                                                                            \\ \cline{3-14} 
                                                                                      &                                   & \multicolumn{1}{c|}{\textbf{All}} & \multicolumn{1}{c|}{\textbf{Unseen}} & \multicolumn{1}{c|}{\textbf{Forget}} & \textbf{Retain} & \multicolumn{1}{c|}{\textbf{All}} & \multicolumn{1}{c|}{\textbf{Unseen}} & \multicolumn{1}{c|}{\textbf{Forget}} & \textbf{Retain} & \multicolumn{1}{c|}{\textbf{All}} & \multicolumn{1}{c|}{\textbf{Unseen}} & \multicolumn{1}{c|}{\textbf{Forget}} & \textbf{Retain} \\ \hline
\multirow{3}{*}{Retrain}                                                              & CIFAR-100                         & \multicolumn{1}{c|}{\gray{79.98}}            & \multicolumn{1}{c|}{\gray{79.10}}           & \multicolumn{1}{c|}{\gray{79.70}}           & \gray{80.83}           & \multicolumn{1}{c|}{62.11}            & \multicolumn{1}{c|}{48.13}           & \multicolumn{1}{c|}{61.43}           & 70.09           & \multicolumn{1}{c|}{56.98}            & \multicolumn{1}{c|}{53.05}           & \multicolumn{1}{c|}{58.30}           & 68.53           \\ \cline{2-14} 
                                                                                      & CINIC-10                          & \multicolumn{1}{c|}{\gray{59.26}}            & \multicolumn{1}{c|}{\gray{57.84}}           & \multicolumn{1}{c|}{\gray{55.28}}           & \gray{63.84}           & \multicolumn{1}{c|}{51.06}            & \multicolumn{1}{c|}{41.96}           & \multicolumn{1}{c|}{43.11}           & 61.39           & \multicolumn{1}{c|}{44.19}            & \multicolumn{1}{c|}{36.48}           & \multicolumn{1}{c|}{39.98}           & 58.72           \\ \cline{2-14} 
                                                                                      & TinyImageNet                      & \multicolumn{1}{c|}{\gray{92.50}}            & \multicolumn{1}{c|}{\gray{93.60}}           & \multicolumn{1}{c|}{\gray{91.24}}           & \gray{92.69}           & \multicolumn{1}{c|}{86.42}            & \multicolumn{1}{c|}{85.56}           & \multicolumn{1}{c|}{84.71}           & 88.61           & \multicolumn{1}{c|}{83.28}            & \multicolumn{1}{c|}{83.81}           & \multicolumn{1}{c|}{84.72}           & 88.54           \\ \hline
\multirow{3}{*}{SISA}                                                                 & CIFAR-100                         & \multicolumn{1}{c|}{\gray{82.33}}            & \multicolumn{1}{c|}{\gray{79.71}}           & \multicolumn{1}{c|}{\gray{85.46}}           & 81.46           & \multicolumn{1}{c|}{78.38}            & \multicolumn{1}{c|}{73.87}           & \multicolumn{1}{c|}{69.80}           & \gray{85.93}           & \multicolumn{1}{c|}{41.17}            & \multicolumn{1}{c|}{46.43}           & \multicolumn{1}{c|}{27.59}           & 47.83           \\ \cline{2-14} 
                                                                                      & CINIC-10                          & \multicolumn{1}{c|}{\gray{59.00}}            & \multicolumn{1}{c|}{\gray{52.42}}           & \multicolumn{1}{c|}{\gray{63.65}}           & 59.28           & \multicolumn{1}{c|}{55.15}            & \multicolumn{1}{c|}{48.30}           & \multicolumn{1}{c|}{39.77}           & \gray{64.76}           & \multicolumn{1}{c|}{33.67}            & \multicolumn{1}{c|}{34.81}           & \multicolumn{1}{c|}{24.38}           & 42.58           \\ \cline{2-14} 
                                                                                      & TinyImageNet                      & \multicolumn{1}{c|}{\gray{95.60}}            & \multicolumn{1}{c|}{\gray{95.99}}           & \multicolumn{1}{c|}{\gray{96.53}}           & \gray{94.32}           & \multicolumn{1}{c|}{80.91}            & \multicolumn{1}{c|}{69.41}           & \multicolumn{1}{c|}{84.75}           & 84.95           & \multicolumn{1}{c|}{70.97}            & \multicolumn{1}{c|}{73.13}           & \multicolumn{1}{c|}{74.64}           & 74.91           \\ \hline
\multirow{3}{*}{Sparsity}                                                             & CIFAR-100                         & \multicolumn{1}{c|}{\gray{79.96}}            & \multicolumn{1}{c|}{\gray{85.41}}           & \multicolumn{1}{c|}{\gray{76.67}}           & \gray{78.56}           & \multicolumn{1}{c|}{68.47}            & \multicolumn{1}{c|}{83.05}           & \multicolumn{1}{c|}{60.81}           & 64.15           & \multicolumn{1}{c|}{60.69}            & \multicolumn{1}{c|}{63.23}           & \multicolumn{1}{c|}{62.24}           & 64.26           \\ \cline{2-14} 
                                                                                      & CINIC-10                          & \multicolumn{1}{c|}{\gray{62.90}}            & \multicolumn{1}{c|}{\gray{58.79}}           & \multicolumn{1}{c|}{\gray{59.49}}           & \gray{67.84}           & \multicolumn{1}{c|}{55.45}            & \multicolumn{1}{c|}{45.29}           & \multicolumn{1}{c|}{48.63}           & 66.52           & \multicolumn{1}{c|}{46.23}            & \multicolumn{1}{c|}{40.85}           & \multicolumn{1}{c|}{46.57}           & 56.52           \\ \cline{2-14} 
                                                                                      & TinyImageNet                      & \multicolumn{1}{c|}{\gray{92.83}}            & \multicolumn{1}{c|}{\gray{97.13}}           & \multicolumn{1}{c|}{\gray{90.24}}           & \gray{91.28}           & \multicolumn{1}{c|}{79.63}            & \multicolumn{1}{c|}{94.52}           & \multicolumn{1}{c|}{70.93}           & 74.34           & \multicolumn{1}{c|}{70.02}            & \multicolumn{1}{c|}{75.22}           & \multicolumn{1}{c|}{70.78}           & 74.86           \\ \hline
\multirow{3}{*}{SCRUB}                                                                & CIFAR-100                         & \multicolumn{1}{c|}{\gray{73.58}}            & \multicolumn{1}{c|}{\gray{83.50}}           & \multicolumn{1}{c|}{\gray{70.74}}           & \gray{68.45}           & \multicolumn{1}{c|}{70.64}            & \multicolumn{1}{c|}{81.00}           & \multicolumn{1}{c|}{67.91}           & 65.20           & \multicolumn{1}{c|}{62.78}            & \multicolumn{1}{c|}{65.41}           & \multicolumn{1}{c|}{64.04}           & 65.97           \\ \cline{2-14} 
                                                                                      & CINIC-10                          & \multicolumn{1}{c|}{\gray{56.95}}            & \multicolumn{1}{c|}{\gray{60.77}}           & \multicolumn{1}{c|}{\gray{49.34}}           & \gray{59.48}           & \multicolumn{1}{c|}{51.07}            & \multicolumn{1}{c|}{45.79}           & \multicolumn{1}{c|}{40.93}           & 58.52           & \multicolumn{1}{c|}{44.77}            & \multicolumn{1}{c|}{42.78}           & \multicolumn{1}{c|}{41.84}           & 53.96           \\ \cline{2-14} 
                                                                                      & TinyImageNet                      & \multicolumn{1}{c|}{\gray{89.31}}            & \multicolumn{1}{c|}{\gray{93.66}}           & \multicolumn{1}{c|}{\gray{85.91}}           & \gray{88.57}           & \multicolumn{1}{c|}{79.34}            & \multicolumn{1}{c|}{82.69}           & \multicolumn{1}{c|}{74.92}           & 80.47           & \multicolumn{1}{c|}{80.27}            & \multicolumn{1}{c|}{82.95}           & \multicolumn{1}{c|}{80.58}           & 84.91           \\ \hline
\multirow{3}{*}{GA}                                                                   & CIFAR-100                         & \multicolumn{1}{c|}{\gray{74.89}}            & \multicolumn{1}{c|}{\gray{83.90}}           & \multicolumn{1}{c|}{\gray{67.81}}           & \gray{71.40}           & \multicolumn{1}{c|}{70.60}            & \multicolumn{1}{c|}{82.49}           & \multicolumn{1}{c|}{64.65}           & 66.95           & \multicolumn{1}{c|}{61.49}            & \multicolumn{1}{c|}{61.50}           & \multicolumn{1}{c|}{61.94}           & 67.57           \\ \cline{2-14} 
                                                                                      & CINIC-10                          & \multicolumn{1}{c|}{\gray{56.88}}            & \multicolumn{1}{c|}{\gray{58.37}}           & \multicolumn{1}{c|}{\gray{51.88}}           & \gray{59.13}           & \multicolumn{1}{c|}{50.02}            & \multicolumn{1}{c|}{51.23}           & \multicolumn{1}{c|}{45.85}           & 52.56           & \multicolumn{1}{c|}{43.95}            & \multicolumn{1}{c|}{41.17}           & \multicolumn{1}{c|}{42.02}           & 52.41           \\ \cline{2-14} 
                                                                                      & TinyImageNet                      & \multicolumn{1}{c|}{\gray{82.92}}            & \multicolumn{1}{c|}{\gray{96.98}}           & \multicolumn{1}{c|}{\gray{78.28}}           & \gray{73.64}           & \multicolumn{1}{c|}{72.12}            & \multicolumn{1}{c|}{94.07}           & \multicolumn{1}{c|}{65.94}           & 56.83           & \multicolumn{1}{c|}{63.68}            & \multicolumn{1}{c|}{77.94}           & \multicolumn{1}{c|}{62.35}           & 62.17           \\ \bottomrule
\end{tabular}
}
\end{table*}

\subsection{Performance of \system\ (RQ1)}  \label{subsec-Performance}

In this section, we report the performance results of \system. 
Table~\ref{tab:Performance} reports these results for both  \system\ and the baseline approaches by using the ResNet-18 model as the target model. The results for the  DenseNet and SimpleCNN models can be found in Appendix~\ref{Performance on Other Models}. The attack accuracy results, evaluated using TPR@5 FPR, are provided in Appendix \ref{appendix:TPR-value}. Besides the attack accuracy results, we measure and report the computational overhead of \system\ and baseline approaches in Appendix~\ref{subsec-Computational overhead}. 

{\bf Overall and Per-class Performance.} 
\system\ demonstrates exceptional performance across all the settings, with the overall F1-score no lower than 56.88\% across all configurations. The overall F1-score can be as high as 95.6\% in some settings (e.g., TinyImageNet with SISA unlearning).  

While \system\ demonstrates strong overall effectiveness, its performance varies across the three membership sets. In most cases, it achieves the highest accuracy on the retain set, demonstrating its effectiveness of exposing privacy risks in the remained data. 
We also observe that it has the lowest accuracy on the forget set, possibly due to the fact that the posterior probabilities of removed samples do not change significantly after unlearning, particularly when only a small portion of the data is removed. As a result, the attack features for the instances in the forget set may closely resemble those in the retain set, making them difficult to distinguish and leading to frequent misclassifications. 

Besides disparate performance across different membership classes,  \system\ exhibits disparate performance across different unlearning models. Specifically, it achieves higher attack accuracy against exact unlearning models (Retrain and SISA) compared to their approximate counterparts. This is because approximate unlearning methods cannot fully eliminate the influence of the forget set, causing some forgotten instances to exhibit posterior probability changes similar to those of the retain set, thereby weakening the model’s ability to separate the two.
However, this advantage does not have equal effect by the exact and approximate unlearning methods. In particular, \system\ performs worse on the unseen set when attacking exact unlearning models than when attacking approximate ones. A plausible explanation is that exact unlearning methods mitigate overfitting more effectively, narrowing the gap between seen and unseen samples and thus making it more challenging for \system\ to reliably identify unseen instances. 

{\bf Comparison with Baselines.}  \system\ consistently outperforms both baseline approaches in terms of both overall F1-score and per-class performance across most settings. For instance, it achieves an average improvement of 8.21\% over U-Leak and 18.72\% over the two-round attack in overall performance. The only two exceptions occur when SISA is used as the unlearning method with CIFAR-100 and CINIC-10 as the training datasets, where U-Leak slightly surpasses \system\ in attack accuracy on the retain set. However, even in these cases, the attack accuracy of \system\ remains close to the strongest baseline, differing by only about 5\%. 
In addition, \system\ shows the most substantial performance gains on the forget set among the three methods. For example, in the setting where retraining is used as the unlearning method with CIFAR-100 as the training dataset, \system\ outperforms U-Leak and the two-round attack by 18.27\% and 21.40\%, respectively.

\begin{table}[!t]
\centering
\caption{Attack accuracy (\%) of ML-leaks \cite{salem2018ml} and \system\  in inferring the retain set before and after unlearning (ResNet-18 model). }
\label{tab:attack_performance_retain}
\scalebox{1}{
  \begin{tabular}{l|c|c|c|c}
    \toprule
    \multirow{2}{*}{\textbf{Method}} & \multirow{2}{*}{\textbf{Dataset}}  
    & \multicolumn{2}{c|}{\bf ML-leaks} & \multirow{2}{*}{\textbf{\system} }\\\cline{3-4}
     & & \textbf{Pre-UL} & \textbf{Post-UL} & \\\hline
    \multirow{3}{*}{Retrain} 
    & CIFAR-10 & 80.27 & 92.87 & 96.13 \\
    & CIFAR-100 & 82.00 & 92.65 & 95.67 \\
    & TinyImageNet & 72.30 & 86.43 & 94.63 \\ \hline
    \multirow{3}{*}{SISA}
    & CIFAR-10 & 72.89 & 79.52 & 84.35 \\
    & CIFAR-100 & 63.25 & 74.41 & 79.65 \\
    & TinyImageNet & 70.01 & 79.34 & 88.65 \\  \hline
    \multirow{3}{*}{Sparsity}
    & CIFAR-10 & 81.33 & 84.13 & 88.33 \\
    & CIFAR-100 & 82.53 & 91.07 & 94.33 \\
    & TinyImageNet & 71.07 & 81.27 & 93.77 \\  \hline
    \multirow{3}{*}{SCRUB}
    & CIFAR-10 & 80.60 & 87.65 & 89.37 \\
    & CIFAR-100 & 81.27 & 86.93 & 87.07 \\
    & TinyImageNet & 70.93 & 85.97 & 89.63 \\  \hline
    \multirow{3}{*}{GA}
    & CIFAR-10 & 79.13 & 88.00 & 92.40 \\
    & CIFAR-100 & 82.43 & 88.40 & 90.33 \\
    & TinyImageNet & 71.07 & 82.93 & 91.43 \\
    \bottomrule
  \end{tabular}
  }
\end{table}

\subsection{Privacy Risks of Retained Data (RQ2)} 

To investigate whether the “privacy onion” effect \cite{onion} extends to the population level, we measure attack accuracy on the retained set using both \system\ and ML-leaks~\cite{salem2018ml} against the original and unlearned models, where the unlearned models are produced by five different unlearning algorithms. The results are summarized in Table~\ref{tab:attack_performance_retain}. We observe that the “privacy onion” effect indeed manifests at the population level for both attacks. Specifically, ML-leaks can expose additional privacy risks in the retained data, as indicated by higher attack accuracy on the post-unlearning model compared to the pre-unlearning model. Furthermore, \system~achieves even higher attack accuracy by leveraging information from both models, improving performance by up to 12\% over ML-leaks. These findings demonstrate that \system\ more effectively uncovers collateral privacy risks in the retained set introduced by unlearning.

\textit{Why does the retain set become more vulnerable after unlearning?} 
From an optimization perspective, removing the forget set effectively changes the empirical risk objective by eliminating its contribution. As a result, gradient updates are reallocated to better fit the retain set but further away from both unseen and forget sets. This leads to larger output-space separations between retain from the other two sets. Empirically, we observe this effect through both increased accuracy gap between the retain and the other two sets and a measurable rise in the Euclidean distance between their outputs (as reported in Appendix \ref{subsec-Retain Set's Increased Vulnerability}). 


\begin{figure*}[t!]
    \centering
    \begin{subfigure}[b]{0.3\textwidth} 
        \includegraphics[width=\textwidth]{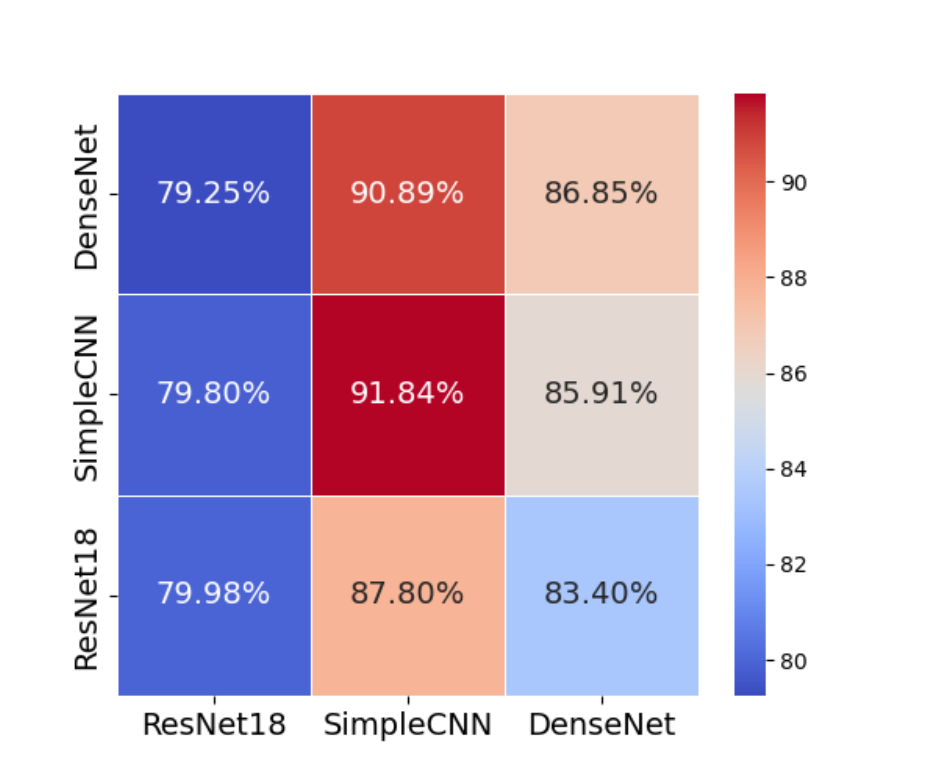} 
        \caption{Across different DNNs}
        \label{fig:model-transfer}
    \end{subfigure}
    \begin{subfigure}[b]{0.3\textwidth}
        \includegraphics[width=\textwidth]{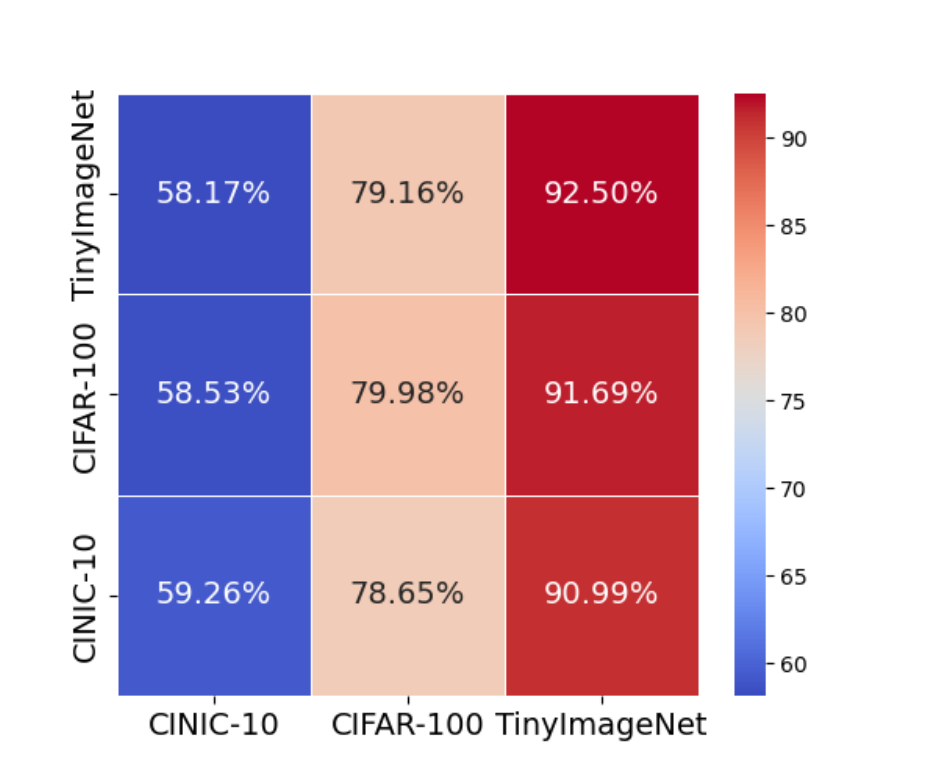} 
        \caption{Across different datasets}
        \label{fig:dataset-transfer}
    \end{subfigure}
     \begin{subfigure}[b]{0.3\textwidth}
        \includegraphics[width=\textwidth]{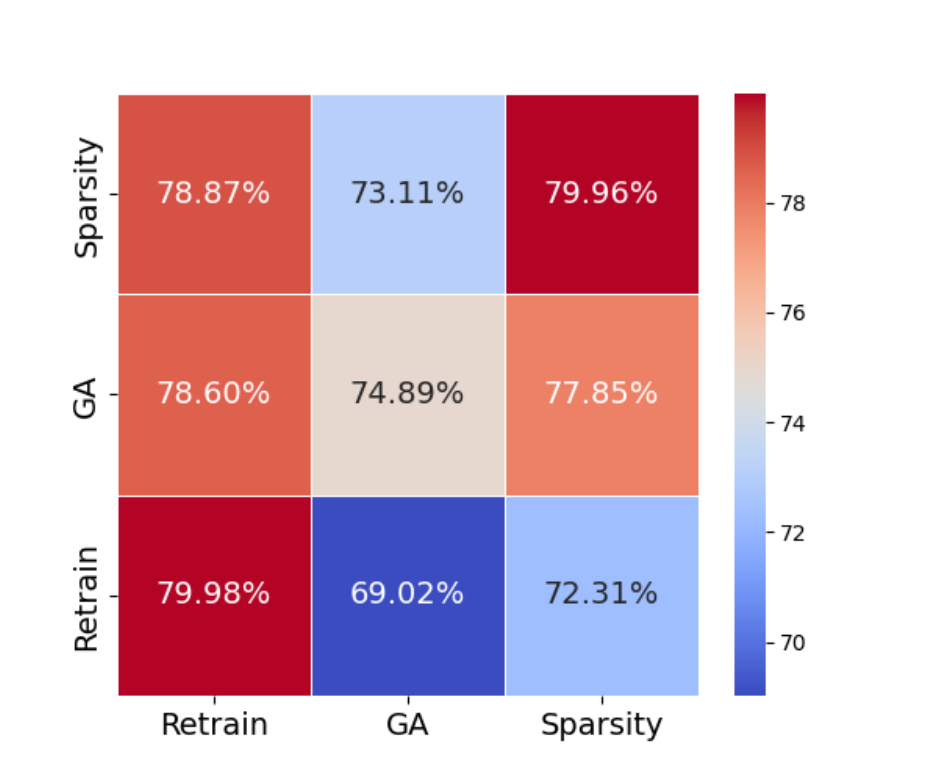}
        \caption{Across different MU methods}
        \label{fig:algorithm-transfer}
    \end{subfigure}
     \caption{Transferability of \system. The horizontal and vertical axes represent the target model/dataset/MU methods and the shadow model/dataset/MU methods, respectively.}
    \label{figure:Attack-transferability}
\end{figure*}

\begin{figure*}[t!]
    \centering
    \begin{subfigure}[b]{0.3\textwidth}
        \includegraphics[width=\textwidth]{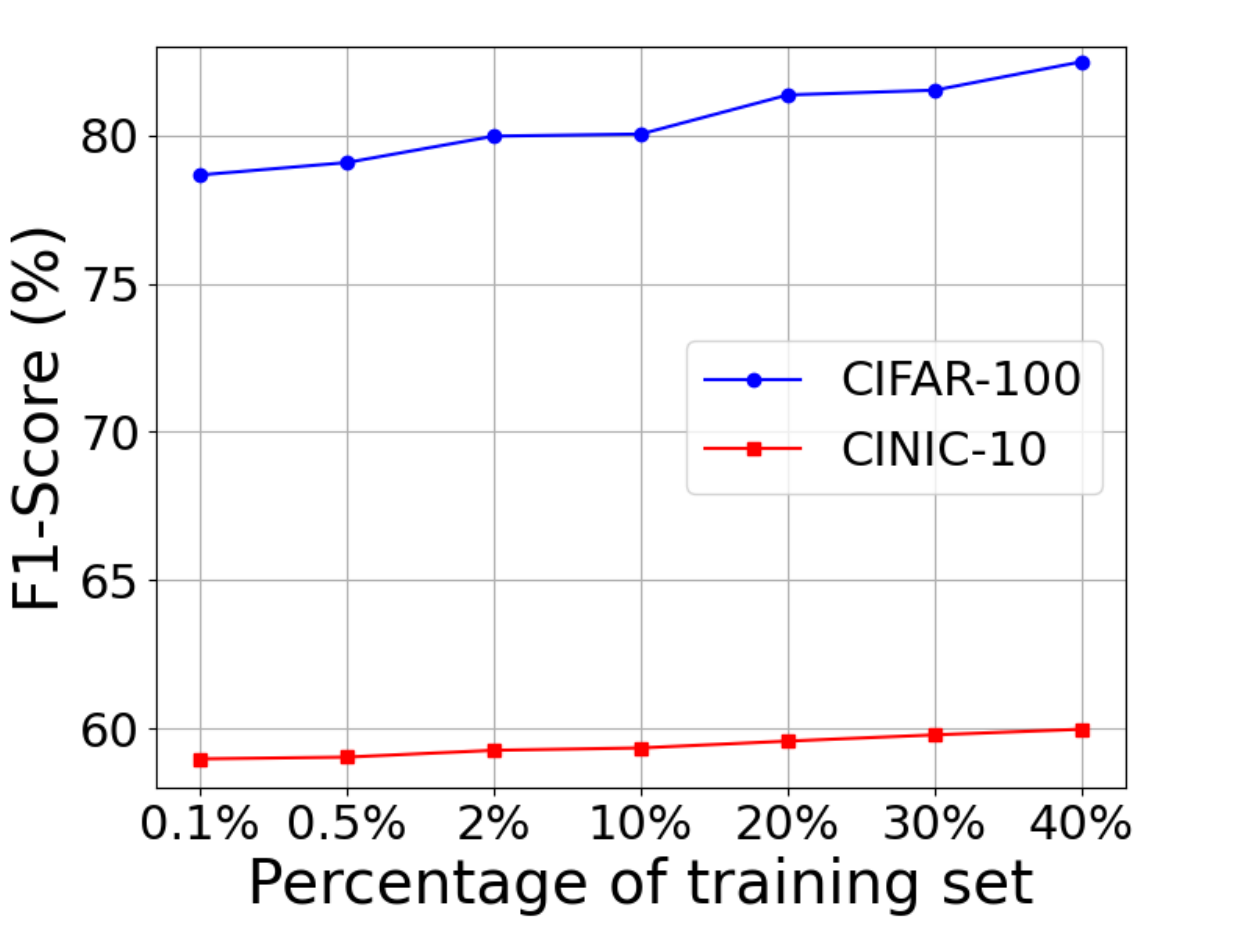} 
        \caption{Size of forget set }
\label{fig:impact_of_various_parameters_size_of_forget_set_retrain}
    \end{subfigure}
     \begin{subfigure}[b]{0.3\textwidth}
        \includegraphics[width=\textwidth]{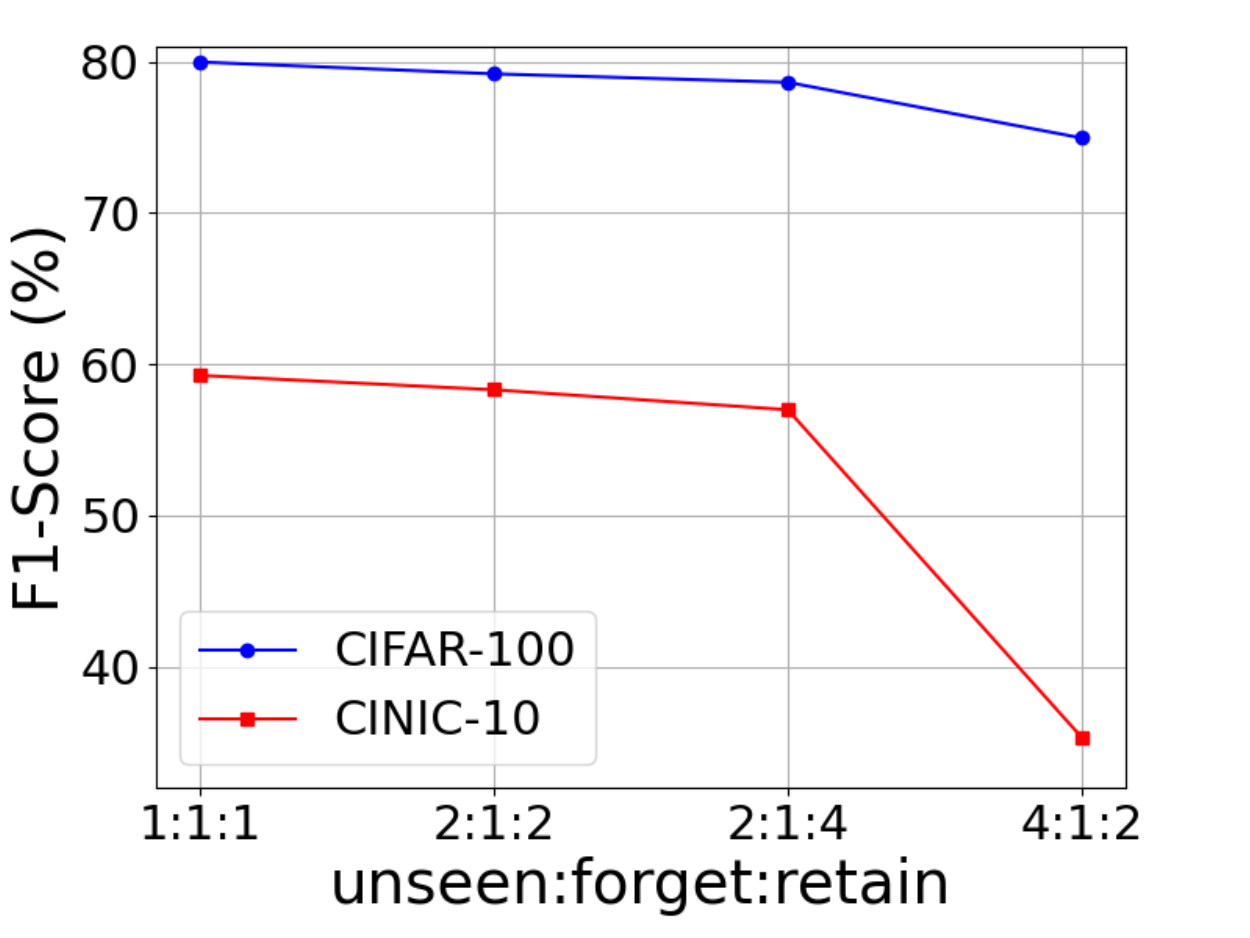} 
        \caption{Size ratio of three classes}
\label{fig:impact_of_various_size_ratios}
    \end{subfigure}
           \begin{subfigure}[b]{0.3\textwidth} 
        \includegraphics[width=\textwidth]{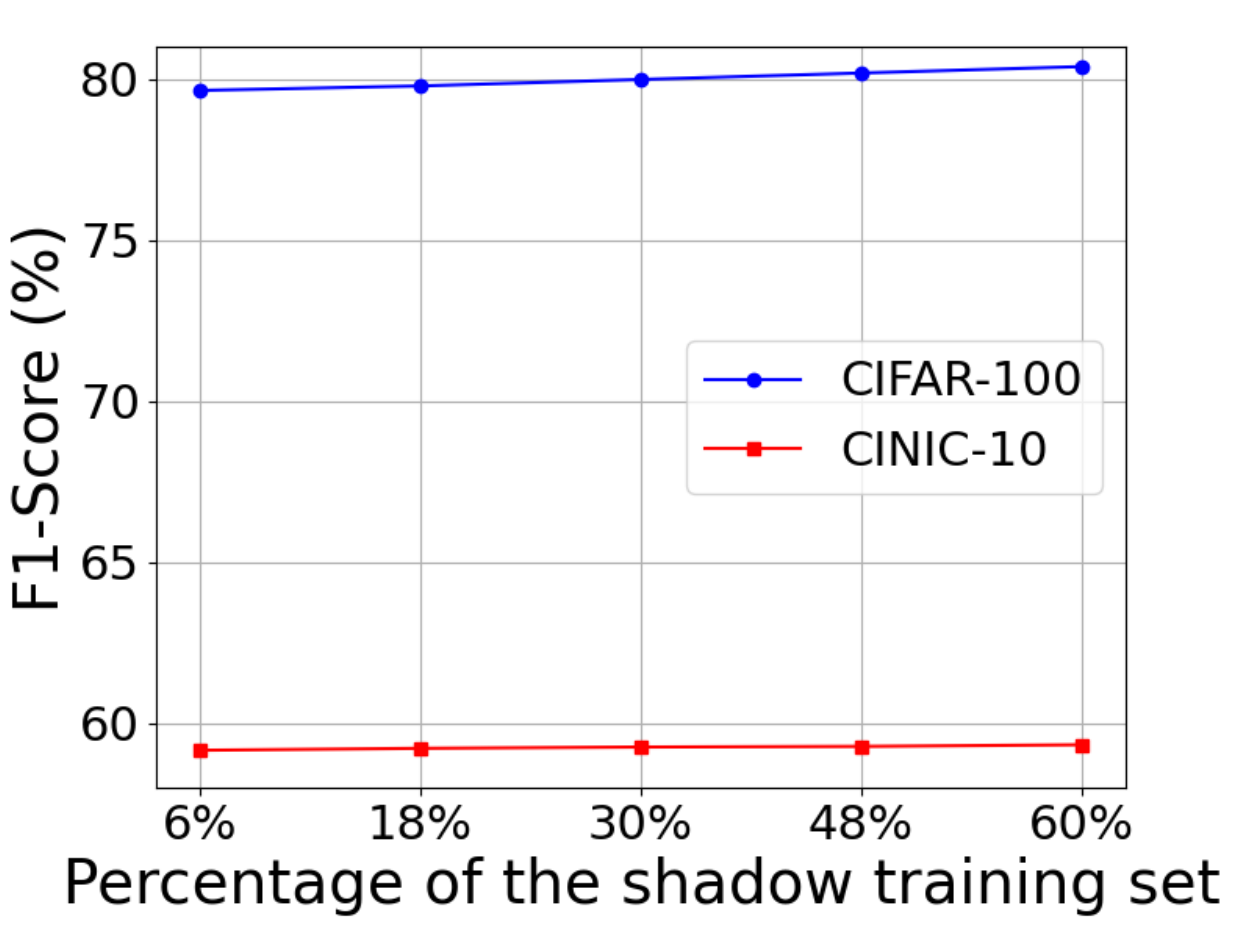} 
        \caption{Size of attack training set}
\label{fig:impact_of_various_parameters_size_of_shadow_set}
    \end{subfigure}
     \caption{Impact of various factors on the performance of \system\ (ResNet-18)} 
    \label{figure:factor-analysis}
\end{figure*}

\subsection{Effects of Unlearning-Induced Overfitting Change on Privacy Risk (RQ3)}
\label{sc:overfitting}

\begin{table}[!t]
\centering
\caption{\label{tab:impact-overfitting}Impact of change in model overfitting on attack accuracy (F1-score in \%). }
\scalebox{0.9}{
\begin{tabular}{c|c|c|c|c|c|c}
\toprule
\multirow{2}{*}{\textbf{Dataset}} & \multicolumn{2}{c|}{{\bf Overfitting}} &\multirow{2}{*}{ \textbf{All}} & \multirow{2}{*}{\textbf{Unseen}} & \multirow{2}{*}{\textbf{Forget}} & \multirow{2}{*}{\textbf{Retain}} \\ \cline{2-3}
& {\bf Pre-UL} & {\bf Post-UL} & & & & \\\hline
\multirow{4}{*}{CIFAR-10} & Low                               & Low                                  & 37.51          & 31.42         & 30.42         & 45.27         \\
                          & Low                                & High                               & 46.13          & 35.00         & 35.14         & 57.60         \\
                          & High                                & Low                                 & 47.60          & 50.59         & 38.10         & 51.17         \\
                          & High                               & High                                 & 53.38          & 52.14         & 43.69         & 59.16         \\ \hline
\multirow{4}{*}{CINIC-10} & Low                               & Low                                & 41.17          & 40.78         & 37.29         & 44.12         \\
                          & Low                                & High                                & 49.60          & 42.83         & 44.21         & 57.82         \\
                          & High                                & Low                                 & 50.54          & 58.51         & 44.80         & 49.06         \\
                          & High                                & High                                 & 60.11          & 60.77         & 54.51         & 63.21         \\ \bottomrule
\end{tabular}
}
\end{table}

Previous research has demonstrated that MIAs are particularly effective against overfitted models \cite{shokri2017membership, carlini2022membership, yeom2018privacy}. Intuitively, removing data samples through unlearning can alter a model’s degree of overfitting, which in turn may affect the performance of \system. In this set of experiments, we investigate how changes in overfitting, caused by unlearning, influence \system's effectiveness.

We quantify the {\itshape overfitting degree} of a given model as the difference between the model’s training and testing accuracy. Specifically, we examine four scenarios: (1) both the original and unlearned models exhibit low overfitting, (2) both exhibit high overfitting, (3) the model transitions from low to high overfitting after unlearning, and (4) the model transitions from high to low overfitting.

Table~\ref{tab:impact-overfitting} presents \system's accuracy under the four different scenarios, using ResNet-18 models trained on the CIFAR-10 and CINIC-10 datasets as the target models. We make the following observations:
First, \system\ achieves the highest performance - both overall and per-class - when both the original and unlearned models exhibit a high degree of overfitting. In contrast, its performance is lowest when both models show minimal overfitting. This is because greater overfitting amplifies the differences in prediction outputs between member samples (i.e., retain and forget sets) and non-member samples (i.e., unseen set), allowing \system\ to more effectively distinguish among the three.
Additionally, \system\ performs better when transitioning from a high-overfitting original model to a low-overfitting unlearned model, compared to the reverse scenario. This suggests that the degree of overfitting in the unlearned model plays a more critical role in \system's effectiveness than the overfitting level of the original model.

\subsection{Attack Transferability (RQ4)} 
\label{subsec-Attack Transferability}
In practice, an adversary's knowledge of the shadow model, shadow dataset, and unlearning algorithms may differ from those used by the server. Therefore, in this set of experiments, we evaluate the performance of \system\ under these settings. We show the overall performance of \system\ as follows. The per-class performance results can be found in Appendix~\ref{appendix-Attack-Transferability}.

{\bf Across different model architectures.} We train three models (ResNet-18, SimpleCNN, and DenseNet) on the CIFAR-100 dataset and use them as target and shadow models alternatively, with retraining as the unlearning algorithm. Figure~\ref{fig:model-transfer} shows the overall F1-score of \system\ across these configurations.
The key observation is that \system\ achieves the highest F1-score when the target and shadow models share the same architecture. However, it remains effective even when transferred across different model architectures, with an F1-score reduction of no more than 4.04\% in all cases.

{\bf Across different datasets.} We use ResNet-18 as the target and shadow models, trained separately on CINIC-10, CIFAR-100, and TinyImageNet datasets, and adopt retraining as the unlearning method. Figure~\ref{fig:dataset-transfer} shows the overall F1-score of \system\ across these settings.  We observe that \system\ remains effective in all cases, with the F1-score decreasing by no more than 1.54\% compared to the setting where the target and shadow models are trained on the same dataset. We attribute this robustness to the similarity in decision boundaries learned by these models, despite the differences in the distribution of their training data.

{\bf Across different unlearning algorithms.}  We consider ResNet-18 trained on the CIFAR-100 dataset as both target model and shadow model, and use three unlearning algorithms (Retrain, GA \cite{graves2021amnesiac}, and Sparsity \cite{liu2024model}) on the target model and shadow model, respectively. Figure~\ref{fig:algorithm-transfer}  presents the overall F1-score of \system\ for these settings. 
The key observation is that while \system\ achieves the highest F1-score when both the original and shadow models use the same unlearning algorithm, it shows only a marginal decrease in accuracy when transferred across different unlearning algorithms, with an F1-score drop of at most 7.65\% across all settings. This demonstrates that \system\ is capable of effectively transferring knowledge across diverse unlearning scenarios, maintaining strong performance even in mismatched conditions.

\subsection{Parameter Sensitivity Analysis (RQ5)} 
\label{sc:factor}

In this section, we evaluate the impact of three parameters: {\itshape size of the forget set}, {\itshape size distribution of the three membership sets}, and {\itshape size of the attack training set}, on the performance of $\system$. Besides these evaluated parameters, we also evaluated the impact of the type of removed samples on the attack performance, and included the results in Appendix \ref{appendix:removedsample}.  
All results reported in this section are evaluated over the ResNet-18 model trained on CINIC-10 and CIFAR-100 datasets.

\textbf{Size of Forget Set.} Intuitively, the number of samples removed from the model may influence the performance of \system. To examine this effect, we vary the size of the forget set to be \{0.1\%, 0.5\%, 2\%, 10\%, 20\%, 30\%, 40\%\} of the training dataset and present the results in Figure~\ref{fig:impact_of_various_parameters_size_of_forget_set_retrain}. We observe that, for both datasets, \system\ achieves higher accuracy as the size of the forget set increases. We hypothesize that this is because removing a larger number of samples induces more significant changes in the model's predictions, thereby allowing \system\ to operate more effectively. 
Notably, when the forget set constitutes an extremely small proportion, removing such a small subset may induce only minimal changes in the model’s outputs, which potentially weakens the attack. However, we observe a slight degradation in attack accuracy when the forget set accounts for only 0.1\% or 0.5\% of the training data, demonstrating its robustness to small-scale data removal.

\begin{figure*}[!t]
    \centering
    \begin{subfigure}[b]{0.245\textwidth} 
        \includegraphics[width=\textwidth]{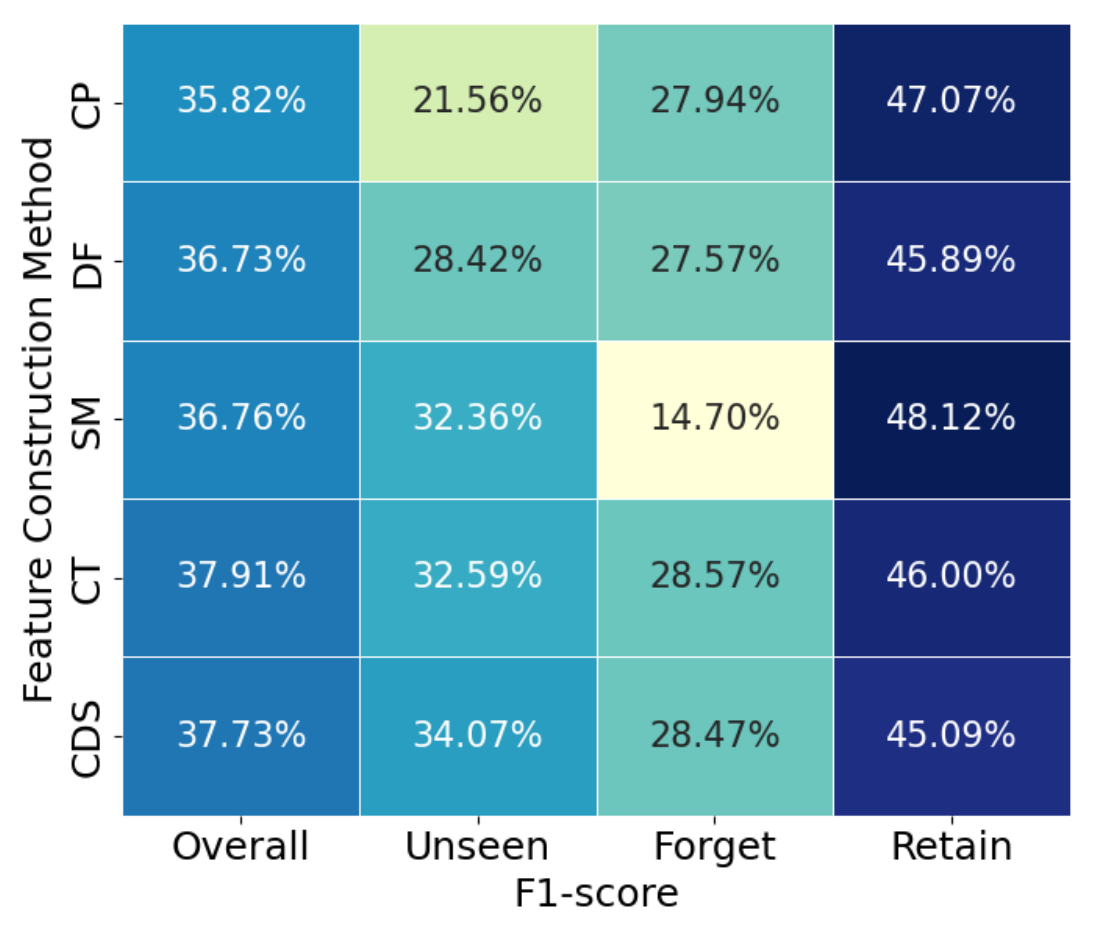} 
        \caption{Well-gen. $\rightarrow$ well-gen.}
        \label{fig:find_optimal_feature_well_to_well}
    \end{subfigure}
    \begin{subfigure}[b]{0.245\textwidth}
        \includegraphics[width=\textwidth]{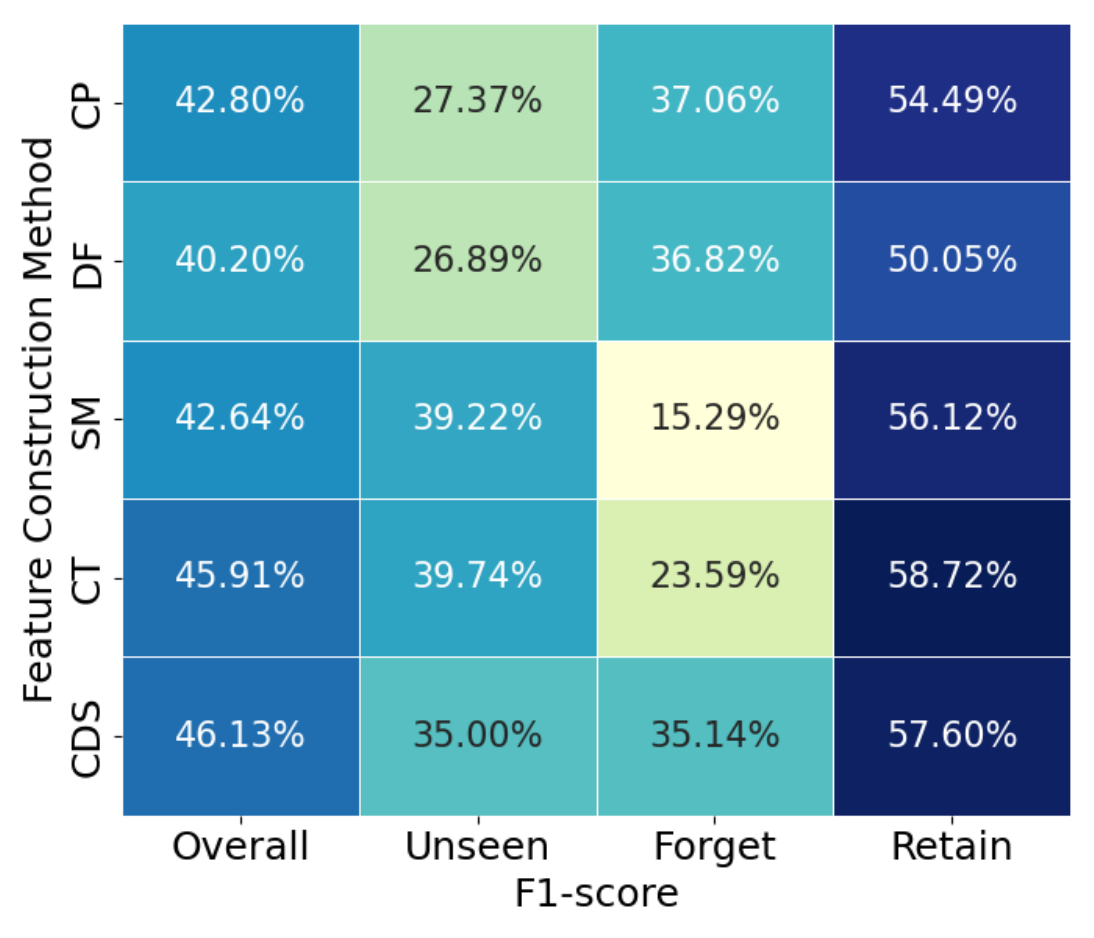} 
        \caption{Well-gen. $\rightarrow$ overfitting}
        \label{fig:find_optimal_feature_well_to_over}
    \end{subfigure}
     \begin{subfigure}[b]{0.245\textwidth}
        \includegraphics[width=\textwidth]{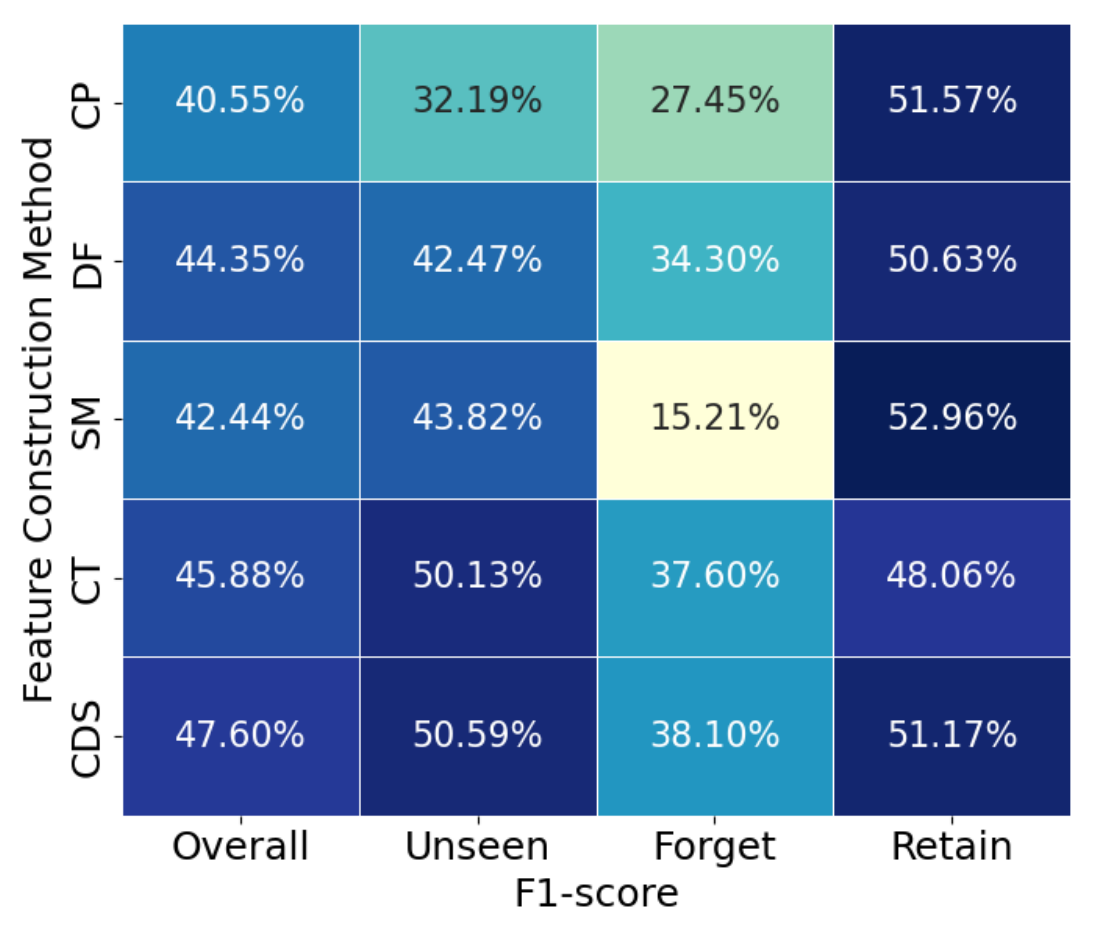}
        \caption{Overfitting $\rightarrow$ well-gen.}
        \label{fig:find_optimal_feature_over_to_well}
    \end{subfigure}
    \begin{subfigure}[b]{0.245\textwidth}
        \includegraphics[width=\textwidth]{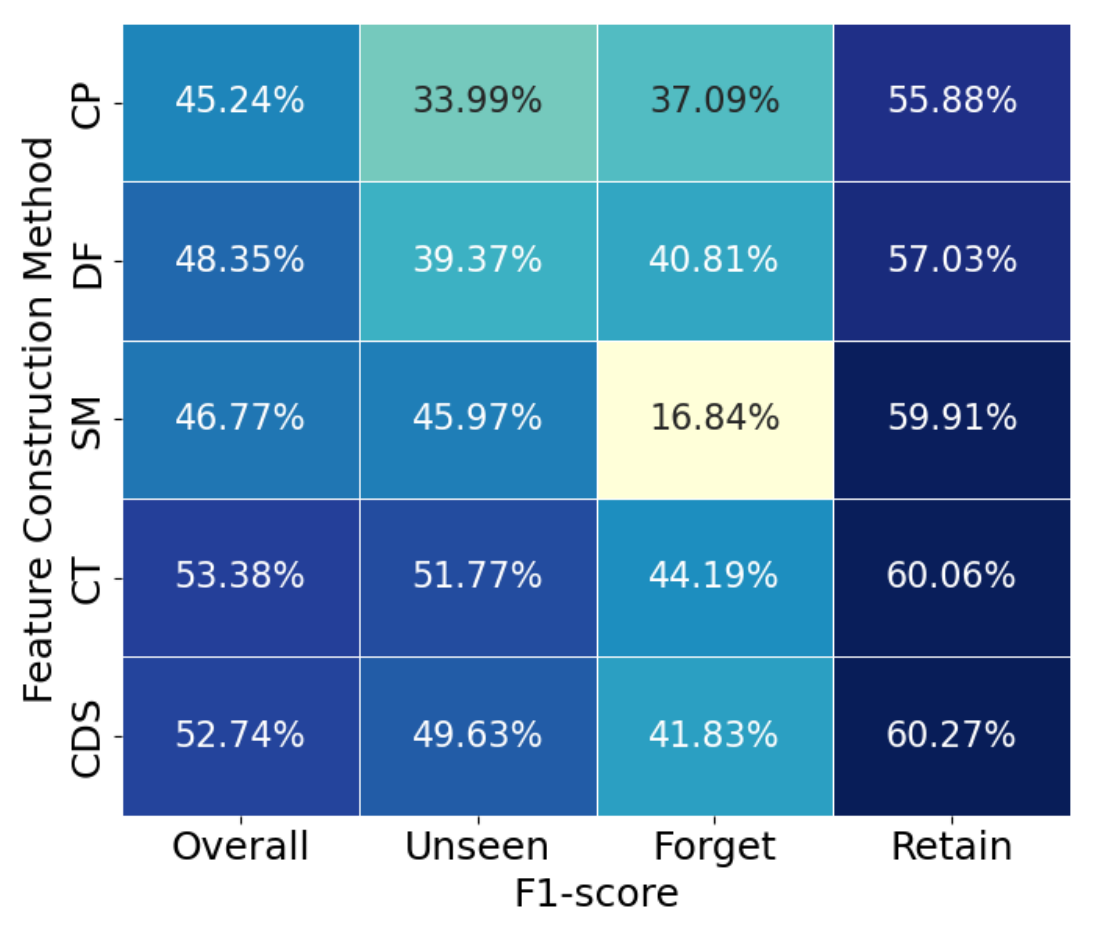}
        \caption{Overfitting $\rightarrow$ overfitting}
        \label{fig:find_optimal_feature_over_to_over}
    \end{subfigure}
    \caption{\system's accuracy (F1-score) using different features. \texttt{CP}, \texttt{DF}, \texttt{SM}, \texttt{CT}, and \texttt{CDS} stand for $\mathbb{P} \parallel \mathbb{P}^-$, $\mathbb{P}_{y} - \mathbb{P}^-_{y}$, $\mathbb{P}_{y} + \mathbb{P}^-_{y}$, $\mathbb{P}_{y} \parallel \mathbb{P}^-_{y}$, $\big(\mathbb{P}_{y} - \mathbb{P}^-_{y}\big) \parallel \big(\mathbb{P}_{y} + \mathbb{P}^-_{y}\big)$, respectively. }
    \label{figure:find_optimal_features}
\end{figure*}

\textbf{Class Distribution.} 
In real-world scenarios, typically only a small number of samples are removed from a trained model. As a result, the three sets are likely to have an imbalanced size distribution, with the forget set significantly smaller than the other two. To examine the impact of this imbalance on \system's performance, we vary the size ratio among the unseen, forget, and retain sets as \{1:1:1, 2:1:2, 4:1:2, 2:1:4\}. As shown in Figure~\ref{fig:impact_of_various_size_ratios}, we observe that increasing the imbalance - particularly when the forget set is relatively much smaller than the other two sets - leads to a noticeable degradation in \system's attack accuracy. This performance drop is likely due to the attack classifier's bias toward the majority class, which diminishes its ability to accurately identify samples from the minority class, namely the forget set.

\textbf{Size of Attack Training Set.} Intuitively, since \system\ is a multi-label classifier, its performance depends on the size of its training data. To investigate this, we vary the size of the attack training set from 6\% to 60\% of the shadow training set and evaluate \system's accuracy. As shown in Figure~\ref{fig:impact_of_various_parameters_size_of_shadow_set}, increasing the size of the attack training set consistently leads to improved accuracy.

\subsection{Finding Optimal Attack Features (RQ6)} \label{subsec-finding-optimal-attack-features}

Recall that we have five alternative methods to construct the attack features (Section \ref{sec-pre-attack}). In this part of experiments, we evaluate the performance of \system\ by using these five types of features. We use ResNet-18 trained on the CIFRA-10 dataset as the target model. 

Since model overfitting significantly impacts \system's performance (Section~\ref{sc:overfitting}), evaluating attack features under a single overfitting condition may yield biased conclusions. To ensure a fair and comprehensive analysis, we examine both overfitting and well-generalized models. Specifically, we consider four scenarios: (1) both the original and unlearned models are well-generalized, (2) both are overfitting, (3) the model transitions from well-generalized to overfitting after unlearning, and (4) the model transitions from overfitting to well-generalized. We then assess the effectiveness of five alternative attack features across these scenarios. Figure~\ref{figure:find_optimal_features} presents the corresponding F1-scores of \system, from which we derive the following observations.

\textbf{Full Posterior (CP) vs. True-label Posterior (CT).} CT consistently outperforms CP in most scenarios. This is likely because using the full posterior introduces additional noise from irrelevant posterior  probabilities, which can degrade attack performance. In contrast, focusing solely on the posterior of the true label enables more effective differentiation among the three sets. 

\textbf{Addition (SM) vs Subtraction (DF) vs Combination of Both (CDS).}
DF performs poorly on both the unseen and retain sets, while SM exhibits lower accuracy on the forget set. In contrast, CDS consistently outperforms both SM and DF across most settings. These results align with our observations in Section~\ref{sec-pre-attack}: the difference in posterior probabilities before and after unlearning is insufficient to distinguish instances in the unseen set from those in the retain set, while the sum of posterior probabilities alone fails to effectively separate unseen instances from those in the forget set. This highlights the necessity of constructing attack features using both the additive and subtractive components of the posterior probabilities before and after unlearning.

\textbf{CT vs. CDS.} Overall, both CT and CDS demonstrate strong performance across all scenarios. Specifically, CT outperforms CDS when both the original and unlearned models are either well-generalized or highly overfitted. In contrast, CDS shows better performance when there is a change in the degree of overfitting - i.e., when the model transitions from well-generalized to overfitted or vice versa. Nevertheless, the performance gap between CT and CDS is marginal, with a maximum difference of no more than 1.72\%.

\textbf{Summary.} Based on our findings, we make the following recommendations:
(1) Use the posterior probability of the true label rather than the full posterior, as it provides more discriminative information with less noise;  (2) Incorporate both additive and subtractive features, rather than relying on only one, to better capture the nuanced changes introduced by unlearning; (3) While both CT and CDS are effective feature construction methods, the choice between them can be guided by the change in the model's overfitting degree.

\section{Potential Defense} \label{subsec-Defense}
In this section, we explore possible defenses to mitigate the privacy risks of \system. 

\begin{table*}[!t]
\center
\caption{Performance of defense mechanisms (CNN model). } 
\label{tab:potential mitigation}
\begin{tabular}{c|l|cccc|c|c}
\toprule
\multirow{2}{*}{\textbf{Dataset}} & \multirow{2}{*}{\textbf{Defense}} & \multicolumn{4}{c|}{{\bf \system}} & \multicolumn{2}{c}{\textbf{Model Utility}} \\ \cline{3-8}
& & \multicolumn{1}{c|}{\textbf{All}} & \multicolumn{1}{c|}{\textbf{Unseen}} & \multicolumn{1}{c|}{\textbf{Forget}} & \textbf{Retain} & $\mathbf{\textbf{Train}_{\textbf{Acc}}}$& $\mathbf{\textbf{Test}_{\textbf{Acc}}}$\\ \hline 
\multirow{5}{*}{CIFAR-10} & W/o defense & \multicolumn{1}{c|}{62.87} & \multicolumn{1}{c|}{63.17} & \multicolumn{1}{c|}{55.42} & 66.52 & \multicolumn{1}{c|}{100} & 68.03 \\ \cline{2-8}
& Label-only output & \multicolumn{1}{c|}{53.33} & \multicolumn{1}{c|}{45.60} & \multicolumn{1}{c|}{42.77} & 60.98 & \multicolumn{1}{c|}{100} & 68.03 \\ \cline{2-8}
& Dropout& \multicolumn{1}{c|}{43.93} & \multicolumn{1}{c|}{48.55} & \multicolumn{1}{c|}{36.31} & 42.23 & \multicolumn{1}{c|}{92.00} & 63.22 \\ \cline{2-8}
& DP$_{\epsilon=5}$ & \multicolumn{1}{c|}{34.60} & \multicolumn{1}{c|}{44.25} & \multicolumn{1}{c|}{35.24} & 12.33 & \multicolumn{1}{c|}{51.02} & 49.77 \\ \cline{2-8}
& DP$_{\epsilon=2}$ & \multicolumn{1}{c|}{33.71} & \multicolumn{1}{c|}{42.85} & \multicolumn{1}{c|}{32.59} & 11.69 & \multicolumn{1}{c|}{48.36} & 47.01 \\ \cline{1-8}
\multirow{5}{*}{CINIC-10} & W/o defense & \multicolumn{1}{c|}{77.19} & \multicolumn{1}{c|}{77.23} & \multicolumn{1}{c|}{75.07} & 78.21 & \multicolumn{1}{c|}{100} & 55.63 \\ \cline{2-8}
& Label-only output & \multicolumn{1}{c|}{62.54} & \multicolumn{1}{c|}{60.54} & \multicolumn{1}{c|}{56.25} & 66.81 & \multicolumn{1}{c|}{100} & 55.63 \\ \cline{2-8}
& Dropout  & \multicolumn{1}{c|}{58.36} & \multicolumn{1}{c|}{69.42} & \multicolumn{1}{c|}{49.14} & 58.52 & \multicolumn{1}{c|}{88.00} & 54.85 \\ \cline{2-8}
& DP$_{\epsilon=5}$ & \multicolumn{1}{c|}{35.11} & \multicolumn{1}{c|}{45.16} & \multicolumn{1}{c|}{34.39} & 13.13 & \multicolumn{1}{c|}{48.94} & 46.78 \\ \cline{2-8}
& DP$_{\epsilon=2}$ & \multicolumn{1}{c|}{34.85} & \multicolumn{1}{c|}{44.51} & \multicolumn{1}{c|}{32.70} & 12.13 & \multicolumn{1}{c|}{41.62} & 40.06 \\ \bottomrule
\end{tabular}
\end{table*}

\subsection{Defense Mechanisms}
We adapt three defense strategies, namely {\itshape label-only output}, {\itshape dropout}, and {\itshape differential privacy}, that have proven effective against MIAs, to our setting.

\smallskip
{\bf Label-only Output}. Returning labels instead of posterior probabilities has been shown to be an effective defense against existing MIAs \cite{chen2021machine, choquette2021label}. However, adapting this defense to counter \system\ requires additional consideration. Specifically, since \system\ relies on changes in the posterior probability of the ground-truth label to construct its attack features, we modify the attack to function in a label-only setting as follows: if the predicted label matches the ground-truth label, we assign a posterior probability of 1; otherwise, we assign 0. \system's attack features are then derived from these binary values. 


\smallskip
{\bf Dropout}. Dropout has been used to defend against MIAs as it can mitigate overfitting ~\cite{kaya2020effectiveness,yin2021defending}. Specifically, we employ dropout regularization by randomly dropping out (temporarily deactivating) $p$ percent of neurons in both the input layer and fully-connected layers of the neural network.  We choose the dropout rate parameter $p=95\%$ by following the prior works \cite{kaya2020effectiveness,srivastava2013improving}. To preserve the model utility, we apply dropout to the fully connected layer only. 

\smallskip
{\bf Differential Privacy}. Differential privacy (DP) \cite{dwork2006differential} has become a standard framework for formal privacy guarantees and has been shown to be effective in defending against MIAs~\cite{fu2024dpsur,jia2019memguard,naseri2020local}. To evaluate its impact on \system, we deploy Differentially Private Stochastic Gradient Descent (DP-SGD)~\cite{abadi2016deep}, a state-of-the-art DP mechanism, to both the original and unlearning models. DP-SGD operates by clipping per-example gradients and adding Gaussian noise to the aggregated gradients during training. The strength of the privacy guarantee is governed by two parameters: $\epsilon$ and $\delta$, with smaller values indicating stronger privacy protection. We use $\epsilon=\{2,5\}$ and $\delta=5*1e-4$  in our experiments. 
\subsection{Results} \label{sec:defense-rusult}

In this part of experiments, we use retraining as the  unlearning algorithm, and evaluate the defense performance over SimpleCNN model.
Besides attack performance, we report {\itshape model utility} in terms of classification accuracy, defined as the percentage of the samples in the testing data that are correctly classified.

{\bf Defense Effectiveness. } 
As reported in Table~\ref{tab:potential mitigation}, all three defense mechanisms reduce \system's attack accuracy, demonstrating their effectiveness in mitigating privacy risks. Among them, the DP-based method yields the lowest overall attack accuracy, indicating the strongest defensive capability. Furthermore, its effectiveness improves as the privacy budget $\epsilon$ decreases, offering increasingly stronger privacy guarantees. In contrast, the label-only output mechanism provides the weakest protection among the three approaches.

We also observe that all three defenses can effectively reduce the attack accuracy across the three membership classes. Among them, the DP-based method performs the best, achieving the lowest attack accuracy on all three classes compared to the other defense mechanisms. 

\smallskip
{\bf Trade-off between Defense Power and Model Accuracy.} The three defense mechanisms introduce varying degrees of loss in model accuracy, reflecting the inherent trade-off between privacy and utility. The label-only output mechanism, which does not interfere with the training process and simply limits the model’s output to predicted labels rather than probabilities, preserves model performance while reducing the information exposed to adversaries. Dropout, by randomly discarding neurons during training, helps mitigate overfitting and improves generalization, but at the cost of a slight reduction in test accuracy. In contrast, DP provides the strongest formal privacy guarantees by injecting noise into the training process; however, this comes at the expense of a substantial decline in model accuracy, especially when the privacy budget is tight.
Taken together, these results highlight the privacy–utility trade-off inherent in defense mechanisms. While DP offers rigorous protection, it significantly undermines predictive performance. Dropout provides moderate protection with relatively minor accuracy degradation. The label-only output approach, although less powerful in terms of theoretical privacy guarantees, causes the least loss in model utility. Overall, among the three defense mechanisms, the dropout approach offers the most favorable balance between privacy protection and model utility.

\vspace{-0.05in}
\section{Attack and Defense over Language Models}
\label{exp:lm}

\definecolor{light-gray}{gray}{0.8}

\begin{table*}[!t]
\centering
\caption{Overall and per-class F1-score (\%) of the attacks over the Pythia-70m model. The best results of the three attacks per evaluation metric are highlighted in gray.}
\label{tab:Performance2}
\scalebox{0.9}{
\begin{tabular}{c|c|cccc|cccc|cccc}
\toprule
\multirow{2}{*}{\textbf{\begin{tabular}[c]{@{}c@{}}Unlearning\\ method\end{tabular}}} & \multirow{2}{*}{\textbf{Dataset}} & \multicolumn{4}{c|}{\textbf{\system\ (Ours)}}                                                                                           & \multicolumn{4}{c|}{\textbf{U-Leak (Best performance)}}                                                                               & \multicolumn{4}{c}{\textbf{Two-round Attack}}                                                                                            \\ \cline{3-14} 
                                                                                      &                                   & \multicolumn{1}{c|}{\textbf{All}} & \multicolumn{1}{c|}{\textbf{Unseen}} & \multicolumn{1}{c|}{\textbf{Forget}} & \textbf{Retain} & \multicolumn{1}{c|}{\textbf{All}} & \multicolumn{1}{c|}{\textbf{Unseen}} & \multicolumn{1}{c|}{\textbf{Forget}} & \textbf{Retain} & \multicolumn{1}{c|}{\textbf{All}} & \multicolumn{1}{c|}{\textbf{Unseen}} & \multicolumn{1}{c|}{\textbf{Forget}} & \textbf{Retain} \\ \hline
\multirow{2}{*}{Retrain}                                                              & SST5                              & \multicolumn{1}{c|}{\gray{73.68}}            & \multicolumn{1}{c|}{\gray{75.99}}           & \multicolumn{1}{c|}{\gray{71.12}}           & \gray{73.99}           & \multicolumn{1}{c|}{62.04}            & \multicolumn{1}{c|}{54.20}           & \multicolumn{1}{c|}{54.30}           & 74.07           & \multicolumn{1}{c|}{46.16}            & \multicolumn{1}{c|}{31.78}           & \multicolumn{1}{c|}{43.15}           & 63.96           \\ \cline{2-14} 
                                                                                      & News20                            & \multicolumn{1}{c|}{\gray{64.11}}            & \multicolumn{1}{c|}{\gray{65.43}}           & \multicolumn{1}{c|}{\gray{58.01}}           & \gray{67.02}           & \multicolumn{1}{c|}{57.42}            & \multicolumn{1}{c|}{58.96}           & \multicolumn{1}{c|}{44.49}           & 63.42           & \multicolumn{1}{c|}{51.62}            & \multicolumn{1}{c|}{45.29}           & \multicolumn{1}{c|}{42.91}           & 63.48           \\ \hline
\multirow{2}{*}{GA}                                                                   & SST5                              & \multicolumn{1}{c|}{\gray{68.52}}            & \multicolumn{1}{c|}{\gray{75.06}}           & \multicolumn{1}{c|}{\gray{56.14}}           & \gray{71.68}           & \multicolumn{1}{c|}{55.95}            & \multicolumn{1}{c|}{42.32}           & \multicolumn{1}{c|}{50.61}           & 65.00           & \multicolumn{1}{c|}{35.45}            & \multicolumn{1}{c|}{26.12}           & \multicolumn{1}{c|}{10.00}           & 56.36           \\ \cline{2-14} 
                                                                                      & News20                            & \multicolumn{1}{c|}{\gray{59.49}}            & \multicolumn{1}{c|}{\gray{64.58}}           & \multicolumn{1}{c|}{\gray{53.38}}           & \gray{60.22}           & \multicolumn{1}{c|}{47.79}            & \multicolumn{1}{c|}{43.32}           & \multicolumn{1}{c|}{31.03}           & 59.17           & \multicolumn{1}{c|}{40.61}            & \multicolumn{1}{c|}{41.61}           & \multicolumn{1}{c|}{26.00}           & 51.76           \\ \hline
\multirow{2}{*}{NPO}                                                                  & SST5                              & \multicolumn{1}{c|}{\gray{70.63}}            & \multicolumn{1}{c|}{\gray{79.06}}           & \multicolumn{1}{c|}{\gray{60.36}}           & \gray{71.26}           & \multicolumn{1}{c|}{62.83}            & \multicolumn{1}{c|}{60.89}           & \multicolumn{1}{c|}{57.96}           & 66.31           & \multicolumn{1}{c|}{45.50}            & \multicolumn{1}{c|}{47.90}           & \multicolumn{1}{c|}{17.36}           & 59.40           \\ \cline{2-14} 
                                                                                      & News20                            & \multicolumn{1}{c|}{\gray{65.98}}            & \multicolumn{1}{c|}{\gray{70.28}}           & \multicolumn{1}{c|}{\gray{58.43}}           & \gray{67.21}           & \multicolumn{1}{c|}{60.57}            & \multicolumn{1}{c|}{59.75}           & \multicolumn{1}{c|}{50.00}           & 66.39           & \multicolumn{1}{c|}{47.70}            & \multicolumn{1}{c|}{51.05}           & \multicolumn{1}{c|}{20.05}           & 59.55           \\ \bottomrule
\end{tabular}
}
\end{table*}

\begin{table}[!t]
\center
\caption{Performance of three defense mechanisms on language model (Pythia-70m, SST5 dataset).} 
\label{tab:potential mitigation-lanuage}
\scalebox{0.9}{
\begin{tabular}{l|cccc|c|c}
\toprule
\multirow{2}{*}{\textbf{Defense}} & \multicolumn{4}{c|}{{\bf Attack accuracy}} & \multicolumn{2}{c}{\textbf{Model Utility}} \\ \cline{2-7}
& \multicolumn{1}{c|}{\textbf{All}} & \multicolumn{1}{c|}{\textbf{Unseen}} & \multicolumn{1}{c|}{\textbf{Forget}} & \textbf{Retain} & $\mathbf{\textbf{Train}_{\textbf{Acc}}}$& $\mathbf{\textbf{Test}_{\textbf{Acc}}}$\\ \hline 
No defense & \multicolumn{1}{c|}{73.68} & \multicolumn{1}{c|}{75.99} & \multicolumn{1}{c|}{71.12} & 73.99 & \multicolumn{1}{c|}{99.21} & 46.33 \\ \hline
Label only& \multicolumn{1}{c|}{66.27} & \multicolumn{1}{c|}{66.49} & \multicolumn{1}{c|}{63.24} & 67.87 & \multicolumn{1}{c|}{99.21} & 46.33 \\ \hline
Dropout & \multicolumn{1}{c|}{49.74} & \multicolumn{1}{c|}{54.80} & \multicolumn{1}{c|}{47.95} & 43.96 & \multicolumn{1}{c|}{77.63} & 42.05 \\ \hline
DP$_{\epsilon=5}$ & \multicolumn{1}{c|}{34.12} & \multicolumn{1}{c|}{43.76} & \multicolumn{1}{c|}{31.52} & 18.18 & \multicolumn{1}{c|}{32.58} & 31.20 \\ \hline
DP$_{\epsilon=2}$ & \multicolumn{1}{c|}{29.37} & \multicolumn{1}{c|}{39.32} & \multicolumn{1}{c|}{24.05} & 17.34 & \multicolumn{1}{c|}{30.46} & 27.76 \\ \bottomrule
\end{tabular}
}
\end{table}

Given the wide use of language models in real-world applications, in this section, 
we extend our studies to language models. 

{\bf Models and Datasets.} In our empirical studies, we consider the Pythia-70m model~\cite{biderman2023pythia}, a compact transformer-based language model with 70 million parameters, with text classification as the downstream task. 
We fine-tune the model on two text datasets, namely,  SST5~\cite{socher2013recursive} and News20~\cite{albishre2015effective} datasets. The details of the model and the two datasets can be found in Appendix~\ref{appendix:expsetup}.

{\bf Unlearning Algorithms.} We consider the following unlearning setting for each dataset, respectively: 
\begin{ditemize}
    \item {\itshape Sentence forgetting}: We randomly select 2\% of sentences from the SST5 dataset for removal; 
    \item {\itshape Document forgetting}: We randomly select 2\% of news documents from the News20 dataset to be removed from the model. 
\end{ditemize}

We employ two state-of-the-art unlearning methods for language models:
\begin{ditemize}
    \item  {\itshape Gradient Ascent (GA)}: it reduces the likelihood of correct predictions on the forget dataset by applying gradient ascent to the cross-entropy loss \cite{jang2023knowledge,ilharco2022editing}; 
    \item {\itshape Negative Preference Optimization (NPO)}: it treats
the forget set as negative preference data, and adapts the offline DPO objective to lower the accuracy of model’s likelihood predictions for this set \cite{zhang2024negative}. 
\end{ditemize}
As GA and NPO do not
inherently preserve the model utility, we follow the literature \cite{liu2022continual,maini2024tofu} and add a gradient descent learning objective to ensure high model accuracy on the retain set. The performance of the two unlearning methods can be found in Appendix \ref{appendix:expsetup}.

{\bf Attack Performance. } Table~\ref{tab:Performance2} reports the F1-score results of \system\ and two baselines against the Pythia-70m model. The TPR@5\%FPR results can be found in Appendix~\ref{appendix:TPR-value}. Overall, \system\ demonstrates strong attack performance, with the overall F1-score as high as 73.68\%. Furthermore, \system\ outperforms the two baseline methods in both overall F1-score and per-class attack accuracy. 
On average, it achieves 9.3\% improvement over U-Leak and 22.56\% improvement over the two-round attack in terms of the overall F1-score. 
This demonstrates that \system\ remains effective against language models. 

{\bf Defense Performance.} Table~\ref{tab:potential mitigation-lanuage} reports the performance of three defense mechanisms (label-only output, dropout, and DP) on the Pythia-70M model. All three mechanisms substantially reduce \system's attack accuracy, with DP achieving the largest decrease and thus offering the strongest protection, while the label-only output provides the weakest mitigation. In addition, each defense incurs a loss in both training and testing accuracy, with DP and label-only output causing the highest and lowest utility degradation, respectively. These findings are consistent with our observations for DNNs trained on image data (Section~\ref{subsec-Defense}) that the dropout approach strikes the most favorable balance between privacy protection and model utility.
\section{Discussions}
\label{sc:discussion}

\subsection{Example-level Privacy Evaluation}

\begin{algorithm}[t!]
\caption{TC-ULiRA}\label{alg:TC-ulira}
\label{alg:TC-ULIRA}
\KwIn{Learning algorithm $A$, unlearning algorithm $U$, number of shadow models $T$, shadow dataset $D'$, logit function $\phi$, target sample ($x^*$,$y^*$), target original model $\simtargetmodeloriginal$, target unlearned model $\simtargetmodelunlearned$, shadow original model $\shadowmodeloriginal$, shadow unlearned model $\shadowmodelunlearned$, function $f(\cdot, \theta)$ returning probabilities given model parameters $\theta$}

\textbf{Observations:} $\mathcal{O} \leftarrow \{\}$, $\hat{\mathcal{O}} \leftarrow \{\}$, $\overline{\mathcal{O}} \leftarrow \{\}$\;
\While{$t \leq T$}{
    $\mathcal{D'} \leftarrow $ \text{sample a shadow dataset }\;
    $\shadowmodeloriginal \leftarrow A(\mathcal{D'})$ \;
    Random sampling a example $(x,y) \in \mathcal{D'}$\;
    $\shadowmodelunlearned \leftarrow U(\shadowmodeloriginal,(x,y))$\;
    Random sampling a example $(\hat{x},\hat{y}) \notin  \mathcal{D'}$\;
    Random sampling a example $(\overline{x}, \overline{y}) \in  \mathcal{D'} (\overline{x} \neq x)$ \;
    $\mathcal{O}[t] \leftarrow \phi( \shadowmodelunlearned (x),y)-\phi( \shadowmodeloriginal(x),y)$ (forget) \; 
    $\hat{\mathcal{O}}[t] \leftarrow \phi( \shadowmodelunlearned(\hat{x}),\hat{y})-\phi(\shadowmodeloriginal(\hat{x}),\hat{y})$ (unseen)\;
    $\overline{\mathcal{O}}[t] \leftarrow \phi(\shadowmodelunlearned(\overline{x}),\overline{y})-\phi(\shadowmodeloriginal(\overline{x}),\overline{y})$ (retain)\;
}
$\mu, \sigma \leftarrow \text{fit Gaussian}(\mathcal{O})$\;
$\hat{\mu}, \hat{\sigma} \leftarrow \text{fit Gaussian}(\hat{\mathcal{O}})$\;
$\overline{\mu}, \overline{\sigma} \leftarrow \text{fit Gaussian}(\overline{\mathcal{O}})$\;

$o \leftarrow \phi(\simtargetmodelunlearned(x^*),y^*)-\phi( \simtargetmodeloriginal(x^*),y^*)$\;

$p \leftarrow N(o; \mu, \sigma^2)$\;
$\hat{p} \leftarrow N(o; \hat{\mu}, \hat{\sigma}^2)$\;
$\overline{p} \leftarrow N(o; \overline{\mu}, \overline{\sigma}^2)$\;

\eIf{$p = \max(p, \hat{p}, \overline{p})$}{
    return $(x^*,y^*)$ is a forget sample\;
}{
    \eIf{$\hat{p} = \max(p, \hat{p}, \overline{p})$}{
        return $(x^*,y^*)$ is an unseen sample\;
    }{
        return $(x^*,y^*)$ is a retain sample\;
    }
}
\end{algorithm}

Currently, \system\ evaluates the privacy risks at the population level. It can be extended to the example-level evaluation. There has been existing work (termed \textit{U-LIRA})~\cite{hayes2024inexact} that adapts the example-level LiRA \cite{carlini2022membership} to the unlearning setting. However, U-LIRA only simulates the shadow model distributions for the forget and unseen sets. 
To address the gap, we extend the simulation to the three-world setting by including the retain set, and propose \textsf{TC-ULiRA}. 
Algorithm \ref{alg:TC-ULIRA} describes the details of \textsf{TC-ULiRA}. The algorithm constructs three shadow-model distributions corresponding to forget, unseen, and retain examples by repeatedly sampling shadow datasets, training shadow models, and applying the unlearning algorithm to remove one example at a time. For each shadow trial, it records the change in logit outputs between the original and unlearned shadow models for the examples in the forget, unseen, and retain sets. These collected logit differences are used to fit three Gaussian distributions that model how each class behaves under unlearning. For a target example, the algorithm computes its logit difference between the original and unlearned target models and evaluates its likelihood under each of the three Gaussian distributions. The class with the highest likelihood determines the predicted status of the target example. It is important to note that this method incurs substantial computational costs, making it less practical than \system\ in real-world scenarios.

\subsection{Evaluating Population-level Unlearning Efficacy by \system}


Unlearning efficacy refers to how effectively a model removes the influence of specific data (e.g., a user’s records) after an unlearning request. It is typically measured by how closely the updated model behaves as if the removed data had never been included during training. 
Prior works have utilized membership inference attacks to evaluate efficacy of unlearning models \cite{ma2022learn,kurmanji2024towards,golatkar2020forgetting,graves2021amnesiac}. Although \system\ primarily focuses on attack-driven U-MIAs, it can be extended as a powerful tool for evaluating unlearning efficacy at the population level. Such an evaluation would assume a strong adversary with complete knowledge of both model versions, including their architectures, parameters, and the specifics of the unlearning algorithms. Leveraging this knowledge, the black-box \system\ can be upgraded to a white-box version, where the attack features can be derived from the loss of individual samples, rather than relying solely on prediction outputs. As a population-level U-MIA, the white-box \system\ would complement existing per-example U-MIAs~\cite{hayes2024inexact,naderloui2025rectifying}, enabling a more comprehensive assessment of unlearning effectiveness. We leave the design and development of this white-box extension to future work. 


\subsection{Attacking Large Language Models}
So far, we only evaluated \system\ over a small language model (Section \ref{exp:lm}). Recently, however, there has been a lot of interest in unlearning of large language models (LLM) \cite{jang2022knowledge,zhang2024negative,eldan2023s,pawelczyk2023context,liu2025rethinking}. While MIAs for language models \cite{shidetectingiclr24,carlini2021extracting,gupta2022recovering,mattern2023membership} can be used to evaluate how much the unlearning model has removed the forget set \cite{shi2024muse}, they were found to be performing poorly due to the high n-gram overlap between members and non-members \cite{duanmembership-colm-24}. This overlap makes it challenging to accurately measure the privacy leakage of LLMs, and highlights the need to extend the membership inference game for such generative models. To extend \system\ to LLMs, several technical challenges need to be addressed. First, the probability-difference features \system\ uses for classification could be generalized to the token-level likelihood shifts between the original and unlearned LLMs; instead of a single score, we can model the full distribution of per-token residuals. Second, we can exploit the next-token predictive dynamics of LLMs by examining how unlearning perturbs the conditional probability trajectories over long contexts. 
Third, the shadow-model simulation used in \system\ must be adapted for LLMs: instead of training full LLMs, we can use parameter-efficient shadow models (e.g., LoRA-based fine-tuned models \cite{han2024parameter}) to emulate the three “worlds” of forget, retain, and unseen samples at scale. Finally, because generative models can leak memorized content through paraphrases, we can design prompt-based probes that condition on semantically equivalent but lexically diverse inputs to amplify differences between member and non-member behavior across the original and unlearned models. 
We leave the completion of these details to the future work. 

\subsection{Attacks using Limited Output}

So far, we have assumed that the adversary has access to the full posterior vector produced by the target model. However, such complete information is often unavailable in practice; instead, the adversary may only observe partial or incomplete outputs. This raises an important question: \textit{How effective is \system\ when only limited output information is accessible?}
To answer this question, we consider two practical scenarios: (\textit{i}) the model exposes only the predicted label, and (\textit{ii}) the model reveals partial probability information, such as the top-$k$ probabilities or rounded values from the posterior vector. We evaluate the attack accuracy of \system\ under these settings using a SimpleCNN model trained on the CINIC-10 dataset as the target model. We consider top-1 and top-3 probabilities in the experiments. 

As shown in Table~\ref{tab:limited_output_attack}, \system\ remains a strong attack even when the adversary can access only restricted model outputs. In particular, using the predicted label alone, \system\ still attains an overall accuracy of 69.23\%. When the adversary is limited to partial probability information, the attack performance decreases only moderately, with overall accuracies of 67.64\%, 68.68\%, and 65.74\% for top-1 probability, top-3 probabilities, and rounded probabilities, respectively. We also observe a consistent trend across all settings: retain samples are the easiest to infer, while forget samples are relatively harder to distinguish. These findings indicate that \system\ does not rely solely on full posterior vectors and can still extract strong membership signals from substantially reduced output information. 

\begin{table}[!t]
\centering
\caption{Attack performance of \system\ using limited outputs (SimpleCNN trained on CINIC-10 dataset as the target model). }
\label{tab:limited_output_attack}
\begin{tabular}{l|c|c|c|c}
\toprule
\textbf{Method} & \textbf{Overall} & \textbf{Unseen} & \textbf{Forget} & \textbf{Retain} \\
\hline
Full post. vector   & 77.19\% & 77.23\% & 75.07\% & 78.21\% \\
Predicted label       & 69.23\% & 69.56\% & 63.14\% & 72.88\% \\
Top-1 Prob.     & 67.64\% & 66.75\% & 61.46\% & 71.94\% \\
Top-3 Prob.   & 68.68\% & 69.45\% & 61.94\% & 72.41\% \\
Rounded Prob. & 65.74\% & 64.34\% & 58.49\% & 70.85\% \\
\bottomrule
\end{tabular}
\end{table}

\subsection{Tailored Defense}

While we have demonstrated the effectiveness of three existing MIA defense mechanisms (label-only output, dropout, and differential privacy) against \system, it is also possible to design alternative defenses specifically tailored to \system.
For example, one can incorporate additional regularization into the unlearning process to suppress overfitting and mitigate the resulting privacy risks. In particular, parameter constraints, output-smoothing terms, or representation-consistency regularizers can be integrated into the unlearning objective to limit overfitting to the retain set and reduce the separability among different data subsets. This, in turn, can lower the model’s vulnerability while preserving unlearning effectiveness. We leave the systematic design and evaluation of such defense mechanisms to future work.

\section{Conclusion}\label{sec-Conclusion}

In this paper, we extend the privacy analysis of machine unlearning to the retained data, an often-overlooked component that may also be vulnerable to leakage. In particular, we extend the analysis beyond the traditional two-class setting by introducing a three-class perspective: the forget, retain, and unseen sets. To this end, we developed \system, a novel black-box attack capable of inferring the membership status of target samples across these classes. Our extensive experiments demonstrate that \system\ is highly effective and consistently outperforms existing baselines, revealing that retained data is particularly vulnerable to privacy leakage resulting from unlearning. Beyond attacks, we evaluated three defense mechanisms against \system, providing the first comprehensive assessment of their effectiveness in mitigating unlearning-related privacy risks. Overall, our findings reveal that current unlearning approaches may expose broader privacy vulnerabilities than previously recognized, highlighting the need for more robust privacy evaluations and stronger defenses in future unlearning methods.

\section*{Acknowledgments}
We thank the reviewers for their feedback. This work was supported by the National Science Foundation (CNS-2029038; CNS-2135988; CNS-2302689; CNS-2308730; CNS-2432533; CNS-2452747). Any opinions, findings, conclusions, or recommendations expressed in this paper are those of the authors and do not necessarily reflect the views of the funding agency.

\newpage


\appendices


\section{Explanation of Attack Features}
\label{appendix:disprob}

\begin{figure*}[t!]
    \centering
    \textbf{Exact unlearning}\\
    \begin{subfigure}[b]{0.26\textwidth}
        \includegraphics[width=\textwidth]{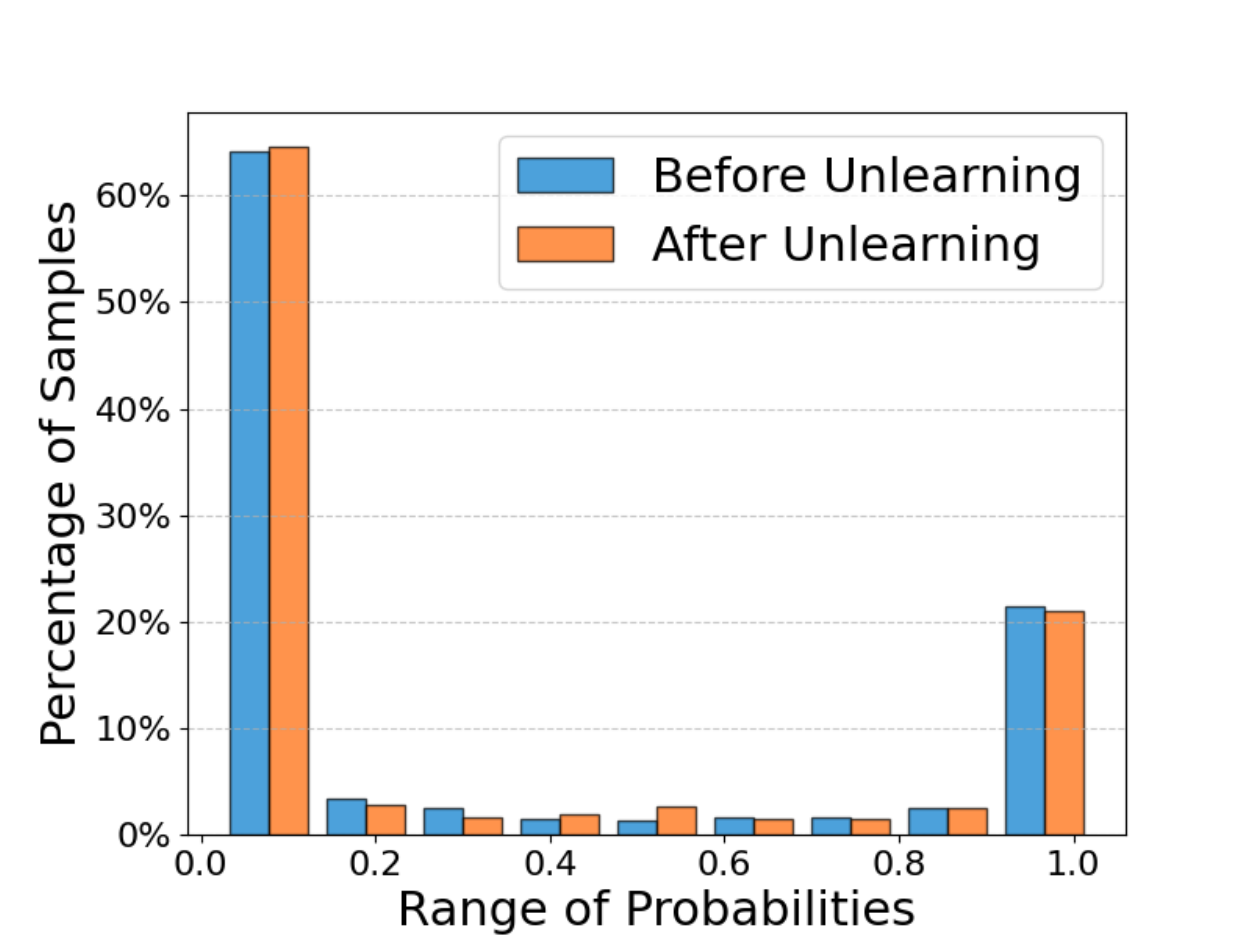} 
        \caption{Unseen set}
        \label{fig:bar_retrain_unseen_prob}
    \end{subfigure}
    \begin{subfigure}[b]{0.26\textwidth}
        \includegraphics[width=\textwidth]{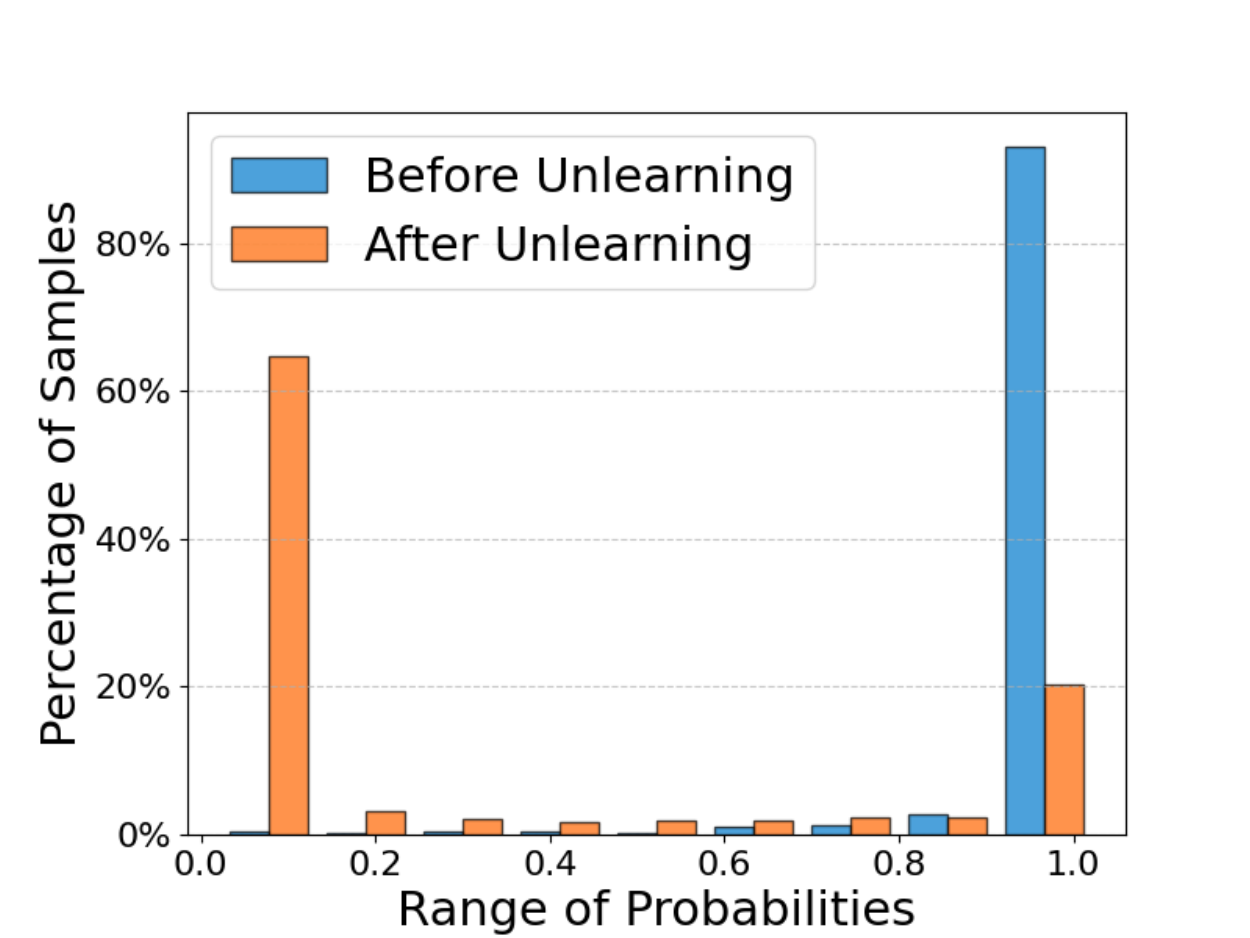} 
        \caption{Forget set}
        \label{fig:bar_retrain_forget_prob}
    \end{subfigure}
    \begin{subfigure}[b]{0.26\textwidth} 
        \includegraphics[width=\textwidth]{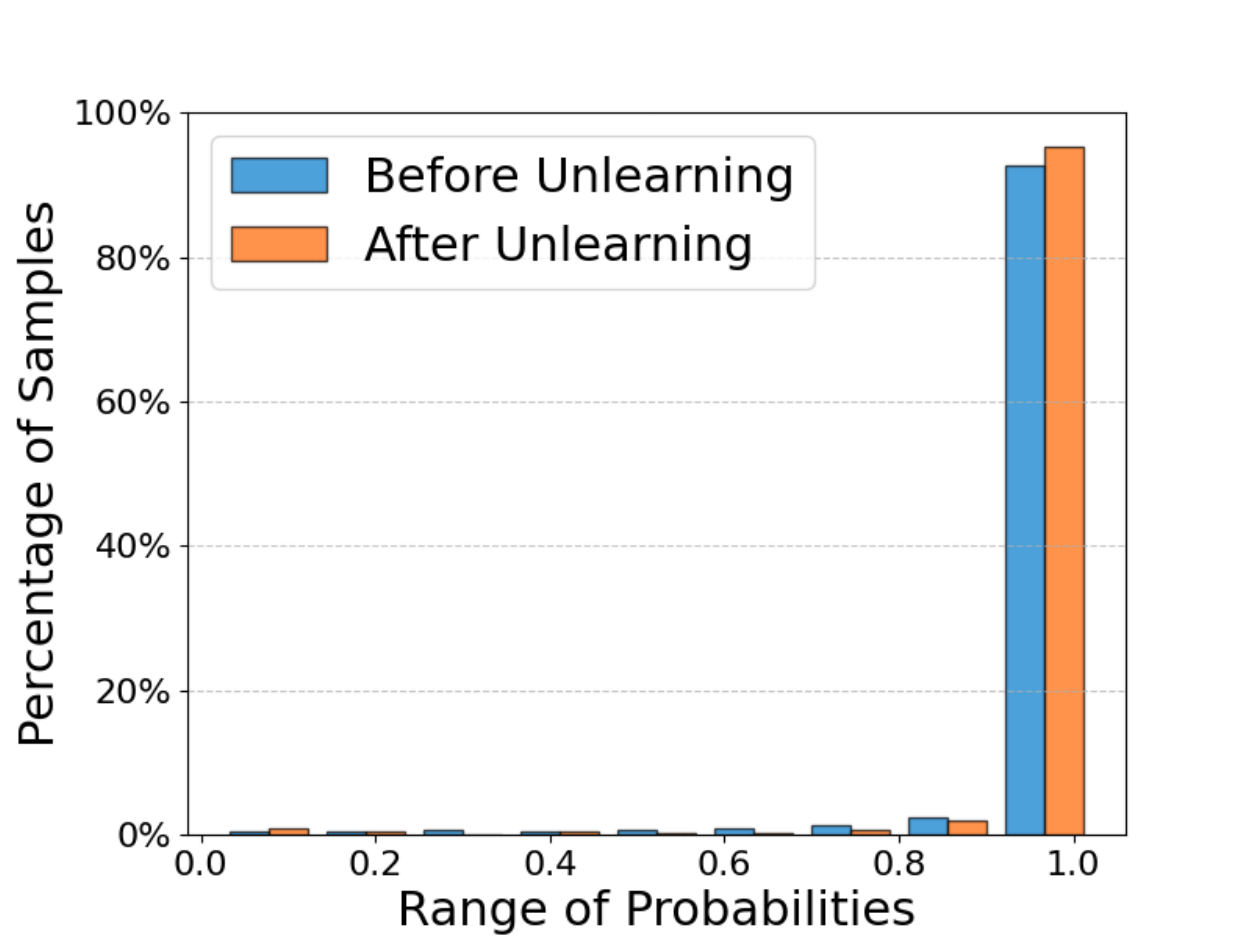} 
        \caption{Retain set}
        \label{fig:bar_retrain_retain_prob}
    \end{subfigure}
    \\
     \textbf{Inexact unlearning}\\
    \begin{subfigure}[b]{0.26\textwidth}
        \includegraphics[width=\textwidth]{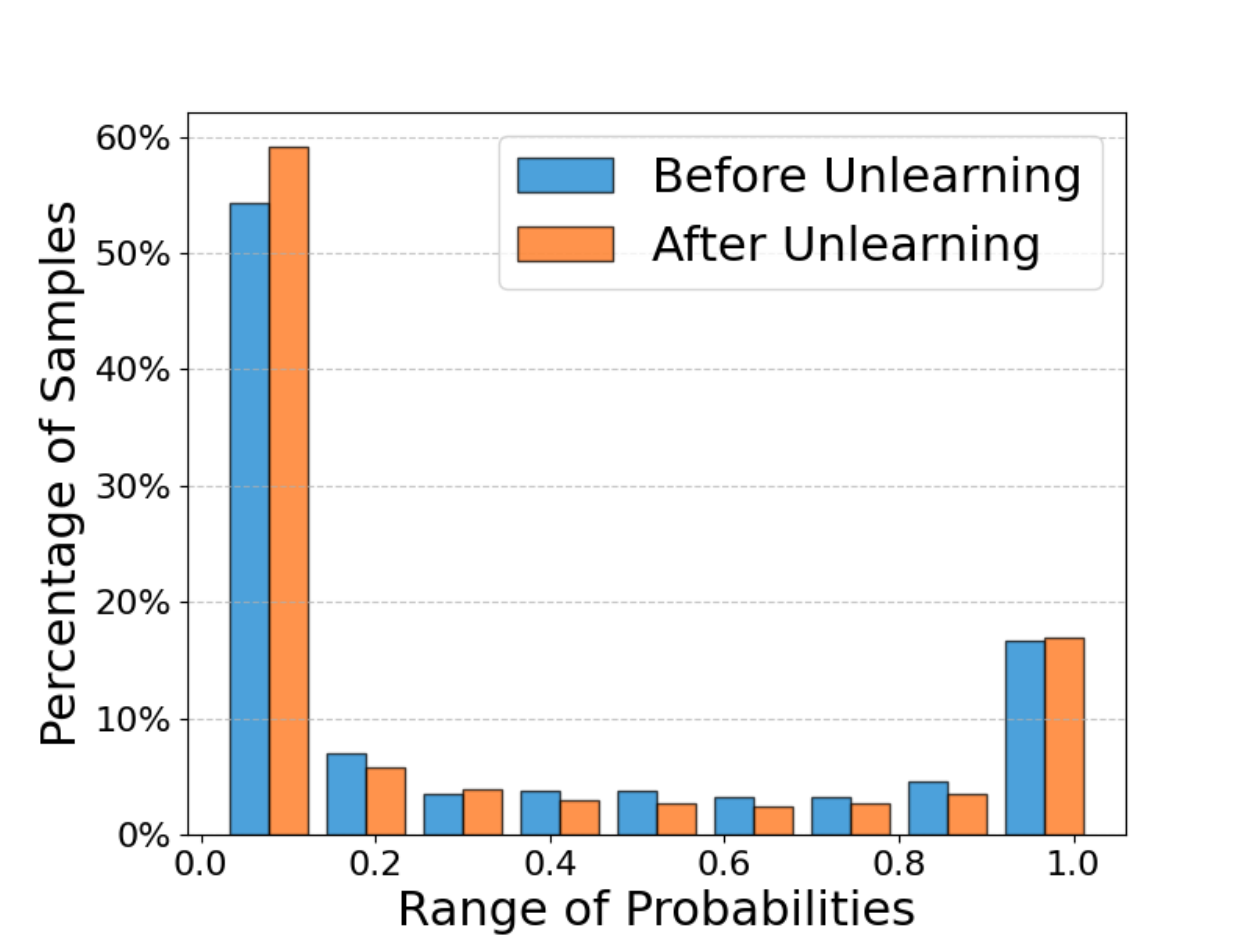} 
        \caption{Unseen set}
        \label{fig:bar_sparsity_unseen_prob}
    \end{subfigure}
    \begin{subfigure}[b]{0.26\textwidth}
        \includegraphics[width=\textwidth]{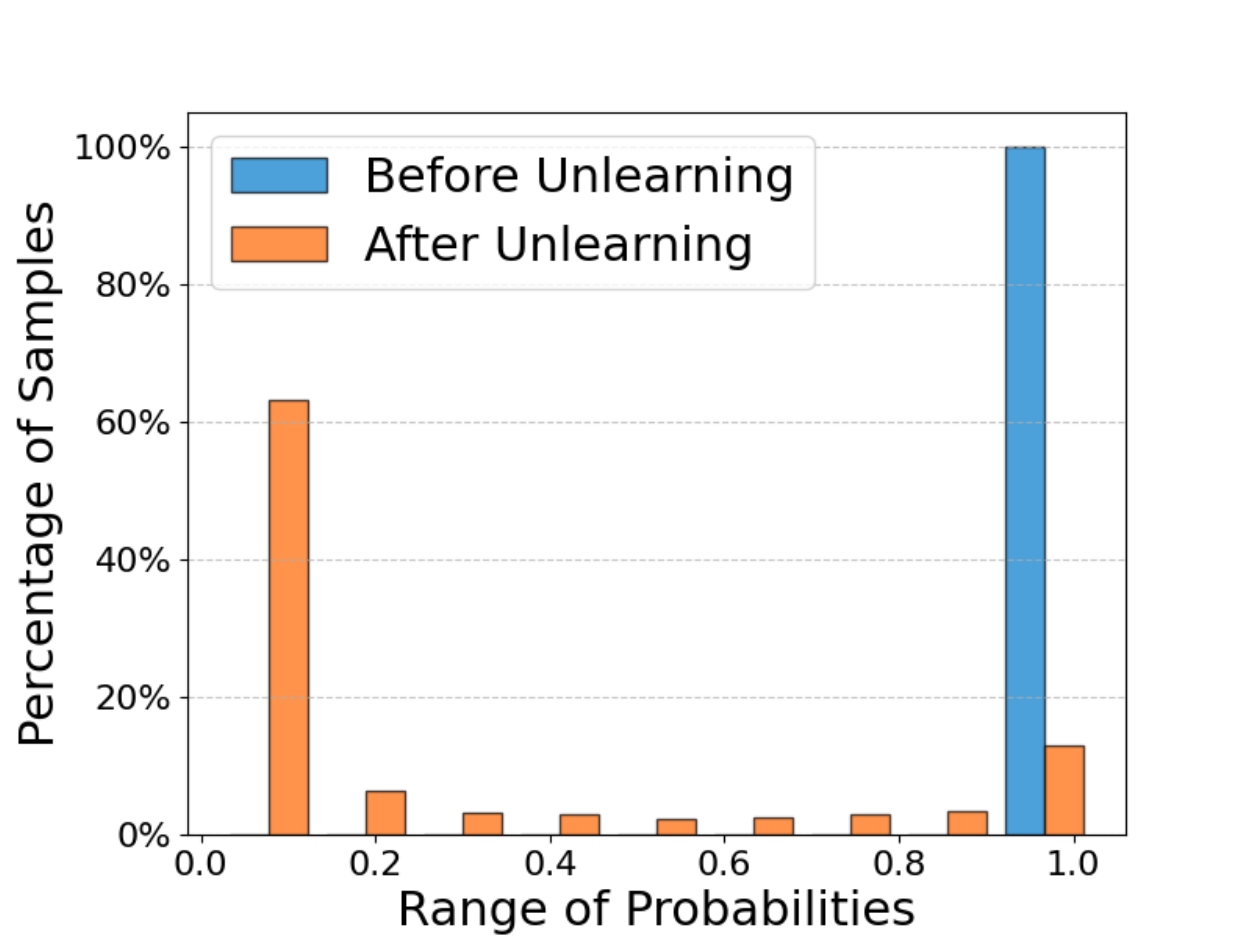} 
        \caption{Forget set}
        \label{fig:bar_sparsity_forget_prob}
    \end{subfigure}
    \begin{subfigure}[b]{0.26\textwidth} 
        \includegraphics[width=\textwidth]{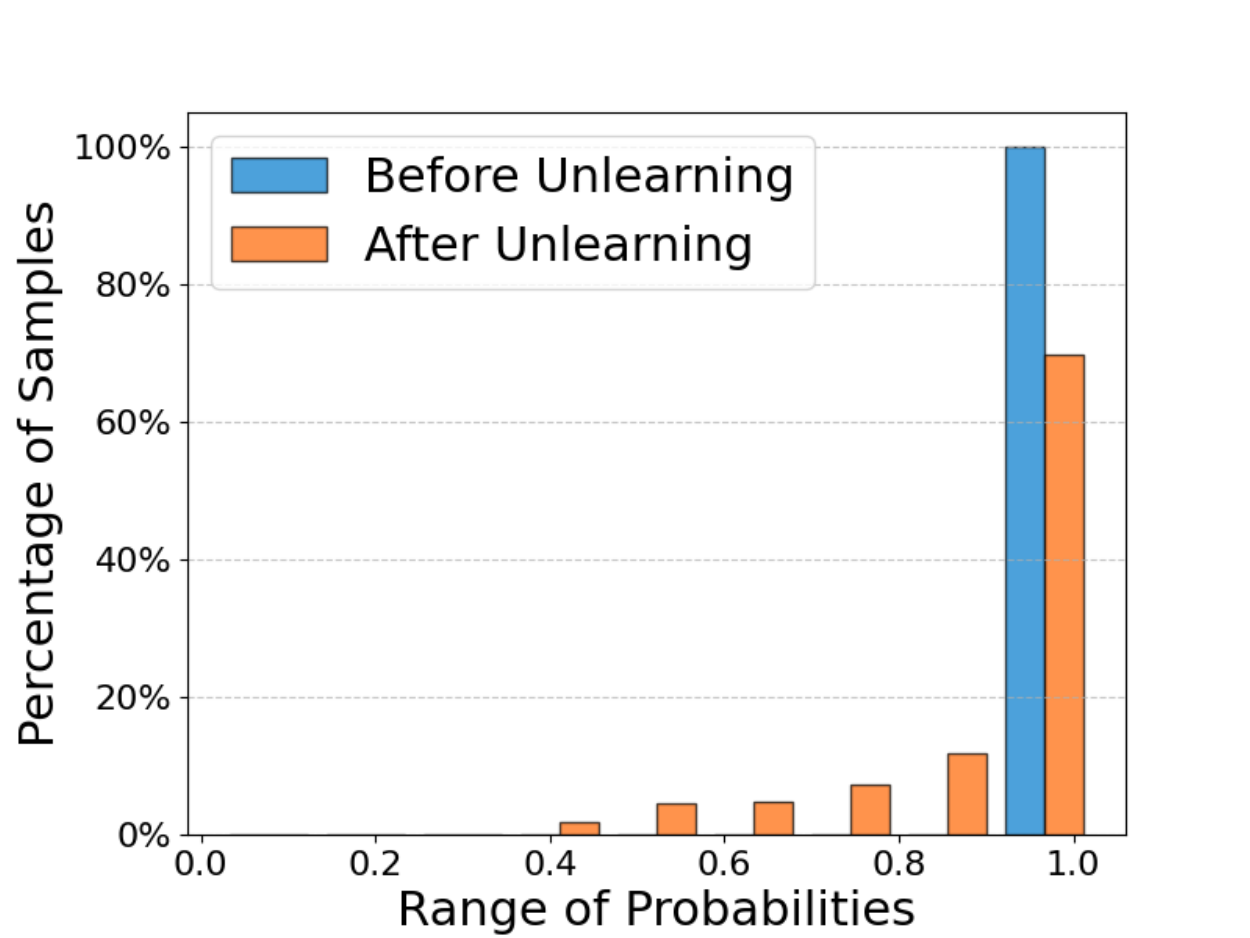} 
        \caption{Retain set}
        \label{fig:bar_sparsity_retain_prob}
    \end{subfigure}
     \caption{The frequency distribution of posterior probability of the true label before and after unlearning (ResNet-18, CIFAR-100 dataset). We use Retraining and Sparsity \cite{liu2024model} as the exact and inexact unlearning methods, respectively. }\vspace{-0.1in}  
    \label{figure:change-of-probability}
\end{figure*}

Figure \ref{figure:change-of-probability} illustrates the posterior probabilities of the three sets output by both models before and after unlearning. There are several important observations. First, using the posteriors of the unlearning model only cannot distinguish the three sets effectively, especially the unseen and forget sets, as a large portion of their instances have similar posteriors in the range [0, 0.2] (as shown in Figure \ref{figure:change-of-probability} (a) \& (b), (d) \& (e)).  Second, only considering difference in the posterior probabilities cannot distinguish unseen and retain sets, which  have marginal difference in the probabilities by unlearning (as shown in Figure \ref{figure:change-of-probability} (a) \& (c), (d) \& (f)). This explains why $\mathbb{P}_{y} -  \mathbb{P}^-_{y}$ or $\mathbb{P}_{y} +   \mathbb{P}^-_{y}$ alone fails to distinguish the three sets.

\section{Pseudo Code of \system}
\label{appendix:pseudo}

\begin{algorithm}
\caption{TC-UMIA}
\label{alg:TC-UMIA}
\SetAlgoLined
\KwIn{
    Target original model $\simtargetmodeloriginal$; 
    target unlearned model $\simtargetmodelunlearned$; target sample $(x,y)$; shadow dataset $\shadowdata$ 
}
\KwOut{Membership label $m \in \{0,1,2\}$}

\textbf{1. Shadow Model Training}

$\mathcal{D}^{\text{R}}_s, \mathcal{D}^{\text{F}}_s, \mathcal{D}^{\text{U}}_s \gets \text{Split}(\mathcal{D}_s)$\;
$f_s \gets \text{Train}(\mathcal{D}^{\text{R}}_s)$\;
$f_s^- \gets \text{Unlearn}(f_s, \mathcal{D}^{\text{F}}_s)$\;

\textbf{2. Attack Classifier Training}

\For {$(\hat{x}_i, \hat{y}_i)$ in $\mathcal{D}^{\text{R}}_S, \mathcal{D}^{\text{F}}_S, \mathcal{D}^{\text{F}}_U$}{
    Get training features $\mathbb{X}_s$ by Eq.$ ~\eqref{eqn:feature1} \& \eqref{eqn:feature2}$\;
    Get corresponding class $\mathbb{Y}_s \in \{0,1,2\}$\;
}
$\attack \gets Train (\mathbb{X}_s, \mathbb{Y}_s)$ \;

\textbf{3. Inference}

Get attack features $\mathbb{X}$ by Eq.$~\eqref{eqn:feature1} \&  \eqref{eqn:feature2}$\;

$m \gets \attack (\mathbb{X})$\;

\Return $m$\;
\end{algorithm}


\begin{table}[t!]
\small
\centering
\caption{SimpleCNN structure and hyperparameter. For RGB images, input\_channel $C_i = 3$, kernel\_size of convolution layer $K_c=3$ and Max-pooling layer $K_m=2$. }
\label{tab:SimpleCNN}
\begin{tabular}{l|l}
\hline
Layer & Hyperparameters \\
\hline
Conv2D\_1 & $(C_i, 32, K_c, 1)$ \\
Tanh & - \\
Conv2D\_2 & $(32, 32, K_c, 1)$ \\
MaxPolling2D & $K_m=2$ \\
Conv2D\_3 & $(32, 64, K_c, 1)$ \\
Tanh & - \\
Conv2D\_4 & $(64, 64, K_c, 1)$ \\
MaxPolling2D & $K_m=2$ \\
Conv2D\_5 & $(64, 128, K_c, 1)$ \\
Tanh & - \\
Conv2D\_6 & $(128, 128, K_c, 1)$ \\
MaxPolling2D & $K_m=2$ \\
Flatten & 1 \\
Linear\_1 & $(128 \times 4 \times 4, 128)$ \\
Tanh & - \\
Linear\_2 & $(128, \text{num\_classes})$ \\
\hline
\end{tabular}
\end{table}

\section{Experimental Setup}
\label{appendix:expsetup}


\subsection{Target Model Setup and Performance}

SimpleCNN is a lightweight 5-layer CNN composed of two convolutional layers (with 32 and 64 filters, respectively), followed by two max-pooling layers and a fully connected layer with 128 neurons.
DenseNet-121 consists of 121 layers organized into four dense blocks containing 6, 12, 24, and 16 layers, respectively. These blocks are connected via transition layers, each employing 1×1 convolution and 2×2 average pooling.
ResNet-18 includes an initial convolutional layer with 64 filters, followed by four residual blocks (each containing 2 layers), and concludes with an output layer.
All three models are trained using a batch size of 256, a learning rate of 0.001, and a weight decay of 1e-4. Cross-entropy is used as the loss function, and optimization is performed using Adam. 

Table~\ref{tab:SimpleCNN} show the SimpleCNN structure and hyperparameter. The structure and hyperparameter of DenseNet and ResNet-18 can refer~\cite{huang2017densely} and~\cite{he2016deep}, respectively. 

Table~\ref{tab:Efficacy of inexact unlearning} shows the unlearned performance of five unlearning methods under ResNet-18 model. We tuned the hyper-parameters to ensure that the unlearning baselines closely matched the performance of the Retrain gold standard.  We follow ~\cite{liu2024model} to evaluate efficacy of unlearning method by unlearned model performance on three sets.

\subsection{Unlearning Model Performance}

We consider the following evaluation metrics of the unlearning models. 
\begin{itemize}
    \item \textbf{Unlearning accuracy (UA):} It measures the classification accuracy of the unlearning model on the forgetting dataset~\cite{graves2021amnesiac,golatkar2020eternal}. 
    
    \item \textbf{Retain accuracy (RA):} It measures the accuracy of the unlearning model on the retain set. 
    
    \item \textbf{Testing accuracy (TA):} This measures the classification accuracy of the unlearning model on the testing dataset.

\end{itemize}

The performance of the unlearning models used in our experiments is reported in Table \ref{tab:Efficacy of inexact unlearning}.

\begin{table}[!h]
\footnotesize
\caption{Performance of unlearning models (classification accuracy in \%).}
\centering
\label{tab:Efficacy of inexact unlearning}
\scalebox{0.9}{
\begin{tabular}{c|c|c|c|c|c}
\hline
\multirow{2}{*}{\textbf{Model}} & \multirow{2}{*}{\textbf{\begin{tabular}[c]{@{}c@{}}Unlearning \\ method\end{tabular}}} & \multirow{2}{*}{\textbf{Dataset}} & \multirow{2}{*}{\textbf{UA}} & \multirow{2}{*}{\textbf{RA}} & \multirow{2}{*}{\textbf{TA}} \\
                                &                                   &                                   &                              &                              &                              \\ \hline
\multirow{20}{*}{ResNet-18}     & \multirow{4}{*}{Retrain}          & CIFAR-10                          & 72.20                        & 98.41                        & 71.68                        \\
                                &                                   & CIFAR-100                         & 31.12                        & 98.20                        & 31.06                        \\
                                &                                   & CINIC-10                          & 61.33                        & 97.81                        & 61.86                        \\
                                &                                   & TinyImageNet                      & 11.60                        & 99.98                        & 12.70                        \\ \cline{2-6}
                                & \multirow{4}{*}{SISA}             & CIFAR-10                          & 64.42                        & 84.72                        & 64.97                        \\
                                &                                   & CIFAR-100                         & 28.72                        & 93.11                        & 28.31                        \\
                                &                                   & CINIC-10                          & 54.67                        & 83.24                        & 55.76                        \\
                                &                                   & TinyImageNet                      & 7.51                         & 93.59                        & 8.44                         \\ \cline{2-6}
                                & \multirow{4}{*}{Sparsity}         & CIFAR-10                          & 60.20                        & 96.09                        & 58.35                        \\
                                &                                   & CIFAR-100                         & 29.35                        & 95.36                        & 29.10                        \\
                                &                                   & CINIC-10                          & 47.87                        & 96.57                        & 48.72                        \\
                                &                                   & TinyImageNet                      & 9.68                         & 91.31                        & 7.63                         \\ \cline{2-6}
                                & \multirow{4}{*}{SCRUB}            & CIFAR-10                          & 74.80                        & 94.41                        & 71.16                        \\
                                &                                   & CIFAR-100                         & 45.73                        & 89.75                        & 35.99                        \\
                                &                                   & CINIC-10                          & 69.00                        & 93.93                        & 60.20                        \\
                                &                                   & TinyImageNet                      & 21.20                        & 99.92                        & 9.83                         \\ \cline{2-6}
                                & \multirow{4}{*}{GA}               & CIFAR-10                          & 74.20                        & 94.37                        & 69.08                        \\
                                &                                   & CIFAR-100                         & 51.40                        & 95.01                        & 35.24                        \\
                                &                                   & CINIC-10                          & 65.88                        & 84.80                        & 58.03                        \\
                                &                                   & TinyImageNet                      & 15.60                        & 78.98                        & 6.59                         \\ \hline
\multirow{6}{*}{Pythia-70m}     & \multirow{2}{*}{Retrain}          & SST5                              & 48.81                        & 99.21                        & 46.33                        \\
                                &                                   & News20                            & 66.08                        & 97.64                        & 58.70                        \\ \cline{2-6}
                                & \multirow{2}{*}{GA}           & SST5                              & 55.16                        & 100                          & 43.24                        \\
                                &                                   & News20                            & 59.88                        & 94.63                        & 58.94                        \\ \cline{2-6}
                                & \multirow{2}{*}{NPO}          & SST5                              & 53.17                        & 99.80                        & 45.76                        \\
                                &                                   & News20                            & 56.93                        & 97.76                        & 59.16                        \\ \hline
\end{tabular}
}
\end{table}

\subsection{Attack Training and Testing Datasets}

The attack training dataset is constructed as follows: for each shadow dataset, we split it into two disjoint subsets using a 4:1 ratio for training and testing the shadow models, respectively. From the shadow training set, we randomly select 2\% of the samples to form the forget set. An equal number of samples is then randomly selected from the remaining shadow training set and the shadow testing set to form the retain set and the unseen set, respectively. We repeat this process five times, resulting in an attack training dataset whose total size is approximately 30\% of the shadow training set.
To construct the attack testing dataset, we generate three disjoint sets (forget, retain, and unseen), each consisting of 2\% of samples randomly selected from the training and testing sets of the target model. The resulting ratio between the sizes of the attack training and testing datasets is 4:1. Specifically, we used 6,000 training and 1,500 testing samples for CIFAR-10 and CIFAR-100 datasets, 10,800 training and 2,700 testing samples for the CINIC-10 dataset, and 12,000 training and 3,000 testing samples for the TinyImageNet dataset.

\section{Additional Experimental Results}

\subsection{Performance of \system\ on SimpleCNN and DenseNet} \label{Performance on Other Models}

\begin{table*}[t!]
\centering
\footnotesize
\caption{Overall F1-score and per-class F1-score of the attacks, with DenseNet and SimpleCNN as the target model. The best results per evaluation metric are marked with the gray color.}
\label{tab:Performance-simplecnn-densenet}
\scalebox{0.9}{
\begin{tabular}{c|c|c|cc|cc|cc}
\hline
\multirow{2}{*}{\textbf{Model}} & \multirow{2}{*}{\textbf{\begin{tabular}[c]{@{}c@{}}Unlearning\\ method\end{tabular}}} & \multirow{2}{*}{\textbf{Dataset}} & \multicolumn{2}{c|}{\textbf{\system\ (Ours)}}                           & \multicolumn{2}{c|}{\textbf{U-leak (best performance)}}               & \multicolumn{2}{c}{\textbf{Two-round Attack}}                            \\ \cline{4-9} 
                                &                                                                                       &                                   & \multicolumn{1}{c|}{\textbf{All}} & \textbf{Unseen/Forget/Retain} & \multicolumn{1}{c|}{\textbf{All}} & \textbf{Unseen/Forget/Retain} & \multicolumn{1}{c|}{\textbf{All}} & \textbf{Unseen/Forget/Retain} \\ \hline
\multirow{15}{*}{DenseNet}      & \multirow{3}{*}{Retrain}                                                              & CIFAR-100                         & \multicolumn{1}{c|}{\gray{86.85}}          & \gray{87.93/86.45/86.33}       & \multicolumn{1}{c|}{84.00}          & 84.48/82.32/85.01       & \multicolumn{1}{c|}{74.40}          & 70.91/77.08/81.04       \\
                                &                                                                                       & CINIC-10                          & \multicolumn{1}{c|}{\gray{65.89}}          & \gray{66.70/61.91/68.18}       & \multicolumn{1}{c|}{55.62}          & 47.02/48.77/65.54       & \multicolumn{1}{c|}{46.89}          & 43.32/45.39/58.57       \\
                                &                                                                                       & TinyImageNet                      & \multicolumn{1}{c|}{\gray{93.76}}          & \gray{97.22/91.93/92.24}       & \multicolumn{1}{c|}{88.28}          & 95.27/84.27/85.70       & \multicolumn{1}{c|}{80.68}          & 84.32/80.88/84.00       \\ \cline{2-9} 
                                & \multirow{3}{*}{SISA}                                                                 & CIFAR-100                         & \multicolumn{1}{c|}{\gray{85.07}}          & \gray{83.10/87.67/84.15}       & \multicolumn{1}{c|}{63.25}          & 48.87/70.14/66.85       & \multicolumn{1}{c|}{38.55}          & 39.76/33.48/45.31       \\
                                &                                                                                       & CINIC-10                          & \multicolumn{1}{c|}{\gray{61.11}}          & \gray{53.35/65.91/62.04}       & \multicolumn{1}{c|}{42.44}          & 40.53/42.91/43.86       & \multicolumn{1}{c|}{33.66}          & 37.09/21.00/42.42       \\
                                &                                                                                       & TinyImageNet                      & \multicolumn{1}{c|}{\gray{94.60}}          & \gray{93.63/96.20/93.94}       & \multicolumn{1}{c|}{70.87}          & 69.37/74.44/67.69       & \multicolumn{1}{c|}{56.96}          & 60.81/56.68/64.97       \\ \cline{2-9} 
                                & \multirow{3}{*}{Sparsity}                                                             & CIFAR-100                         & \multicolumn{1}{c|}{\gray{73.80}}          & 88.81/\gray{67.37/67.38}       & \multicolumn{1}{c|}{64.09}          & \gray{89.22}/52.96/53.56       & \multicolumn{1}{c|}{51.20}          & 60.53/51.27/52.87       \\
                                &                                                                                       & CINIC-10                          & \multicolumn{1}{c|}{\gray{54.03}}          & \gray{67.59/45.97/50.64}       & \multicolumn{1}{c|}{46.60}          & 52.25/42.61/45.90       & \multicolumn{1}{c|}{36.84}          & 36.78/39.70/41.93       \\
                                &                                                                                       & TinyImageNet                      & \multicolumn{1}{c|}{\gray{73.34}}          & \gray{96.46/61.69/62.90}       & \multicolumn{1}{c|}{62.41}          & 92.71/57.20/34.80       & \multicolumn{1}{c|}{51.73}          & 70.07/52.78/44.96       \\ \cline{2-9} 
                                & \multirow{3}{*}{SCRUB}                                                                & CIFAR-100                         & \multicolumn{1}{c|}{\gray{70.89}}          & \gray{88.27/67.92/58.15}       & \multicolumn{1}{c|}{68.91}          & 87.77/66.18/53.52       & \multicolumn{1}{c|}{60.13}          & 71.58/61.44/56.44       \\
                                &                                                                                       & CINIC-10                          & \multicolumn{1}{c|}{\gray{60.48}}          & \gray{66.81/53.89/60.91}       & \multicolumn{1}{c|}{52.36}          & 50.96/44.01/58.88       & \multicolumn{1}{c|}{44.99}          & 44.81/43.37/53.22       \\
                                &                                                                                       & TinyImageNet                      & \multicolumn{1}{c|}{\gray{87.41}}          & \gray{96.64/83.73/82.20}       & \multicolumn{1}{c|}{79.20}          & 92.34/74.55/71.85       & \multicolumn{1}{c|}{73.58}          & 82.57/73.19/73.51       \\ \cline{2-9} 
                                & \multirow{3}{*}{GA}                                                                   & CIFAR-100                         & \multicolumn{1}{c|}{\gray{74.45}}          & \gray{86.02/67.61/71.21}       & \multicolumn{1}{c|}{68.02}          & 82.31/61.88/62.67       & \multicolumn{1}{c|}{60.71}          & 67.57/58.99/63.92       \\
                                &                                                                                       & CINIC-10                          & \multicolumn{1}{c|}{\gray{58.72}}          & \gray{65.74/51.12/59.52}       & \multicolumn{1}{c|}{51.33}          & 56.74/44.26/52.79       & \multicolumn{1}{c|}{44.28}          & 44.53/42.51/52.27       \\
                                &                                                                                       & TinyImageNet                      & \multicolumn{1}{c|}{\gray{79.20}}          & \gray{86.62/74.48/75.30}       & \multicolumn{1}{c|}{63.50}          & 70.01/56.51/62.70       & \multicolumn{1}{c|}{55.99}          & 65.84/57.24/59.21       \\ \hline
\multirow{15}{*}{SimpleCNN}     & \multirow{3}{*}{Retrain}                                                              & CIFAR-100                         & \multicolumn{1}{c|}{\gray{91.84}}          & \gray{92.44/92.21/91.02}       & \multicolumn{1}{c|}{89.47}          & 89.86/88.44/90.05       & \multicolumn{1}{c|}{81.07}          & 79.62/83.82/85.84       \\
                                &                                                                                       & CINIC-10                          & \multicolumn{1}{c|}{\gray{77.19}}          & \gray{77.23/75.07/78.21}       & \multicolumn{1}{c|}{68.99}          & 69.28/62.56/72.87       & \multicolumn{1}{c|}{63.41}          & 54.30/62.97/73.06       \\
                                &                                                                                       & TinyImageNet                      & \multicolumn{1}{c|}{\gray{96.32}}          & \gray{97.65/95.10/96.03}       & \multicolumn{1}{c|}{86.07}          & 89.04/81.59/87.40       & \multicolumn{1}{c|}{80.22}          & 81.95/81.66/86.86       \\ \cline{2-9} 
                                & \multirow{3}{*}{SISA}                                                                 & CIFAR-100                         & \multicolumn{1}{c|}{\gray{89.62}}          & \gray{89.58/91.63/87.77}       & \multicolumn{1}{c|}{75.50}          & 70.82/78.72/75.76       & \multicolumn{1}{c|}{53.82}          & 59.60/49.78/61.42       \\
                                &                                                                                       & CINIC-10                          & \multicolumn{1}{c|}{\gray{61.78}}          & \gray{55.41/64.46/63.21}       & \multicolumn{1}{c|}{38.78}          & 5.02/45.56/48.04        & \multicolumn{1}{c|}{35.22}          & 39.86/24.53/44.64       \\
                                &                                                                                       & TinyImageNet                      & \multicolumn{1}{c|}{\gray{98.37}}          & \gray{98.42/98.64/98.04}       & \multicolumn{1}{c|}{90.79}          & 92.46/90.57/89.58       & \multicolumn{1}{c|}{83.88}          & 87.24/83.79/86.32       \\ \cline{2-9} 
                                & \multirow{3}{*}{Sparsity}                                                             & CIFAR-100                         & \multicolumn{1}{c|}{\gray{84.09}}          & \gray{93.81/79.53/79.72}       & \multicolumn{1}{c|}{79.49}          & 91.69/74.11/73.96       & \multicolumn{1}{c|}{73.80}          & 79.97/73.36/75.86       \\
                                &                                                                                       & CINIC-10                          & \multicolumn{1}{c|}{\gray{69.67}}          & \gray{73.29/65.64/70.26}       & \multicolumn{1}{c|}{63.88}          & 58.87/60.50/69.59       & \multicolumn{1}{c|}{52.03}          & 49.60/50.72/61.41       \\
                                &                                                                                       & TinyImageNet                      & \multicolumn{1}{c|}{\gray{85.66}}          & \gray{96.96/80.72/79.69}       & \multicolumn{1}{c|}{66.40}          & 82.00/62.68/55.66       & \multicolumn{1}{c|}{59.29}          & 69.45/60.12/59.07       \\ \cline{2-9} 
                                & \multirow{3}{*}{SCRUB}                                                                & CIFAR-100                         & \multicolumn{1}{c|}{\gray{88.22}}          & \gray{93.28/86.50/85.44}       & \multicolumn{1}{c|}{86.04}          & 91.69/83.60/83.46       & \multicolumn{1}{c|}{80.65}          & 84.14/81.04/82.38       \\
                                &                                                                                       & CINIC-10                          & \multicolumn{1}{c|}{\gray{59.97}}          & \gray{59.19/57.79/61.86}       & \multicolumn{1}{c|}{46.46}          & 20.60/50.30/54.17       & \multicolumn{1}{c|}{53.17}          & 49.57/50.47/60.77       \\
                                &                                                                                       & TinyImageNet                      & \multicolumn{1}{c|}{\gray{88.83}}          & \gray{98.02/84.70/83.97}       & \multicolumn{1}{c|}{77.91}          & 90.69/73.29/70.86       & \multicolumn{1}{c|}{71.67}          & 80.44/71.26/72.95       \\ \cline{2-9} 
                                & \multirow{3}{*}{GA}                                                                   & CIFAR-100                         & \multicolumn{1}{c|}{\gray{88.07}}          & \gray{93.24/86.29/85.28}       & \multicolumn{1}{c|}{85.51}          & 91.35/82.56/83.17       & \multicolumn{1}{c|}{80.27}          & 83.15/81.04/82.29       \\
                                &                                                                                       & CINIC-10                          & \multicolumn{1}{c|}{\gray{66.22}}          & \gray{77.30/58.15/64.63}       & \multicolumn{1}{c|}{60.59}          & 64.28/54.01/62.91       & \multicolumn{1}{c|}{54.41}          & 54.81/51.66/60.42       \\
                                &                                                                                       & TinyImageNet                      & \multicolumn{1}{c|}{\gray{87.37}}          & \gray{97.29/83.65/81.40}       & \multicolumn{1}{c|}{72.89}          & 84.82/70.66/64.19       & \multicolumn{1}{c|}{69.14}          & 65.84/57.24/59.21       \\ \hline
\end{tabular}
}
\end{table*}

Table~\ref{tab:Performance-simplecnn-densenet} reports overall and per-class performance of \system\ and the two baseline approaches on attacking the DenseNet and SimpleCNN models. First, \system\ achieves impressive overall and per-class performance on DenseNet and SimpleCNN models. For instance, the overall F1-score of \system\ can be as high as 98.37\% when simpleCNN trained on the TinyImageNet dataset is used as the target model with SISA as the unlearning method. Furthermore, \system\ consistently outperforms both baseline approaches in terms of both overall and per-class performance across all settings. 
This demonstrates that \system\ achieves strong attack performance regardless of the network architecture, highlighting the robustness of \system.

\subsection{TPR@5\%FPR Results of \system}
\label{appendix:TPR-value}

\begin{table*}[!t]
\footnotesize
\centering
\caption{Attack Accuracy (TPR@5\% FPR).}
\label{tab:TPR-restnet}
\scalebox{0.95}{
\begin{tabular}{c|c|c|c|c|c}
\hline
\multirow{2}{*}{\textbf{Target Model}} & \multirow{2}{*}{\textbf{Unlearning Method}} & \multirow{2}{*}{\textbf{Dataset}} & \textbf{\system\ (Ours)} & \textbf{U-Leak (Best performance)} & \textbf{Two-round Attack} \\
\cline{4-6}
& & & \textbf{Unseen/Forget/Retain} & \textbf{Unseen/Forget/Retain} & \textbf{Unseen/Forget/Retain} \\ \hline
\multirow{15}{*}{ResNet-18} & \multirow{3}{*}{Retrain} & CIFAR-100 & 74.80/69.00/44.33 & 44.60/33.73/10.13 & 38.87/33.73/3.07 \\
& & CINIC-10 & 44.07/31.96/12.67 & 13.00/15.59/16.41 & 20.41/15.89/6.33 \\
& & TinyImageNet & 96.37/92.30/99.97 & 93.33/80.07/99.87 & 84.17/83.37/0.65 \\ \cline{2-6}
& \multirow{3}{*}{SISA} & CIFAR-100 & 68.27/82.33/23.29 & 54.82/37.35/12.05 & 31.33/42.77/9.04 \\
& & CINIC-10 & 37.11/28.33/10.89 & 9.67/9.00/13.00 & 8.33/10.00/7.33 \\
& & TinyImageNet & 95.10/95.90/88.39 & 78.38/75.38/11.41 & 58.56/77.48/0.90 \\ \cline{2-6}
& \multirow{3}{*}{Sparsity} & CIFAR-100 & 79.87/68.40/24.27 & 79.33/22.13/21.07 & 58.33/24.73/0.00 \\
& & CINIC-10 & 45.07/34.70/16.26 & 20.19/18.00/20.93 & 24.15/15.74/3.93 \\
& & TinyImageNet & 97.10/90.20/92.57 & 96.13/43.70/55.03 & 77.80/42.40/0.72 \\ \cline{2-6}
& \multirow{3}{*}{SCRUB} & CIFAR-100 & 77.93/37.13/31.13 & 77.40/36.40/26.47 & 58.87/19.33/0.30 \\
& & CINIC-10 & 48.19/18.11/10.70 & 29.11/13.15/10.34 & 28.52/12.93/3.52 \\
& & TinyImageNet & 95.83/86.37/93.20 & 89.77/62.00/91.00 & 86.33/81.27/0.00 \\ \cline{2-6}
& \multirow{3}{*}{GA} & CIFAR-100 & 79.00/18.60/14.93 & 78.73/41.87/14.47 & 55.47/36.20/0.00 \\
& & CINIC-10 & 45.30/23.96/10.52 & 24.48/10.74/9.70 & 22.59/12.37/4.37 \\
& & TinyImageNet & 97.80/62.97/60.57 & 96.47/13.90/41.17 & 84.03/16.47/2.03 \\ \hline
\multirow{6}{*}{Pythia-70m} & \multirow{2}{*}{Retrain} & SST5 & 70.63/50.79/26.59 & 12.30/19.44/34.92 & 23.81/26.19/2.38 \\
& & News20 & 57.23/44.84/9.73 & 48.67/26.84/5.90 & 38.05/26.55/0.29 \\ \cline{2-6}
& \multirow{2}{*}{GA+GDR} & SST5 & 67.86/41.27/8.73 & 32.94/35.32/26.19 & 45.63/4.76/1.98 \\
& & News20 & 52.21/40.41/19.17 & 4.13/18.88/5.60 & 25.07/11.50/1.77 \\ \cline{2-6}
& \multirow{2}{*}{NPO+GDR} & SST5 & 77.38/48.02/24.60 & 52.78/44.84/18.65 & 55.16/13.89/1.19 \\
& & News20 & 64.60/46.61/23.01 & 34.51/36.58/5.01 & 44.84/10.91/0.29 \\ \hline
\end{tabular}
}
\end{table*}
Table~\ref{tab:TPR-restnet} shows the per-class TPR results when FPR=5\% on both ResNet-18 and Pythia-70m. 
We observe that, first, the per-class's TPR of \system\ significantly outperforms both baseline methods in most cases. Second, the TPR value vary across the three sets. Such discrepancy reflects the attack model's significantly varying discriminative capability across different classes. In particular, the retain set demonstrates the lowest TPR value. This is because the samples from both unseen and forget sets are frequently misclassified to the retain set \footnote{As shown in Figure~\ref{fig:conprob} and \ref{fig:difsum}, the feature space where the retain set region contains significant overlap with samples from the other two sets)}. Consequently, when maintaining the FPR as low as 5\%, the TPR for the retain set exhibits a notable decline. Conversely, the unseen set achieves the highest value, since the samples from both retain and forget sets are rarely misclassified.
Finally, we observe that the two-round attack has the TPR value for the retain set as low as zero in some settings, demonstrating its ineffectiveness for the three-set inference.

\subsection{Computational Overhead} \label{subsec-Computational overhead}
We compare the computational overhead of \system\ against the two baselines, namely U-Leak and the two-round attack, using ResNet-18 trained on the CIFAR-100 dataset as the target model. As shown in Table~\ref{tab:overhead_comparison}, the training time of \system\ is comparable to that of the faster baseline, i.e., the two-round attack (2.45s vs.\ 2.32s). Meanwhile, \system\  achieves substantially lower inference time than both baselines. In particular, its inference time is reduced by 68.6\% compared with U-Leak (49.6ms vs.\ 81.5ms), and by 53.7\% compared with the two-round attack (49.6ms vs.\ 55.3ms). These results demonstrate that \system\  maintains competitive training efficiency while offering a clear advantage in inference efficiency.

\begin{table}[t]
\centering
\caption{Computational overhead of \system\ and the baseline attacks (ResNet-18 trained on CIFAR-100 as the target model). }
\label{tab:overhead_comparison}
\begin{tabular}{lcc}
\toprule
\textbf{Method} & \textbf{Training time (s)} & \textbf{Inference time (ms)} \\
\midrule
\system\          & 2.45 & 49.6 \\
U-Leak          & 8.12 & 81.5 \\
Two-round Attack & 2.32 & 55.3 \\
\bottomrule
\end{tabular}
\end{table}

\subsection{Explanation of Retain Set's Increased Vulnerability After Unlearning} \label{subsec-Retain Set's Increased Vulnerability}

From an optimization perspective, removing the forget set changes the empirical risk, reallocating gradients to better fit the retain set and increasing the output-space separation between the retain set and the unseen/forget sets. Table~\ref{tab:retain_separation} presents the change in separability of the retain set relative to the unseen and forget sets before and after unlearning. For the accuracy-gap analysis, we separately evaluate the classification accuracy of the pre-unlearning and post-unlearning models on the retain, unseen, and forget sets. We define the retain--unseen accuracy gap as the difference between the accuracy on the retain set and that on the unseen set, and define the retain--forget accuracy gap analogously. For the output-distance analysis, we collect the outputs of the pre-unlearning and post-unlearning models on all samples in each set and compute the average output representation for each set. We then measure the separation in the output space using the Euclidean distance between the average outputs of different sets, and use this value as the output distance. Empirically, we observe that after unlearning, both the accuracy gaps and the output distances increase, indicating that the retain set becomes more distinguishable from the unseen and forget sets.

\begin{table}[t]
\centering
\caption{Comparison of separability between the retain set and the other two sets before and after unlearning. }
\label{tab:retain_separation}
\begin{tabular}{lcc}
\toprule
\textbf{Metric} & \textbf{Pre-unlearning} & \textbf{Post-unlearning} \\
\midrule
Retain--unseen accuracy gap         & 24.21\% & 26.43\% \\
Retain--forget accuracy gap         & 1.12\%  & 23.43\% \\
Retain--unseen output distance      & 0.03    & 0.05    \\
Retain--forget output distance      & 0.02    & 0.04    \\
\bottomrule
\end{tabular}
\end{table}

\subsection{Per-Class Attack Transferability of \system}\label{appendix-Attack-Transferability}

\begin{figure*}[!t]
    \centering
    {\bf Across Models} 
    \\
    \begin{subfigure}[b]{0.26\textwidth} 
        \includegraphics[width=\textwidth]{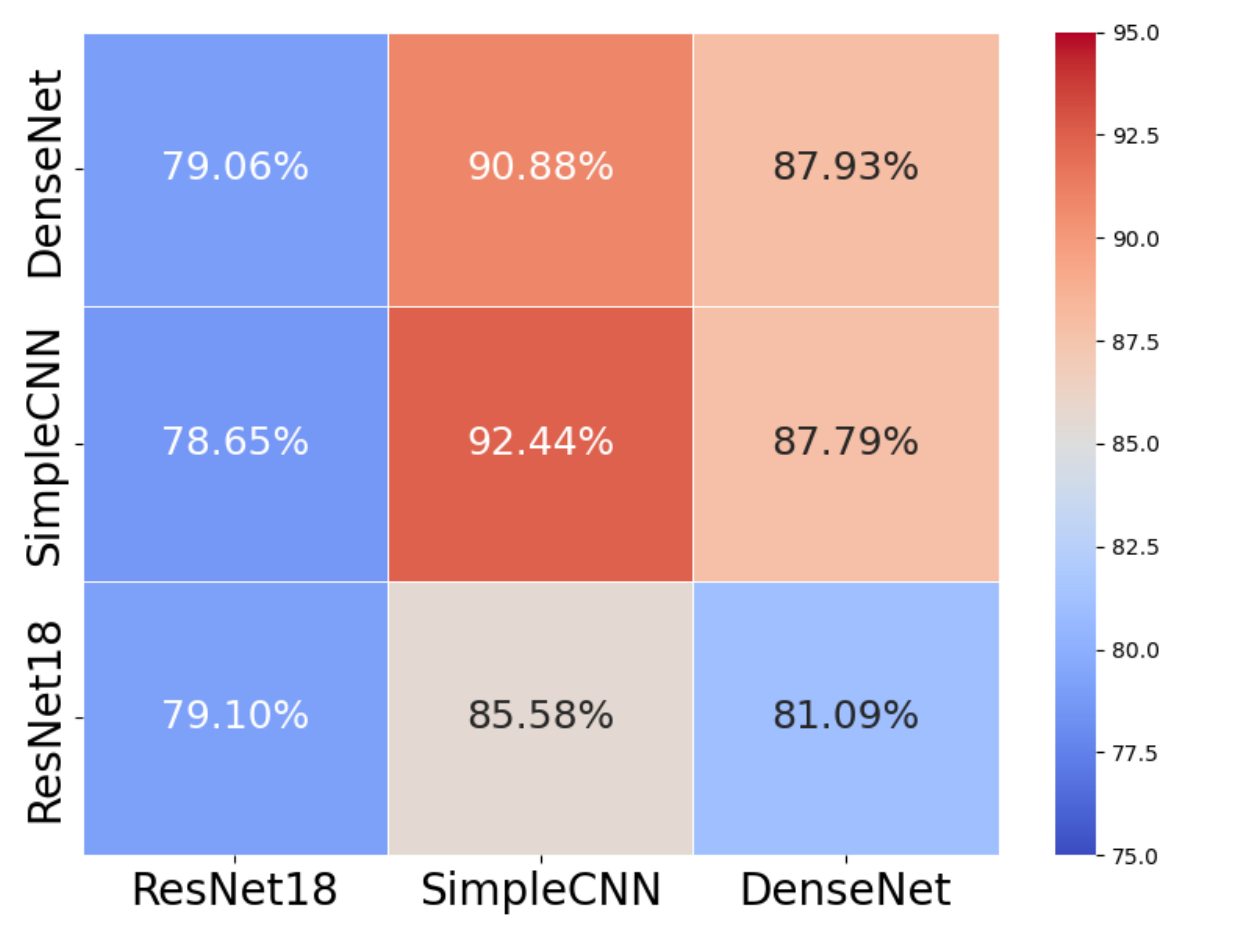} 
        \caption{Unseen set}
        \label{fig:transfer_model_prec_unseen}
    \end{subfigure}
    \begin{subfigure}[b]{0.26\textwidth}
        \includegraphics[width=\textwidth]{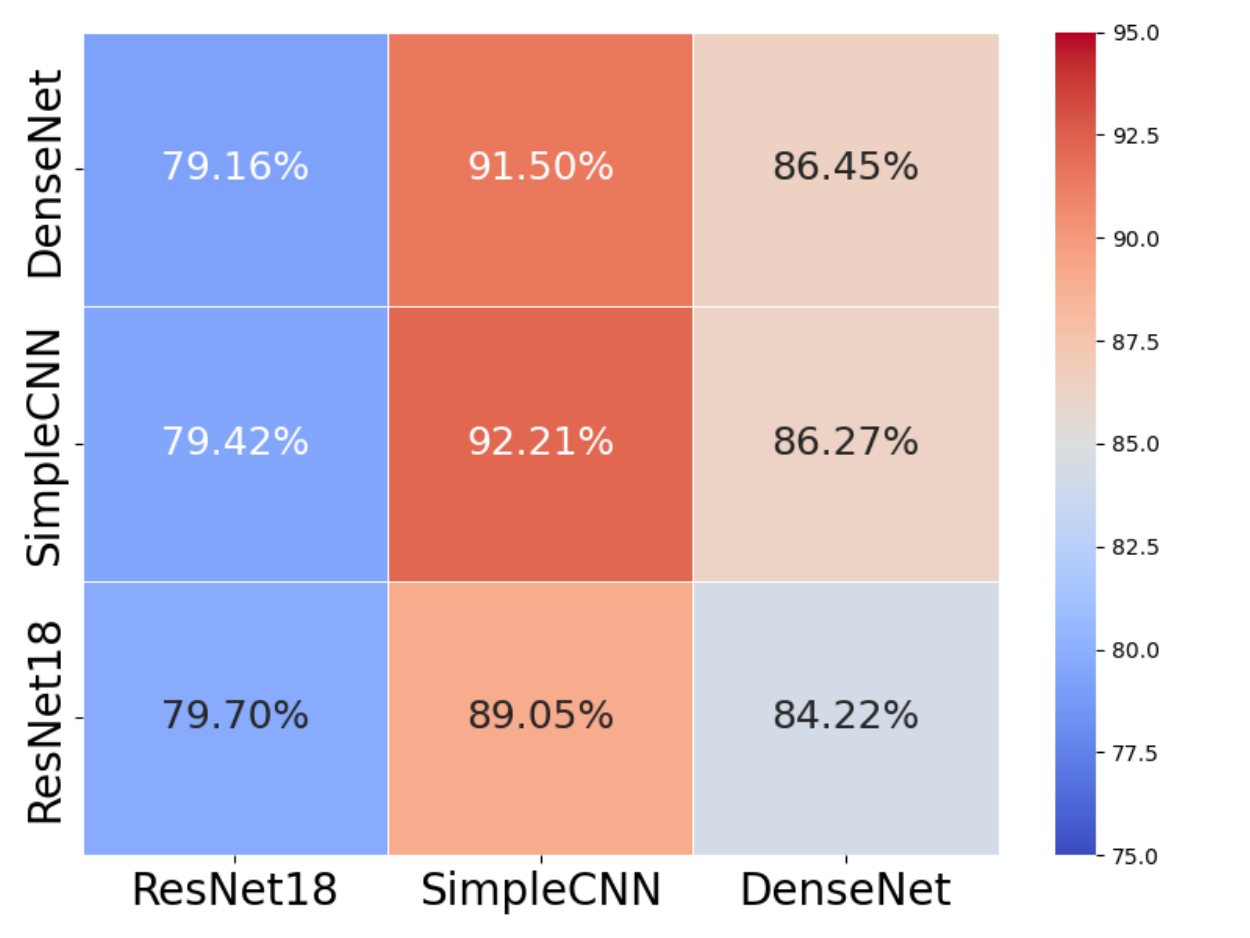} 
        \caption{Forget set}
        \label{fig:transfer_model_prec_forget}
    \end{subfigure}
     \begin{subfigure}[b]{0.26\textwidth}
        \includegraphics[width=\textwidth]{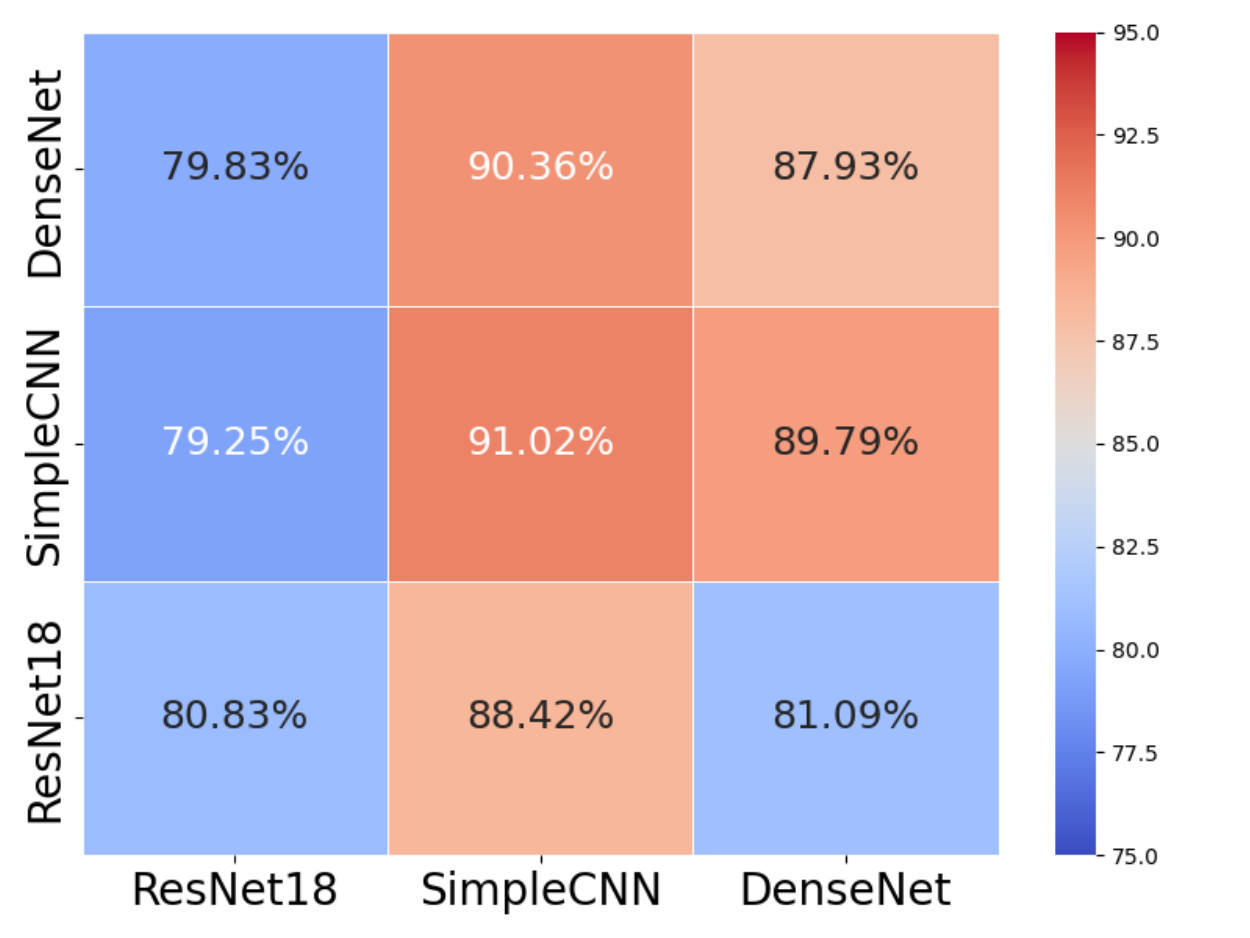}
        \caption{Retain set}
        \label{fig:transfer_model_prec_retain}
    \end{subfigure}
    \\
    {\bf Across Datasets}
    \\
       \begin{subfigure}[b]{0.26\textwidth} 
        \includegraphics[width=\textwidth]{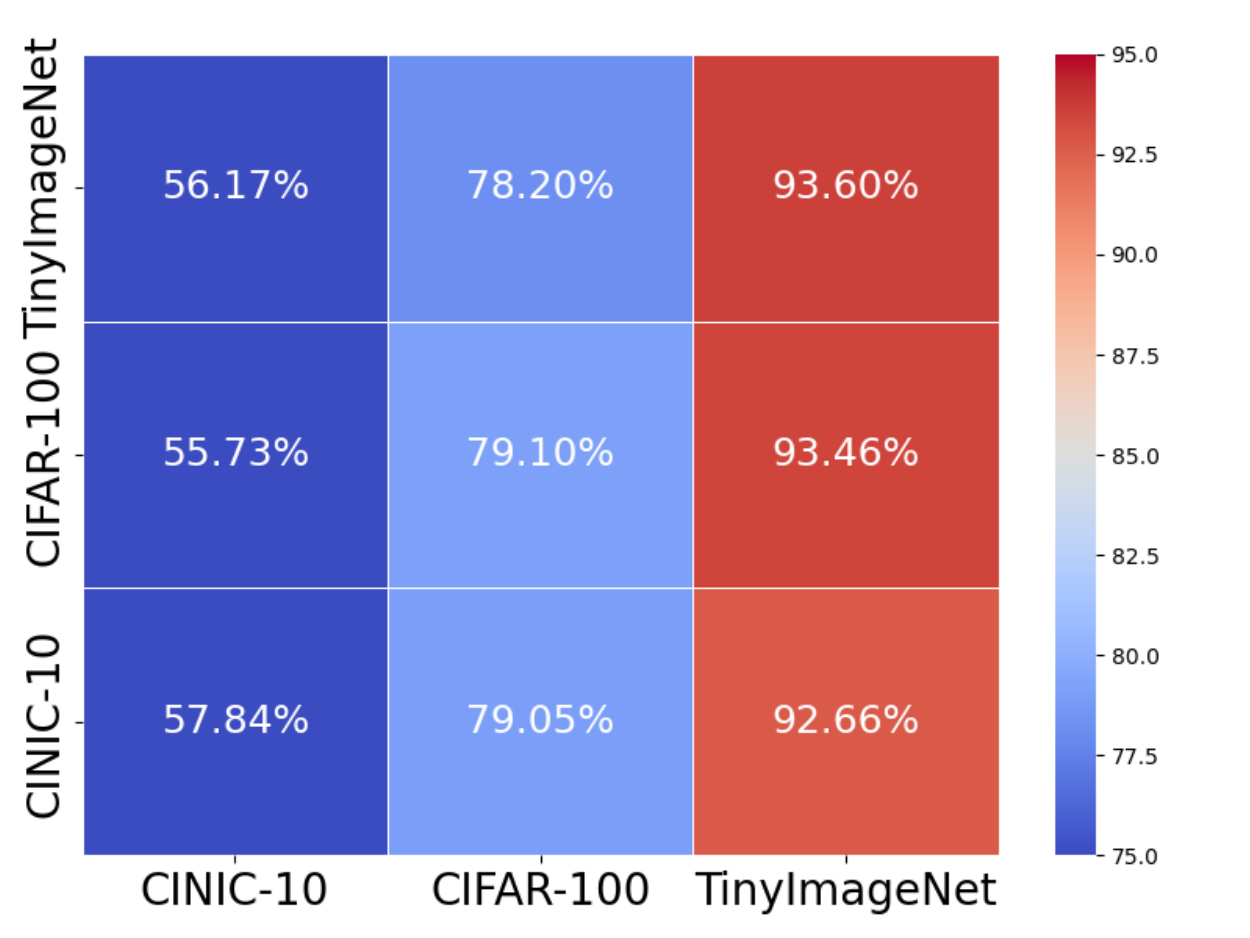} 
        \caption{Unseen set}
        \label{fig:transfer_data_prec_unseen}
    \end{subfigure}
    \begin{subfigure}[b]{0.26\textwidth}
        \includegraphics[width=\textwidth]{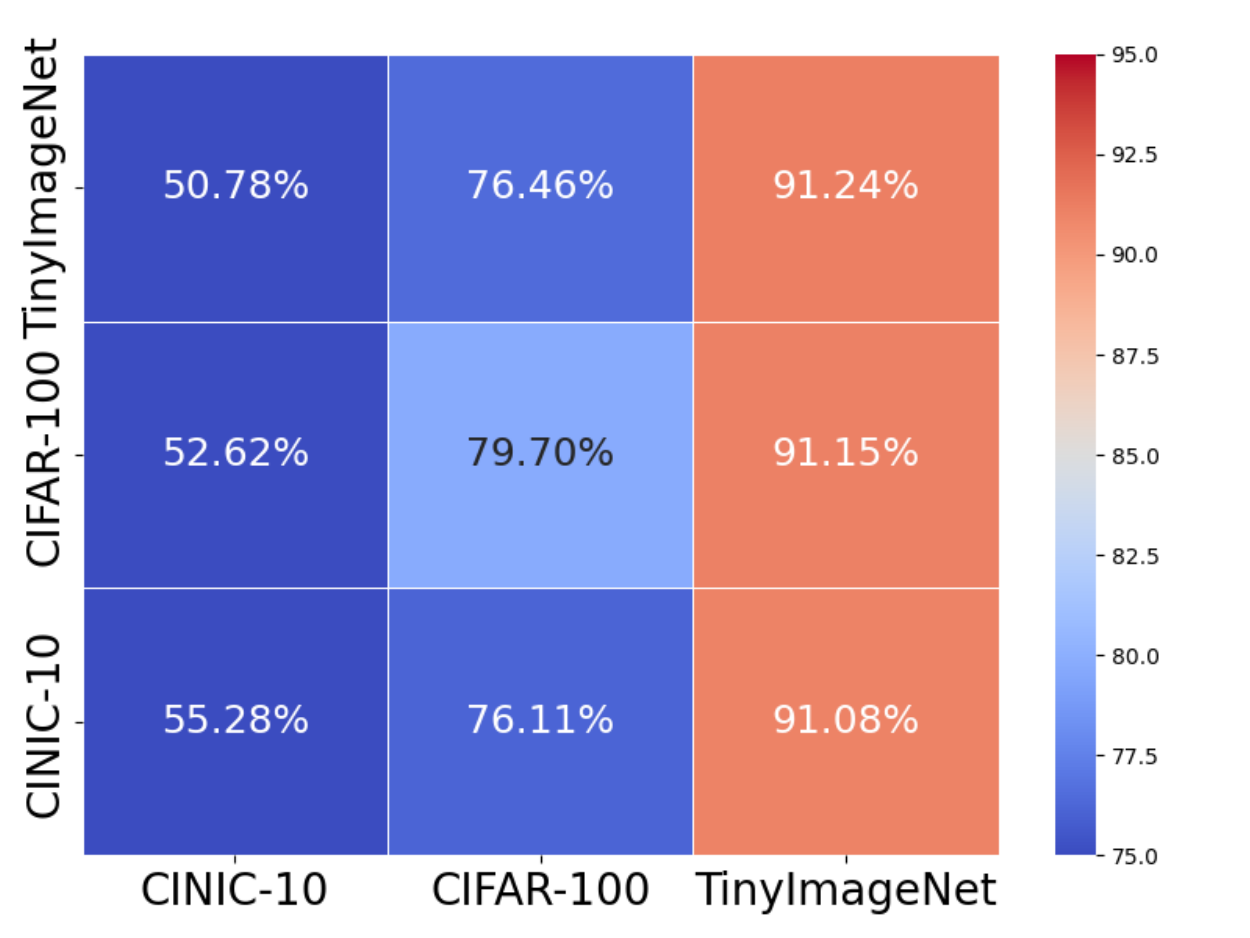} 
        \caption{Forget set}
        \label{fig:transfer_data_prec_forget}
    \end{subfigure}
     \begin{subfigure}[b]{0.26\textwidth}
        \includegraphics[width=\textwidth]{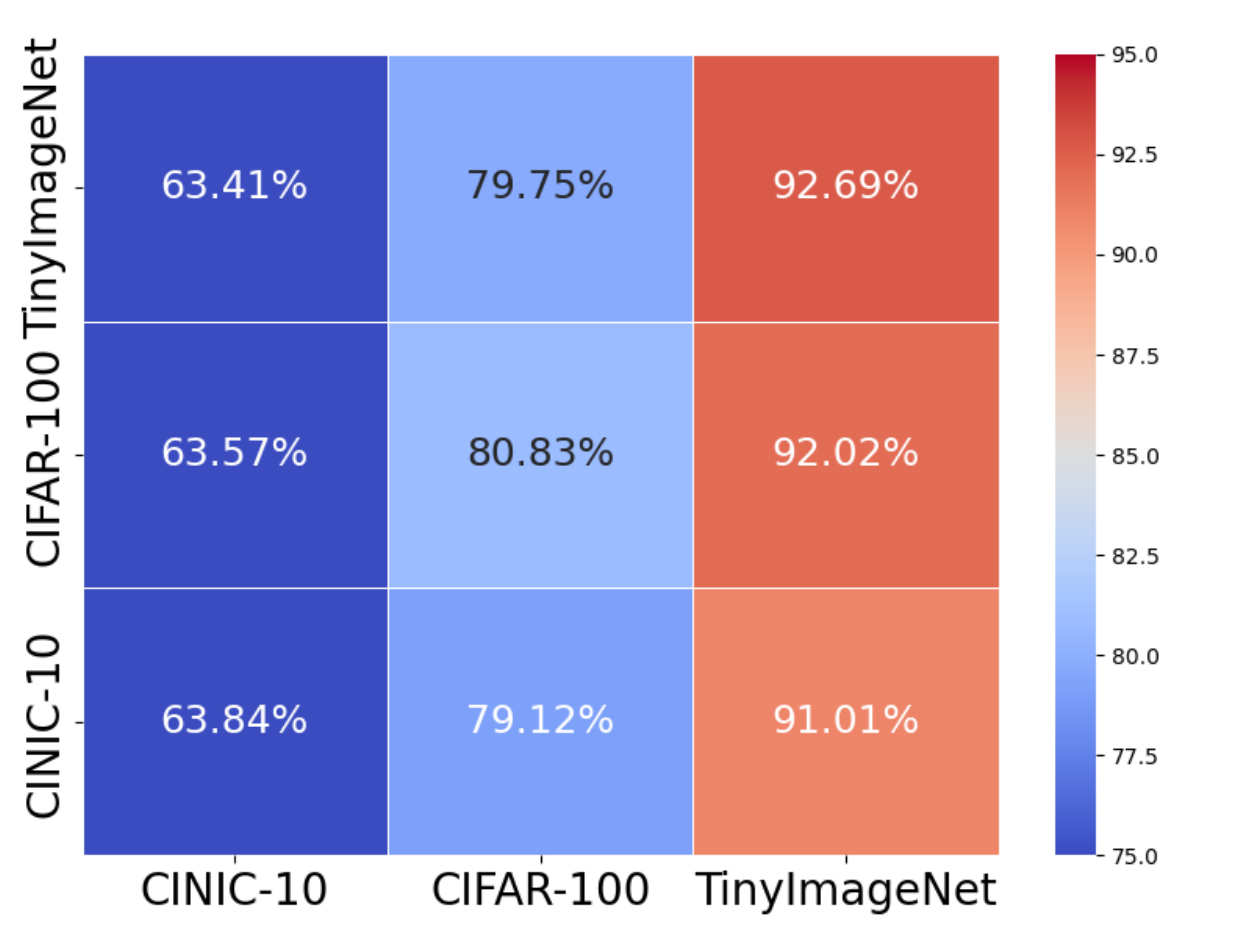}
        \caption{Retain set}
        \label{fig:transfer_data_prec_retain}
    \end{subfigure}
    \\
    {\bf Across Unlearning Methods}
    \\
       \begin{subfigure}[b]{0.26\textwidth} 
        \includegraphics[width=\textwidth]{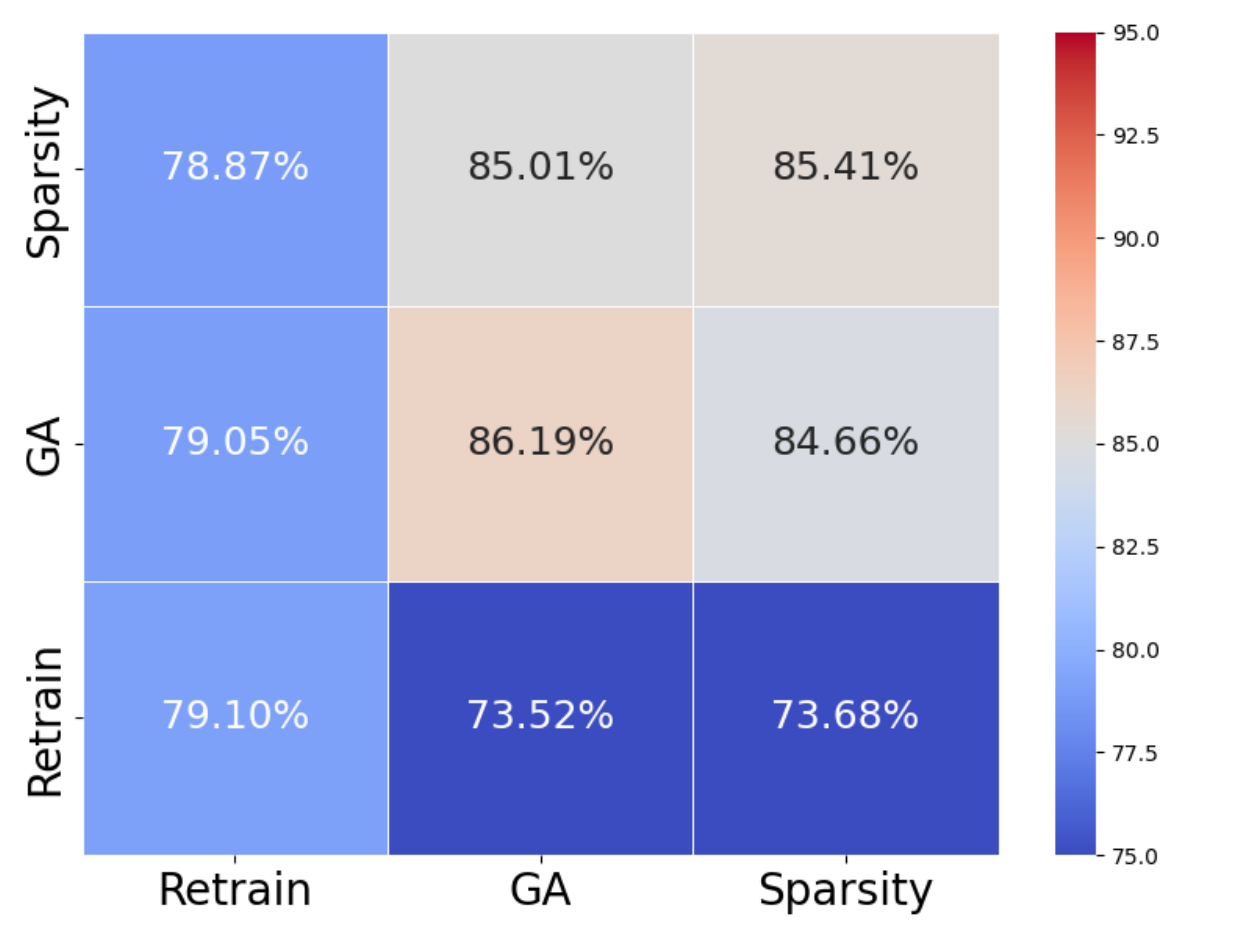} 
        \caption{Unseen set}
        \label{fig:transfer_algo_prec_unseen}
    \end{subfigure}
    \begin{subfigure}[b]{0.26\textwidth}
        \includegraphics[width=\textwidth]{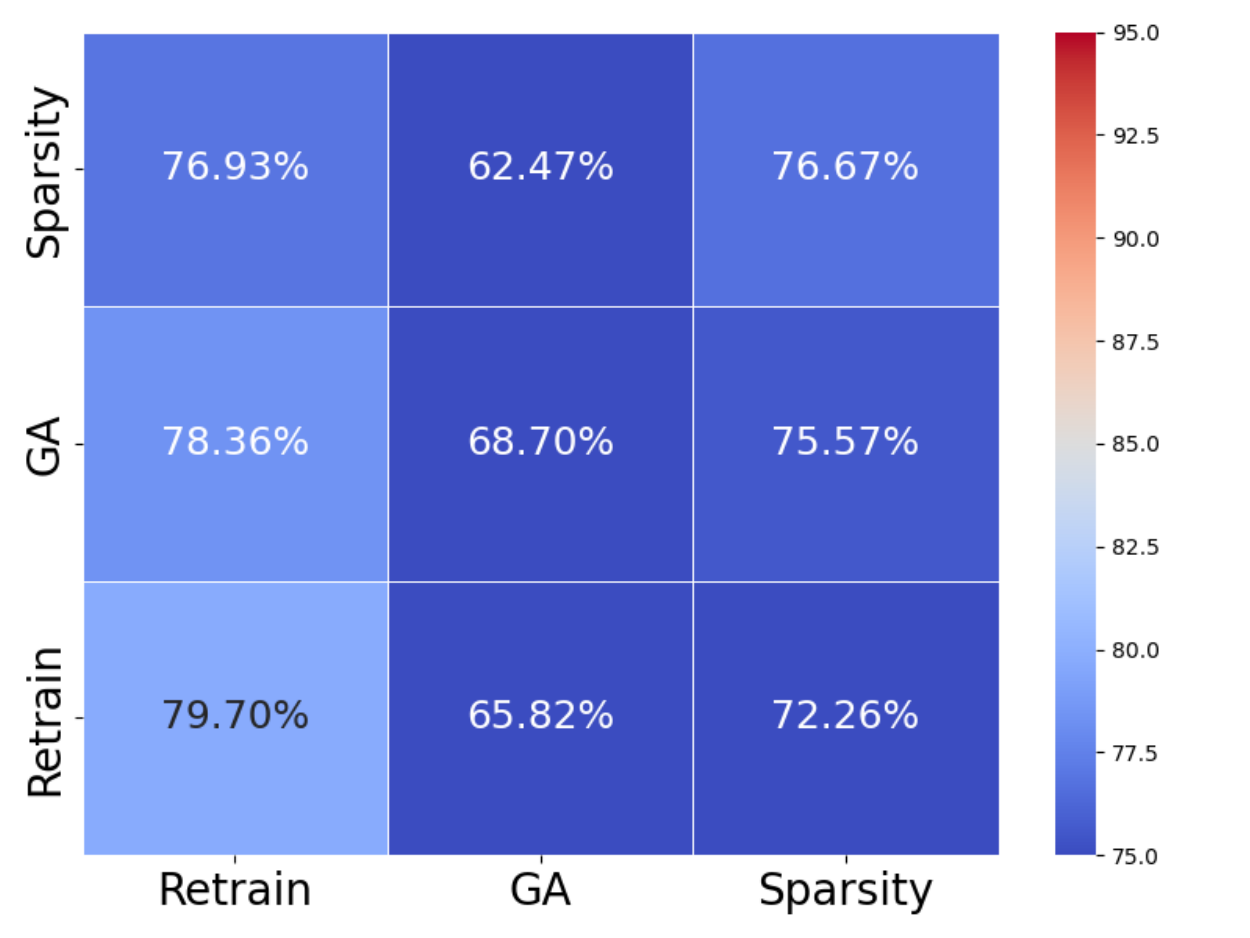} 
        \caption{Forget set}
        \label{fig:transfer_algo_prec_forget}
    \end{subfigure}
     \begin{subfigure}[b]{0.26\textwidth}
        \includegraphics[width=\textwidth]{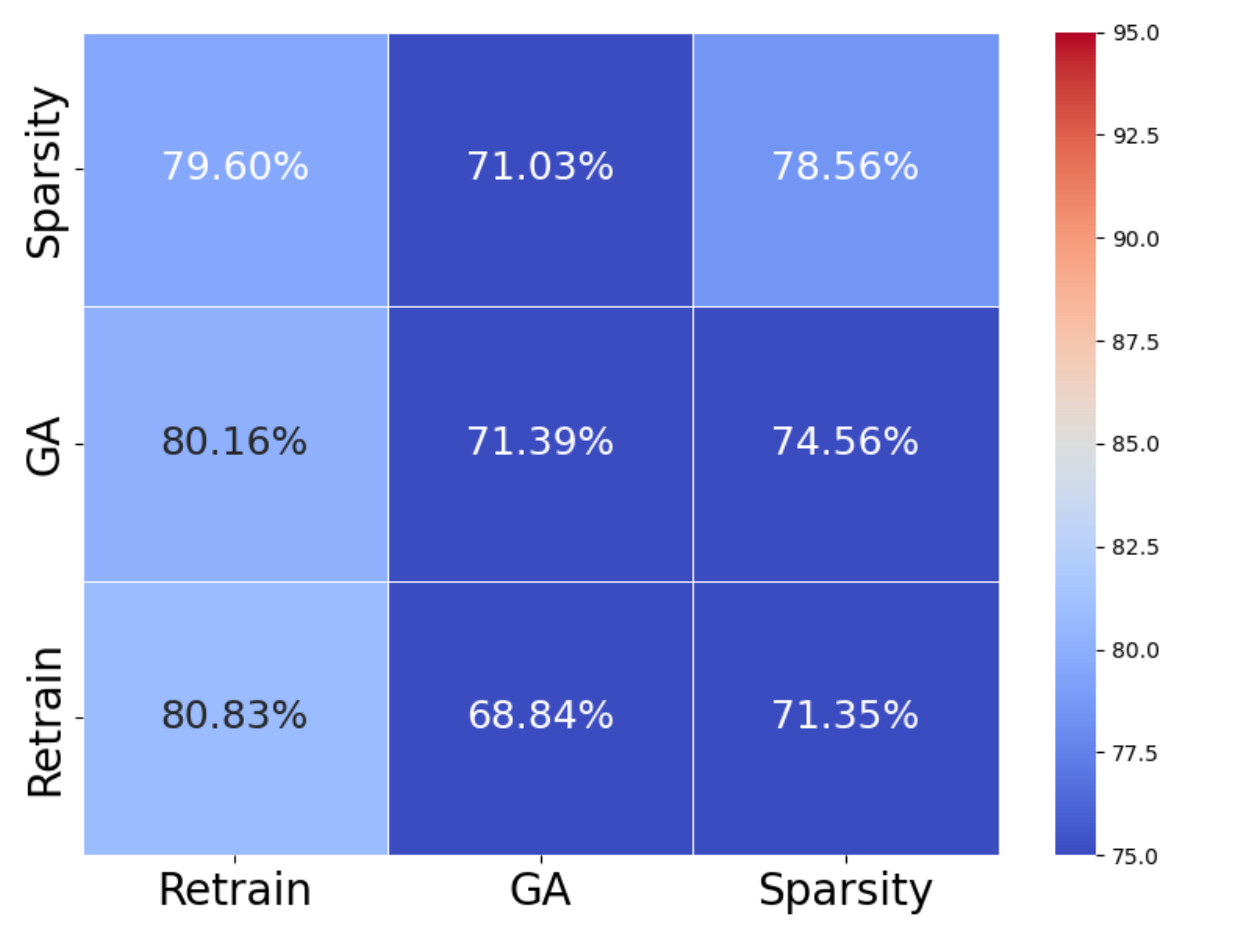}
        \caption{Retain set}
        \label{fig:transfer_algo_prec_retain}
    \end{subfigure}
     \caption{Per-class transferability of \system}
  \label{figure:Attack-transferability-per-class}
\end{figure*}

Figure~\ref{figure:Attack-transferability-per-class} shows the F1-score of three classes under the transfer setting (as described in Section~\ref{subsec-Attack Transferability}). 
The key observation is that \system\ remains effective in per-class when transferred across different model architectures. The same conclusion can be obtained for both settings of transfer datasets and transfer unlearning algorithms. 

\subsection{Ablation Studies: Various Types of Removed Samples}
\label{appendix:removedsample}
Intuitively, the success of \system\ relies on the fact that removing samples leads to disparate amounts of change in the predictions across different classes. For example, the forget set witnesses more significant change in their predictions than the other two sets.  
As removing different samples from the model can lead to different changes in model predictions, we investigate how different characteristics of removed samples affect \system's performance.

\subsubsection{Characteristics of Removed Samples}

We consider three types of characteristics, namely 
{\itshape uncertainty},  {\itshape per-sample privacy vulnerability}, and {\itshape outlierness}, that have been commonly considered for privacy analysis \cite{feldman2020neural,carlini2022membership,onion}. 

\nop{
{\bf Margin to decision boundary.}
Given a classification model $\learningalgo$, the margin of an sample $x$ to the decision boundary of $\learningalgo$ is measured as \cite{pengcheng2018query}:
\[Dis_x=p_{\hat{y}}(x) - p_{\tilde{y}}(x),\]
where $p_{\hat{y}}(x)$ and $p_{\tilde{y}}(x)$ denote the highest and the second highest posterior probability output by $\learningalgo$.  A higher (lower, resp.) $Dis$ value indicates that $x$ is more far away from (closer to, resp.) the decision boundary. 
}

{\bf Uncertainty.} We use the Max Entropy  \cite{berger1996maximum} to measure the {\itshape uncertainty} of the prediction of any given sample $x$: 
$\mathcal{H}(x) = -\sum_{y\in Y}P(y|x)log P(y|x)$,
where $P(y|x)$ is the posterior probability of $y$ given $x$.  A higher entropy value indicates a higher uncertainty of $x$'s prediction.   

{\bf Privacy Vulnerability. } Prior studies have shown that the examples do not have equal vulnerabilities to MIAs  \cite{carlini2022membership,yeom2018privacy,tramer2022truth,feldman2020neural,onion}. We measure the privacy vulnerability of a given sample $x$ as the average attack success rate  (ASR) \cite{onion} for $x$. A higher ASR value indicates that $x$ is more vulnerable to MIAs. 

{\bf Outlierness.} As shown by prior works, the {\itshape outlier} samples are frequently subject to privacy leakage \cite{feldman2020neural,carlini2022membership,onion,leino2020stolen}. Therefore, outlierness is one of the important data characteristics for privacy analysis. We define out-of-distribution (OOD) samples as outliers. To evaluate OOD, we utilize the model's softmax probability distribution and compute its Kullback-Leibler (KL) divergence from a uniform distribution~\cite{hendrycks2017baseline}. Samples with lower KL divergence values are more likely to be outlier samples. 

For both uncertainty and privacy vulnerability, we rank samples based on their measurement values, and pick the top 2\% as well as the bottom 2\% as two different sets of forget set for comparison. For outlierness, we construct a set of outliers that takes 2\% of the training dataset. These outliers are constructed by adding zero-mean Gaussian noise with a variance of 5 to the original images~\cite{hendrycks2017baseline}, so that they are OOD samples. The noise-corrupted images were labeled as outliers, while the original images were treated as inliers. 


\begin{table}[!t]
\footnotesize
\centering
\caption{Effect of the type of removed samples on attack accuracy (F1-score in \%).}
\label{tab:Various Unlearned Samples}
\scalebox{0.9}{
\begin{tabular}{c|c|c|c|c|c}
\hline
\textbf{Dataset}          & \textbf{\begin{tabular}[c]{@{}c@{}}Removed \\ samples\end{tabular}} & \textbf{Overall} & \textbf{Unseen} & \textbf{Forget} & \textbf{Retain} \\ \hline
\multirow{6}{*}{CIFAR-10} & Hi unc.                                                             & 56.89          & 49.70         & 30.09         & 82.14         \\
                          & Lo unc.                                                             & 70.22          & 80.60         & 65.04         & 66.32         \\ \cline{2-6} 
                          & Hi vuln.                                                            & 63.55          & 82.61         & 49.21         & 59.14         \\
                          & Lo  vuln.                                                           & 69.54          & 62.26         & 62.87         & 81.29         \\ \cline{2-6} 
                          & Inlier                                                         & 52.57          & 62.59         & 46.07         & 45.28         \\
                          & Outlier                                                             & 84.53          & 74.77         & 98.43         & 79.56         \\ \hline
\multirow{6}{*}{CINIC-10} & Hi unc.                                                             & 66.42          & 64.03         & 50.22         & 82.14         \\
                          & Lo unc.                                                             & 76.48          & 85.98         & 74.63         & 76.39         \\ \cline{2-6} 
                          & Hi vuln.                                                            & 69.14          & 82.44         & 57.27         & 66.67         \\
                          & Lo  vuln.                                                           & 76.10          & 72.23         & 69.20         & 84.40         \\ \cline{2-6} 
                          & Inlier                                                         & 59.11          & 63.52         & 48.99         & 61.50         \\
                          & Outlier                                                             & 89.48          & 84.44         & 98.04         & 85.17         \\ \hline
\end{tabular}
}
\end{table}

\subsubsection{Key findings} 
Table~\ref{tab:Various Unlearned Samples} presents the micro F1-score of \system\ when removing samples of different characteristics. We have the following findings.

{\bf Uncertainty. } First, \system\ is more effective when removing samples with low uncertainty compared to those with high uncertainty. Specifically, \system\ achieves higher attack accuracy on both the unseen and forget sets, while exhibiting slightly worse performance on the retain set when low-uncertainty instances are removed. 
Such disparate effects across the three sets are likely due to the differing influence these samples have on the model. Low-uncertainty samples (i.e., those the model predicts with high confidence) tend to have a stronger impact on the model's decision boundaries. As a result, their removal introduces more noticeable changes in the model's behavior, particularly for the forget and unseen sets, which aids \system\ in distinguishing these categories more effectively. In contrast, high-uncertainty samples contribute less to the model behaviors, and their removal leads to subtler changes, making it harder for \system\ to infer the membership of these samples.

{\bf Privacy Vulnerability.} First, \system\ is more effective when removing less vulnerable samples than more vulnerable ones. Furthermore, \system\ exhibits disparate effects across the three sets when removing samples of different vulnerabilities. Specifically, removing less vulnerable samples results in a lower attack accuracy on the unseen set, but a higher accuracy on both the forget and retain sets. This trend can be explained by how vulnerability relates to posterior probability behaviors. Less vulnerable samples typically introduce smaller differences in posterior probabilities between the original and unlearning models. When such samples are removed, the resulting changes in model outputs are more subtle, making it harder for \system\ to detect derivations for unseen samples, whose outputs are already less affected by unlearning. On the other hand, as the forget and retain sets consist of samples in the training data, even minor changes due to the removal of less vulnerable samples may still amplify distinctions between them. This enables \system\ to better distinguish between these two sets.

{\bf Outlierness.} \system\ is more effective when removing outliers than inliers in terms of both overall and per-class accuracy. This is because outliers tend to have a more distinct influence on the model's decision boundary, making the change in the posterior probabilities more significant when they are removed. Such significant changes enable \system\ to better distinguish the three sets. In contrast, inliers are typically well-represented by the rest of the training data, so their removal causes subtler changes to the model's behavior, making it harder for \system\ to infer accurately.


\begin{thebibliography}{00}

\bibitem{gdpr}
G.~D.~P. Regulation, ``Gdpr,'' \url{https://gdpr-info.eu/}, 2016.

\bibitem{CCPA}
T.~C. C. P.~A. (CCPA), ``Ccpa,'' \url{https://oag.ca.gov/privacy/ccpa}, 2018.


\bibitem{carlini2022privacy}
N.~Carlini, M.~Jagielski, C.~Zhang, N.~Papernot, A.~Terzis, and F.~Tramèr, 
``The privacy onion effect: Memorization is relative,'' 
Advances in Neural Information Processing Systems, 
vol.~35, pp. 13\,263--13\,276, 2022.

\bibitem{sekhari2021remember}
A.~Sekhari, J.~Acharya, G.~Kamath, and A.~T. Suresh, ``Remember what you want to forget: Algorithms for machine unlearning,'' Advances in Neural Information Processing Systems, vol.~34, pp. 18\,075--18\,086, 2021.

\bibitem{kurmanji2024towards}
M.~Kurmanji, P.~Triantafillou, J.~Hayes, and E.~Triantafillou, ``Towards unbounded machine unlearning,'' Advances in neural information processing systems, vol.~36, 2024.

\bibitem{thudi2022unrolling}
A.~Thudi, G.~Deza, V.~Chandrasekaran, and N.~Papernot, ``Unrolling sgd: Understanding factors influencing machine unlearning,'' in 2022 IEEE 7th European Symposium on Security and Privacy (EuroS\&P).\hskip 1em plus 0.5em minus 0.4em\relax IEEE, 2022, pp. 303--319.

\bibitem{wu2023certified}
K.~Wu, J.~Shen, Y.~Ning, T.~Wang, and W.~H. Wang, ``Certified edge unlearning for graph neural networks,'' in Proceedings of the 29th ACM SIGKDD Conference on Knowledge Discovery and Data Mining, 2023, pp. 2606--2617.

\bibitem{wu2023gif}
J.~Wu, Y.~Yang, Y.~Qian, Y.~Sui, X.~Wang, and X.~He, ``Gif: A general graph unlearning strategy via influence function,'' in Proceedings of the ACM Web Conference 2023, 2023, pp. 651--661.

\bibitem{chen2022recommendation}
C.~Chen, F.~Sun, M.~Zhang, and B.~Ding, ``Recommendation unlearning,'' in Proceedings of the ACM web conference 2022, 2022, pp. 2768--2777.

\bibitem{li2024making}
Y.~Li, C.~Chen, X.~Zheng, J.~Liu, and J.~Wang, ``Making recommender systems forget: Learning and unlearning for erasable recommendation,'' Knowledge-Based Systems, vol. 283, p. 111124, 2024.

\bibitem{liu2025rethinking}
S.~Liu, Y.~Yao, J.~Jia, S.~Casper, N.~Baracaldo, P.~Hase, Y.~Yao, C.~Y. Liu, X.~Xu, H.~Li et~al., ``Rethinking machine unlearning for large language models,'' Nature Machine Intelligence, pp. 1--14, 2025.

\bibitem{yao2024large}
Y.~Yao, X.~Xu, and Y.~Liu, ``Large language model unlearning,'' Advances in Neural Information Processing Systems, vol.~37, pp. 105\,425--105\,475, 2024.

\bibitem{chen2021machine}
M.~Chen, Z.~Zhang, T.~Wang, M.~Backes, M.~Humbert, and Y.~Zhang, ``When machine unlearning jeopardizes privacy,'' in Proceedings of the 2021 ACM SIGSAC conference on computer and communications security, 2021, pp. 896--911.

\bibitem{golatkar2020forgetting}
A.~Golatkar, A.~Achille, and S.~Soatto, ``Forgetting outside the box: Scrubbing deep networks of information accessible from input-output observations,'' in European Conference on Computer Vision.\hskip 1em plus 0.5em minus 0.4em\relax Springer, 2020, pp. 383--398.

\bibitem{hayes2024inexact}
J.~Hayes, I.~Shumailov, E.~Triantafillou, A.~Khalifa, and N.~Papernot, ``Inexact unlearning needs more careful evaluations to avoid a false sense of privacy,'' arXiv preprint arXiv:2403.01218, 2024.

\bibitem{naderloui2025rectifying}
N.~Naderloui, S.~Yan, B.~Wang, J.~Fu, W.~H. Wang, W.~Liu, and Y.~Hong, ``Rectifying privacy and efficacy measurements in machine unlearning: A new inference attack perspective,'' in 34th USENIX security symposium (USENIX Security 25), 2025.

\bibitem{shokri2017membership}
R.~Shokri, M.~Stronati, C.~Song, and V.~Shmatikov, ``Membership inference attacks against machine learning models,'' in 2017 IEEE symposium on security and privacy (SP).\hskip 1em plus 0.5em minus 0.4em\relax IEEE, 2017, pp. 3--18.

\bibitem{salem2018ml}
A.~Salem, Y.~Zhang, M.~Humbert, P.~Berrang, M.~Fritz, and M.~Backes, ``Ml-leaks: Model and data independent membership inference attacks and defenses on machine learning models,'' arXiv preprint arXiv:1806.01246, 2018.

\bibitem{ma2022learn}
Z.~Ma, Y.~Liu, X.~Liu, J.~Liu, J.~Ma, and K.~Ren, ``Learn to forget: Machine unlearning via neuron masking,'' IEEE Transactions on Dependable and Secure Computing, vol.~20, no.~4, pp. 3194--3207, 2022.

\bibitem{graves2021amnesiac}
L.~Graves, V.~Nagisetty, and V.~Ganesh, ``Amnesiac machine learning,'' in Proceedings of the AAAI Conference on Artificial Intelligence, vol.~35, no.~13, 2021, pp. 11\,516--11\,524.

\bibitem{florida2023sb264}
T.~F. Senate, ``Senate bill 264: Interests of foreign countries,'' \url{https://www.flsenate.gov/Session/Bill/2023/264}, 2023.

\bibitem{texas2025sb17}
T.~T. Senate, ``Senate bill 17: Relating to the purchase or acquisition of an interest in real property by certain aliens or foreign entities,'' \url{https://legiscan.com/TX/text/SB17/2025}, 2025.

\bibitem{hu2022membership}
H.~Hu, Z.~Salcic, L.~Sun, G.~Dobbie, P.~S. Yu, and X.~Zhang, ``Membership inference attacks on machine learning: A survey,'' ACM Computing Surveys (CSUR), vol.~54, no. 11s, pp. 1--37, 2022.

\bibitem{carlini2022membership}
N.~Carlini, S.~Chien, M.~Nasr, S.~Song, A.~Terzis, and F.~Tramer, ``Membership inference attacks from first principles,'' in 2022 IEEE Symposium on Security and Privacy (SP).\hskip 1em plus 0.5em minus 0.4em\relax IEEE, 2022, pp. 1897--1914.

\bibitem{long2018understanding}
Y.~Long, V.~Bindschaedler, L.~Wang, D.~Bu, X.~Wang, H.~Tang, C.~A. Gunter, and K.~Chen, ``Understanding membership inferences on well-generalized learning models,'' arXiv preprint arXiv:1802.04889, 2018.

\bibitem{truex2019demystifying}
S.~Truex, L.~Liu, M.~E. Gursoy, L.~Yu, and W.~Wei, ``Demystifying membership inference attacks in machine learning as a service,'' IEEE transactions on services computing, vol.~14, no.~6, pp. 2073--2089, 2019.

\bibitem{bourtoule2021machine}
L.~Bourtoule, V.~Chandrasekaran, C.~A. Choquette-Choo, H.~Jia, A.~Travers, B.~Zhang, D.~Lie, and N.~Papernot, ``Machine unlearning,'' in 2021 IEEE Symposium on Security and Privacy (SP).\hskip 1em plus 0.5em minus 0.4em\relax IEEE, 2021, pp. 141--159.

\bibitem{nguyen2022survey}
T.~T. Nguyen, T.~T. Huynh, Z.~Ren, P.~L. Nguyen, A.~W.-C. Liew, H.~Yin, and Q.~V.~H. Nguyen, ``A survey of machine unlearning,'' arXiv preprint arXiv:2209.02299, 2022.

\bibitem{shaik2024exploring}
T.~Shaik, X.~Tao, H.~Xie, L.~Li, X.~Zhu, and Q.~Li, ``Exploring the landscape of machine unlearning: A comprehensive survey and taxonomy,'' IEEE Transactions on Neural Networks and Learning Systems, 2024.

\bibitem{guo2020certified}
C.~Guo, T.~Goldstein, A.~Hannun, and L.~Van Der~Maaten, ``Certified data removal from machine learning models,'' in Proceedings of the 37th International Conference on Machine Learning, 2020, pp. 3832--3842.

\bibitem{liu2024model}
J.~Liu, P.~Ram, Y.~Yao, G.~Liu, Y.~Liu, P.~Sharma, and S.~Liu, ``Model sparsity can simplify machine unlearning,'' Advances in Neural Information Processing Systems, vol.~36, 2024.

\bibitem{baumhauer2022machine}
T.~Baumhauer, P.~Sch{\"o}ttle, and M.~Zeppelzauer, ``Machine unlearning: Linear filtration for logit-based classifiers,'' Machine Learning, vol. 111, no.~9, pp. 3203--3226, 2022.

\bibitem{liu2022membership}
Y.~Liu, Z.~Zhao, M.~Backes, and Y.~Zhang, ``Membership inference attacks by exploiting loss trajectory,'' in Proceedings of the 2022 ACM SIGSAC Conference on Computer and Communications Security, 2022, pp. 2085--2098.

\bibitem{ye2022enhanced}
J.~Ye, A.~Maddi, S.~K. Murakonda, V.~Bindschaedler, and R.~Shokri, ``Enhanced membership inference attacks against machine learning models,'' in Proceedings of the 2022 ACM SIGSAC Conference on Computer and Communications Security, 2022, pp. 3093--3106.

\bibitem{wen2022canary}
Y.~Wen, A.~Bansal, H.~Kazemi, E.~Borgnia, M.~Goldblum, J.~Geiping, and T.~Goldstein, ``Canary in a coalmine: Better membership inference with ensembled adversarial queries,'' arXiv preprint arXiv:2210.10750, 2022.

\bibitem{zarifzadeh2024low}
S.~Zarifzadeh, P.~Liu, and R.~Shokri, ``Low-cost high-power membership inference attacks,'' in Forty-first International Conference on Machine Learning, 2024.

\bibitem{goel2022towards}
S.~Goel, A.~Prabhu, A.~Sanyal, S.-N. Lim, P.~Torr, and P.~Kumaraguru, ``Towards adversarial evaluations for inexact machine unlearning,'' arXiv preprint arXiv:2201.06640, 2022.

\bibitem{xu2024machine}
J.~Xu, Z.~Wu, C.~Wang, and X.~Jia, ``Machine unlearning: Solutions and challenges,'' IEEE Transactions on Emerging Topics in Computational Intelligence, 2024.

\bibitem{datarobot}
``Datarobot inc.'' \url{https://www.datarobot.com}, 2017.

\bibitem{h2oai}
``H2o.ai ai platform,'' \url{https://h2o.ai}, 2015.

\bibitem{krizhevsky2009learning}
A.~Krizhevsky, G.~Hinton, ``Learning multiple layers of features from tiny images,'' 2009.


\bibitem{darlow2018cinic}
L.~N. Darlow, E.~J. Crowley, A.~Antoniou, and A.~J. Storkey, ``Cinic-10 is not imagenet or cifar-10,'' arXiv preprint arXiv:1810.03505, 2018.

\bibitem{deng2009imagenet}
J.~Deng, W.~Dong, R.~Socher, L.-J. Li, K.~Li, and L.~Fei-Fei, ``Imagenet: A large-scale hierarchical image database,'' in 2009 IEEE conference on computer vision and pattern recognition.\hskip 1em plus 0.5em minus 0.4em\relax Ieee, 2009, pp. 248--255.

\bibitem{socher2013recursive}
R.~Socher, A.~Perelygin, J.~Wu, J.~Chuang, C.~D. Manning, A.~Y. Ng, and C.~Potts, ``Recursive deep models for semantic compositionality over a sentiment treebank,'' in Proceedings of the 2013 conference on empirical methods in natural language processing, 2013, pp. 1631--1642.

\bibitem{albishre2015effective}
K.~Albishre, M.~Albathan, and Y.~Li, ``Effective 20 newsgroups dataset cleaning,'' in 2015 IEEE/WIC/ACM International Conference on Web Intelligence and Intelligent Agent Technology (WI-IAT), vol.~3.\hskip 1em plus 0.5em minus 0.4em\relax IEEE, 2015, pp. 98--101.

\bibitem{huang2017densely}
G.~Huang, Z.~Liu, L.~Van Der~Maaten, and K.~Q. Weinberger, ``Densely connected convolutional networks,'' in Proceedings of the IEEE conference on computer vision and pattern recognition, 2017, pp. 4700--4708.

\bibitem{he2016deep}
K.~He, X.~Zhang, S.~Ren, and J.~Sun, ``Deep residual learning for image recognition,'' in Proceedings of the IEEE conference on computer vision and pattern recognition, 2016, pp. 770--778.

\bibitem{biderman2023pythia}
S.~Biderman, H.~Schoelkopf, Q.~G. Anthony, H.~Bradley, K.~O’Brien, E.~Hallahan, M.~A. Khan, S.~Purohit, U.~S. Prashanth, E.~Raff et~al., ``Pythia: A suite for analyzing large language models across training and scaling,'' in International Conference on Machine Learning.\hskip 1em plus 0.5em minus 0.4em\relax PMLR, 2023, pp. 2397--2430.

\bibitem{golatkar2020eternal}
A.~Golatkar, A.~Achille, and S.~Soatto, ``Eternal sunshine of the spotless net: Selective forgetting in deep networks,'' in Proceedings of the IEEE/CVF Conference on Computer Vision and Pattern Recognition, 2020, pp. 9304--9312.

\bibitem{grandini2020metrics}
M.~Grandini, E.~Bagli, and G.~Visani, ``Metrics for multi-class classification: an overview,'' arXiv preprint arXiv:2008.05756, 2020.

\bibitem{yeom2018privacy}
S.~Yeom, I.~Giacomelli, M.~Fredrikson, and S.~Jha, ``Privacy risk in machine learning: Analyzing the connection to overfitting,'' in 2018 IEEE 31st computer security foundations symposium (CSF).\hskip 1em plus 0.5em minus 0.4em\relax IEEE, 2018, pp. 268--282.

\bibitem{jang2023knowledge}
J.~Jang, D.~Yoon, S.~Yang, S.~Cha, M.~Lee, L.~Logeswaran, and M.~Seo, ``Knowledge unlearning for mitigating privacy risks in language models,'' in 61st Annual Meeting of the Association for Computational Linguistics, ACL 2023.\hskip 1em plus 0.5em minus 0.4em\relax Association for Computational Linguistics (ACL), 2023, pp. 14\,389--14\,408.

\bibitem{ilharco2022editing}
G.~Ilharco, M.~T. Ribeiro, M.~Wortsman, L.~Schmidt, H.~Hajishirzi, and A.~Farhadi, ``Editing models with task arithmetic,'' in The Eleventh International Conference on Learning Representations, 2022.

\bibitem{zhang2024negative}
R.~Zhang, L.~Lin, Y.~Bai, and S.~Mei, ``Negative preference optimization: From catastrophic collapse to effective unlearning,'' arXiv preprint arXiv:2404.05868, 2024.

\bibitem{liu2022continual}
B.~Liu, Q.~Liu, and P.~Stone, ``Continual learning and private unlearning,'' in Conference on Lifelong Learning Agents.\hskip 1em plus 0.5em minus 0.4em\relax PMLR, 2022, pp. 243--254.

\bibitem{maini2024tofu}
P.~Maini, Z.~Feng, A.~Schwarzschild, Z.~C. Lipton, and J.~Z. Kolter, ``Tofu: A task of fictitious unlearning for llms,'' in First Conference on Language Modeling, 2024.

\bibitem{choquette2021label}
C.~A. Choquette-Choo, F.~Tramer, N.~Carlini, and N.~Papernot, ``Label-only membership inference attacks,'' in International conference on machine learning.\hskip 1em plus 0.5em minus 0.4em\relax PMLR, 2021, pp. 1964--1974.

\bibitem{kaya2020effectiveness}
Y.~Kaya, S.~Hong, and T.~Dumitras, ``On the effectiveness of regularization against membership inference attacks,'' arXiv preprint arXiv:2006.05336, 2020.

\bibitem{yin2021defending}
Y.~Yin, K.~Chen, L.~Shou, and G.~Chen, ``Defending privacy against more knowledgeable membership inference attackers,'' in Proceedings of the 27th ACM SIGKDD Conference on Knowledge Discovery \& Data Mining, 2021, pp. 2026--2036.

\bibitem{srivastava2013improving}
N.~Srivastava, ``Improving neural networks with dropout,'' University of Toronto, vol. 182, no. 566, p.~7, 2013.

\bibitem{dwork2006differential}
C.~Dwork, ``Differential privacy,'' in International colloquium on automata, languages, and programming.\hskip 1em plus 0.5em minus 0.4em\relax Springer, 2006, pp. 1--12.

\bibitem{fu2024dpsur}
J.~Fu, Q.~Ye, H.~Hu, Z.~Chen, L.~Wang, K.~Wang, and X.~Ran, ``Dpsur: Accelerating differentially private stochastic gradient descent using selective update and release,'' Proceedings of the VLDB Endowment, vol.~17, no.~6, pp. 1200--1213, 2024.

\bibitem{jia2019memguard}
J.~Jia, A.~Salem, M.~Backes, Y.~Zhang, and N.~Z. Gong, ``Memguard: Defending against black-box membership inference attacks via adversarial examples,'' in Proceedings of the 2019 ACM SIGSAC conference on computer and communications security, 2019, pp. 259--274.

\bibitem{naseri2020local}
M.~Naseri, J.~Hayes, and E.~De~Cristofaro, ``Local and central differential privacy for robustness and privacy in federated learning,'' arXiv preprint arXiv:2009.03561, 2020.

\bibitem{abadi2016deep}
M.~Abadi, A.~Chu, I.~Goodfellow, H.~B. McMahan, I.~Mironov, K.~Talwar, and L.~Zhang, ``Deep learning with differential privacy,'' in Proceedings of the 2016 ACM SIGSAC conference on computer and communications security, 2016, pp. 308--318.

\bibitem{jang2022knowledge}
J.~Jang, D.~Yoon, S.~Yang, S.~Cha, M.~Lee, L.~Logeswaran, and M.~Seo, ``Knowledge unlearning for mitigating privacy risks in language models,'' arXiv preprint arXiv:2210.01504, 2022.

\bibitem{eldan2023s}
R.~Eldan and M.~Russinovich, ``Who's harry potter? approximate unlearning in llms,'' arXiv preprint arXiv:2310.02238, 2023.

\bibitem{pawelczyk2023context}
M.~Pawelczyk, S.~Neel, and H.~Lakkaraju, ``In-context unlearning: Language models as few shot unlearners,'' arXiv preprint arXiv:2310.07579, 2023.

\bibitem{shidetectingiclr24}
W.~Shi, A.~Ajith, M.~Xia, Y.~Huang, D.~Liu, T.~Blevins, D.~Chen, and L.~Zettlemoyer, ``Detecting pretraining data from large language models,'' in The Twelfth International Conference on Learning Representations (ICLR), 2024.

\bibitem{carlini2021extracting}
N.~Carlini, F.~Tramer, E.~Wallace, M.~Jagielski, A.~Herbert-Voss, K.~Lee, A.~Roberts, T.~Brown, D.~Song, U.~Erlingsson et~al., ``Extracting training data from large language models,'' in 30th USENIX security symposium (USENIX Security 21), 2021, pp. 2633--2650.

\bibitem{gupta2022recovering}
S.~Gupta, Y.~Huang, Z.~Zhong, T.~Gao, K.~Li, and D.~Chen, ``Recovering private text in federated learning of language models,'' Advances in neural information processing systems, vol.~35, pp. 8130--8143, 2022.

\bibitem{mattern2023membership}
J.~Mattern, F.~Mireshghallah, Z.~Jin, B.~Sch{\"o}lkopf, M.~Sachan, and T.~Berg-Kirkpatrick, ``Membership inference attacks against language models via neighbourhood comparison,'' arXiv preprint arXiv:2305.18462, 2023.

\bibitem{shi2024muse}
W.~Shi, J.~Lee, Y.~Huang, S.~Malladi, J.~Zhao, A.~Holtzman, D.~Liu, L.~Zettlemoyer, N.~A. Smith, and C.~Zhang, ``Muse: Machine unlearning six-way evaluation for language models,'' arXiv preprint arXiv:2407.06460, 2024.

\bibitem{duanmembership-colm-24}
M.~Duan, A.~Suri, N.~Mireshghallah, S.~Min, W.~Shi, L.~Zettlemoyer, Y.~Tsvetkov, Y.~Choi, D.~Evans, and H.~Hajishirzi, ``Do membership inference attacks work on large language models?'' in Conference on Language Modeling (COLM), 2024.

\bibitem{feldman2020neural}
V.~Feldman and C.~Zhang, ``What neural networks memorize and why: Discovering the long tail via influence estimation,'' Advances in Neural Information Processing Systems, vol.~33, pp. 2881--2891, 2020.

\bibitem{onion}
N.~Carlini, M.~Jagielski, C.~Zhang, N.~Papernot, A.~Terzis, and F.~Tramer, ``The privacy onion effect: Memorization is relative,'' in Advances in Neural Information Processing Systems, S.~Koyejo, S.~Mohamed, A.~Agarwal, D.~Belgrave, K.~Cho, and A.~Oh, Eds., vol.~35, 2022, pp. 13\,263--13\,276.

\bibitem{berger1996maximum}
A.~Berger, S.~A. Della~Pietra, and V.~J. Della~Pietra, ``A maximum entropy approach to natural language processing,'' Computational linguistics, vol.~22, no.~1, pp. 39--71, 1996.

\bibitem{tramer2022truth}
F.~Tram{\`e}r, R.~Shokri, A.~San~Joaquin, H.~Le, M.~Jagielski, S.~Hong, and N.~Carlini, ``Truth serum: Poisoning machine learning models to reveal their secrets,'' in Proceedings of the 2022 ACM SIGSAC Conference on Computer and Communications Security, 2022, pp. 2779--2792.

\bibitem{leino2020stolen}
K.~Leino and M.~Fredrikson, ``Stolen memories: Leveraging model memorization for calibrated white-box membership inference,'' in 29th USENIX security symposium (USENIX Security 20), 2020, pp. 1605--1622.

\bibitem{hendrycks2017baseline}
D.~Hendrycks and K.~Gimpel, ``A baseline for detecting misclassified and out-of-distribution examples in neural networks,'' in International Conference on Learning Representations, 2017.

\bibitem{han2024parameter}
Z.~Han, C.~Gao, J.~Liu, J.~Zhang, and S.~Q.~Zhang, ``Parameter-efficient fine-tuning for large models: A comprehensive survey,'' \textit{arXiv preprint arXiv:2403.14608}, 2024


\end{thebibliography}
\end{document}